\UseRawInputEncoding

\documentclass[aip,graphicx]{revtex4-1}

\draft 

\usepackage{graphicx}
\usepackage{dcolumn}
\usepackage{bm}

\usepackage[utf8]{inputenc}
\usepackage[T1]{fontenc}
\usepackage{mathptmx}
\usepackage{etoolbox}
\usepackage[dvipsnames]{xcolor}
\usepackage{makecell}

\begin{document}


\title{Quantum metaphotonics: recent advances and perspective}

\author{Jihua Zhang}
\email{zhangjihua@sslab.org.cn}
\affiliation{Songshan Lake Materials Laboratory, Dongguan, Guangdong 523808, P. R. China}
\affiliation{Research School of Physics, The Australian National University, Canberra, ACT 2601, Australia}
 
\author{Yuri Kivshar}%
\email{yuri.kivshar@anu.edu.au}
\affiliation{Research School of Physics, The Australian National University, Canberra, ACT 2601, Australia}


\date{\today}

\begin{abstract}
Quantum metaphotonics has emerged as a cutting-edge subfield of meta-optics employing subwavelength resonators and their planar structures such as metasurfaces to generate, manipulate, and detect quantum states of light. It holds a great potential for the miniaturization of current bulky quantum optical elements by developing a design of on-chip quantum systems for various applications of quantum technologies. Over the past few years, this field has witnessed a surge of intriguing theoretical ideas, groundbreaking experiments, and novel application proposals. This perspective paper aims to summarize the most recent advancements and also provide a perspective on the further progress in this rapidly developing field of research. 
\end{abstract}

\pacs{}

\maketitle 

\section{Introduction}

Recent development of advanced photonic technologies underpins the rapid progress in efficient generation, manipulation, and detection of quantum states. One of the breakthroughs of modern photonics is the development of compact subwavelength-thick optical structures such as {\it optical metasurfaces}. Metasurface physics can be viewed as the realisation of a broader field of {\it metaphotonics} that enrich traditional optics with new concepts and functionalities provided by metamaterials. Being seemingly close to the two-dimensional realisation of metamaterials, metasurface physics brings many novel physical concepts~\cite{yuri2021,fan2022EmergingTrend} and provides a planar platform to the design of compact, multifunctional, broadband, and fast optical components with tiny thickness, giving a birth to a new generation of achromatic metalenses, diffractive elements, filters, polarization converters, and many others. Some years ago, it became clear that these achievements can be useful for the functionality and scalability required by large-scale quantum information processing. The use of subwavelength nanostructures can dramatically speed up quantum photonic processes such as the spontaneous emission rate to enable high-speed single photon sources and overcome the rate of decoherence processes~\cite{wang2022MetasurfacesQuantum}. 

Advanced concepts of metamaterials applied to quantum technologies hold great promise for the development of quantum metaphotonics. They offer the dramatic enhancement
of single-photon emission from solid-state quantum emitters. The dramatic speed-up of the spontaneous emission may also
allow quantum decoherence to be overcome and the generation of indistinguishable photons even outside of cryostats. The use of resonant dielectric structures with low losses would allow to open full potential of meta-optics for applications in the quantum domain.

This progress has been recognized in many recent publications being summarized in several review papers on this topic~\cite{solntsev2021MetasurfacesQuantum, liu2021QuantumPhotonics, wang2022MetasurfacesQuantum, sharapova2023NonlinearDielectric, kan2023AdvancesMetaphotonics, ding2023AdvancesQuantum, xu2023MetasurfacesOptical, parry2024ChapterNonlineara}. However, the potential of quantum metaphotonics and quantum metasurfaces is still being explored, and ongoing research continues to unlock new functionalities and applications.  Over the past few years, we observe rapid developments of intriguing theoretical ideas, realisations of groundbreaking experiments, and proposals for novel applications. 

The use of metaphotonic structures is expected to move the generation of quantum states of light to a new level. For quantum communication, metasurfaces can provide the states with enormous information capacity and fast switching capabilities. For quantum imaging, unprecedentedly tight photon-photon correlations will lead to the realisation of super-resolution effects. Both quantum spectroscopy and quantum sensing may benefit from the availability of states with practically unlimited spectral bandwidth. On the other hand, resonant metasurfaces may produce very narrowband quantum states for coupling to other quantum systems. Finally, the research on quantum metaphotonic is expected to stimulate other fields like material science and quantum information.

\begin{figure}
    \includegraphics[width=0.8\linewidth]{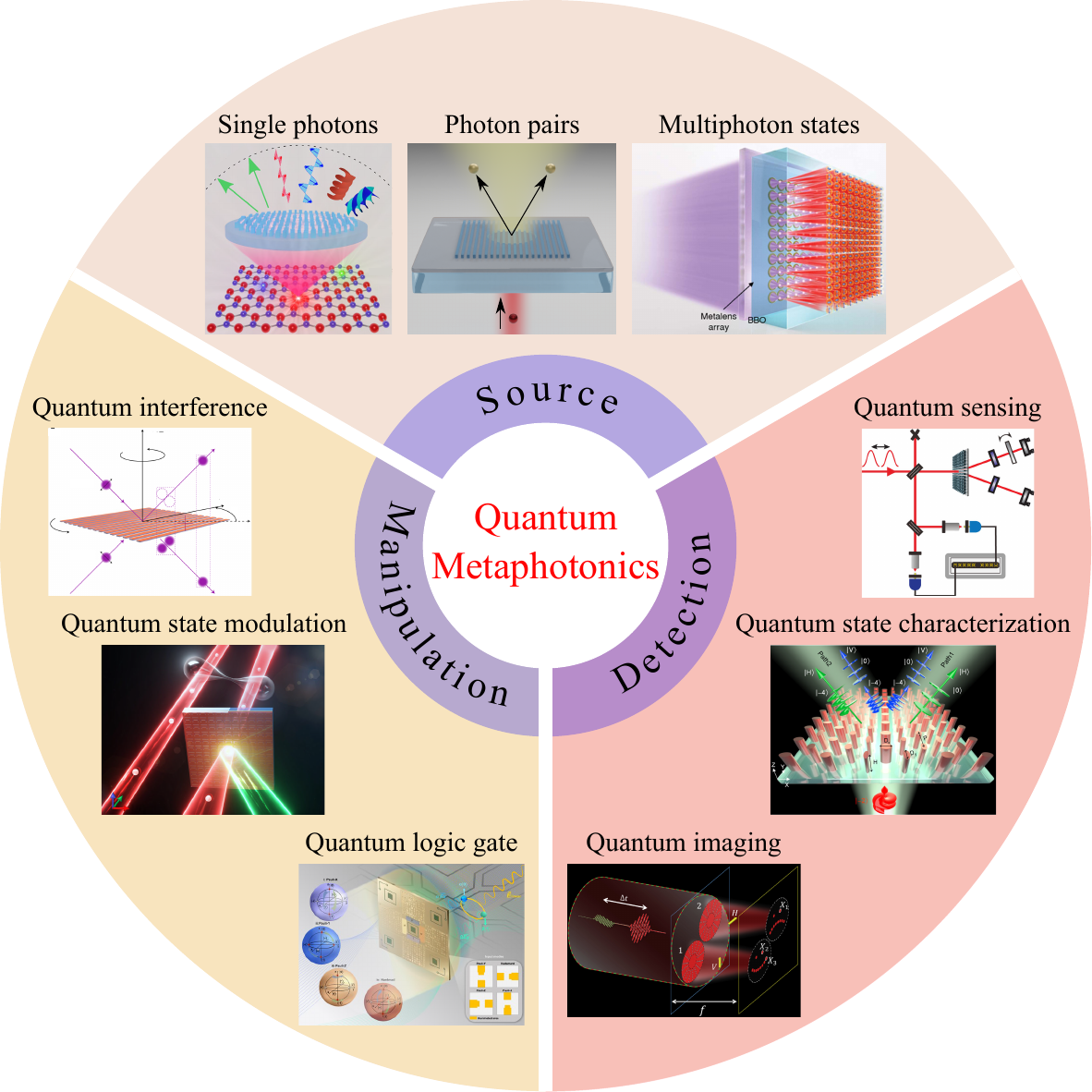}
    \caption{{\bf Functionalities of quantum metaphotonics}. Top: Quantum light sources: examples of single-photon source~\cite{li2023ArbitrarilyStructured}, two-photon source~\cite{zhang2022SpatiallyEntangled}, and multi-photon source~\cite{li2020MetalensarrayBased}, respectively. Left: Quantum light manipulation: interference~\cite{li2021NonunitaryMetasurface}, state modulation~\cite{zhang2022AllopticalModulation}, and logic gates~\cite{ding2023MetasurfacebasedOptical}. Right: Quantum light detection: sensing~\cite{georgi2019MetasurfaceInterferometry}, state characterization~\cite{wang2023CharacterizationOrbital}, and imaging~\cite{yung2022PolarizationCoincidence}. 
    Li \textit{et al.} eLight 3, 19 (2023); licensed under a Creative Commons Attribution (CC BY) license. 
    Zhang \textit{et al.} Sci. Adv. 8 eabq4240 (2022); licensed under a Creative Commons Attribution NonCommercial License 4.0 (CC BY-NC).
    Reproduced from Li \textit{et al.} Science 368, 1487 (2020); Copyright 2020 AAAS.
    Reproduced with permission from Li \textit{et al.} Nat. Photon. 15, 267 (2021). Copyright 2021 Springer Nature.
    Zhang \textit{et al.} Light Sci. Appl. 11, 58 (2022); licensed under a Creative Commons Attribution (CC BY) license.
    Ding \textit{et al.} Adv. Mater. 36, 2308993 (2023); Copyright 2023 John Wiley and Sons.
    Georgi \textit{et al.} Light Sci Appl. 8, 70 (2019); licensed under a Creative Commons Attribution (CC BY) license. 
    Reproduced with permission from Wang \textit{et al.} Nano Lett. 23, 3921 (2023). Copyright 2023 American Chemical Society.
    Yung \textit{et al.} iScience 25, 104155 (2022); licensed under a Creative Commons Attribution (CC BY) license.
    }
    \label{fig1:general}
\end{figure}

This {\em Perspective} aims to summarize the most recent advancements and provide our personal view on the further development of this rapidly developing field of research. We shape our presentation of the most recent results around three major areas driven by selective functionalities of quantum metasurfaces and quantum metaphotonics elements, as depicted in  Fig.~\ref{fig1:general}.  First, we discuss the use of flat optics for quantum light sources, including single-photon sources created by integrating metasurfaces with quantum emitters~\cite{li2023ArbitrarilyStructured}, two-photon sources, by employing nonlinear metasurfaces~\cite{zhang2022SpatiallyEntangled}, and multi-photon sources, by integrating a metalens array with a nonlinear crystal~\cite{li2020MetalensarrayBased}. Next, we discuss the use of metasurfaces for quantum light manipulation, including but not limited to 
quantum interference~\cite{li2021NonunitaryMetasurface}, quantum state modulation~\cite{zhang2022AllopticalModulation}, and quantum logic gates~\cite{ding2023MetasurfacebasedOptical}. Next, we overview several ideas for the use of metasurfaces for quantum light detection including quantum sensing~\cite{georgi2019MetasurfaceInterferometry}, quantum state characterization~\cite{wang2023CharacterizationOrbital}, and quantum imaging~\cite{yung2022PolarizationCoincidence}.
Accordingly, Secs. II-IV are devoted to the discussions of results in these three major directions. Finally, in Sec.~V we provide our perspective on the anticipated future developments in this rapidly growing field.

\section{Metaphotonics for quantum light sources}
In quantum photonics, non-classical light sources are needed for the implementation of quantum protocols and algorithms. These sources include the most commonly used single-photon and two-photon states, as well as more complicated multi-photon, squeezing, and cluster states. Most of these source states have been generated through quantum spontaneous transitions, such as single photons by spontaneous emission (SE) from a single quantum emitter (QE) and photon pairs by spontaneous parametric down-conversion (SPDC) or spontaneous four-wave mixing from nonlinear materials. It is well understood that the rate of spontaneous transition can be effectively controlled by the photonic environment through the Purcell effect~\cite{lodahl2015InterfacingSingle, pelton2015ModifiedSpontaneousa, qian2021SpontaneousEmission}. 
However, the accelerated photon emission may go to unwanted channels such as non-radiative absorption, surface and guided modes, or radiation of photons with unusable properties in polarization, direction, and orbital angular momentum. Fortunately, it is possible to use the same photonic structure to route the emitted photons into the desired channels. 
Based on this fundamental principle, metaphotonics have been widely utilized to develop miniaturized and integrated quantum light sources with enhanced emission rate to well-defined modes in different degrees of freedoms such as wavelength, polarization, and wavefront.

In a general figure, for a spontaneous transition process (e.g. SE from a QE and SPDC from a nonlinear material) in a photonic environment, both its transition rate/probability in the near field and emission profiles in the far field are determined by the Green function supported by the photonic structure. The Green function $\textbf{G}(\textbf{\textit{r}},\textbf{\textit{r}}_0)$ quantifies the electric field at a position $\textbf{\textit{r}}$ emitted by a dipole at another position $\textbf{\textit{r}}_0$. According to the Fermi's golden rule, the SE rate is proportional to the local density of states (LDOS), which is determined by the imaginary part of the trace of the Green tensor at the same location of the QE~\cite{barnes2020ClassicalAntennas}. On the other hand, a QE can be considered as a dipole with a strength proportional to the pump field at the position of the QE. Obviously, the far field emitted by a QE in a photonic structure is directly governed by the Green function supported by the structure. Interestingly, based on the quantum classical correspondence theory~\cite{poddubny2016GenerationPhotonPlasmon}, SPDC can be treated as an extended dipole array in the nonlinear region at two daughter (signal and idler) photon frequencies, and their joint dipole strength or SPDC generation matrix is proportional to the product of the pump field and nonlinear coefficient. Following this, the far-field two-photon transition amplitude of SPDC can be explicitly expressed by the integration in the nonlinear region over the product of two Green functions at the signal and idler frequencies with their joint dipole strength~\cite{poddubny2016GenerationPhotonPlasmon}. Finally, the pump field distribution in the photonic structures, which determines the dipole strength at the emission frequencies, depends on the external source profile and the Green function at the pump frequency due to the reciprocity of the Green function.
Therefore, the key problem in a photonic structure engineered QE and SPDC quantum light sources is to design the Green function. Metaphotonics, which employs single nanoparticle, finite-size cluster of properly designed nanoparticles, and infinite-size periodic planar array of nanoparticles and clusters, have the full degree of freedom to design the photonic structures in the 2D space with subwavelength resolution and thus engineer the Green function with unprecedented capability. With other advantages such as ultracompact size and ease of fabrication and integration, metaphotonics empowered QE single-photon source and SPDC photon-pair source have recently attracted intensive research (Fig. \ref{fig1:general})~\cite{kan2023AdvancesMetaphotonics, sharapova2023NonlinearDielectric, parry2024ChapterNonlinear}.

It is worthy noting that except some regular shaped structures such as multilayer thin films and sphere, most photonic structures and almost all metaphotonic structures have no analytical form of Green function. In these structures, the Green function can only be calculated by numerical simulations. An efficient and powerful way to accomplish this is based on the quasi-normal modes (QNMs) supported by the photonic structures. In the QNM theory, the Green function is simply related to the tensor product of the near and far fields of each QNM, and the frequency detuning of the eigenfrequency of the QNM with respect to the frequency of interest~\cite{lalanne2018LightInteraction}. With this theory and tool, one can transfer the design goal from the complex Green function to the simple QNMs through the complex eigenfrequency of the QNMs and their electric field distributions in both the near and far fields. Both quantities can be easily obtained through an eigenfrequency study of the structure. It has been employed to model the SPDC from single nanoparticles~\cite{weissflog2021ModellingPhotonpair, weissflog2024NonlinearNanoresonators}. Note that, the QNM method can not only predict the behaviour of the metaphotonics quantum light source, but also tell you the contribution from each QNM and thus provide a clear physical insight of the enhanced quantum light source. By designing the dominant QNMs and their weighted sum (i.e. interferences), metaphotonics is able to enhance the emission rate of QEs or SPDC in the near field and in the same time control the far-field emission profiles in different degrees of freedom. 


\subsection{Single-photon sources}
Solid-state QEs such as quantum dots, molecules, color centers in diamond, and defects in 2D materials are natural source of single photons~\cite{aharonovich2016SolidstateSinglephoton,ristori2023SinglePhoton}. Under excitation by a pump laser whose photon energy is larger than the transition energy of the QE, the electron transits from the ground state to the excited state. Within a certain amount of time, the electron goes back to the ground state and the QE has a chance to emit a photon through the SE which is caused by the interaction with the quantum vacuum state (i.e. zero photon Fock state $|0\rangle$). In the time window between the excitation of the electron and its going back to the ground state, the QE cannot absorb pump photons and thus cannot emit photons. Therefore, the radiative lifetime of SE intrinsically determines the maximum photon emission rate of a QE. As mentioned above, photonic structures, e.g. metasurfaces (MSs), are able to reduce this lifetime by increasing the LDOS at the position of the QE and improve the SE rate. 
On top of the rate enhancement, MSs can simultaneously increase the radiative emission to the far field through a specific radiation mode with pre-defined direction, polarization, or orbital angular momentum. 

One efficient way to increase the SE rate is to put the QE close to a metal surface by exciting the high-$k$ surface plasmon polariton (SPP) mode propagating along the metal surface. However, such SPP mode is non-radiative. MSs realized by fabrication of nanoholes in the metal or dielectric nanostructures on top of the metal can couple the SPP into far-field radiation. Importantly, with the capability of MSs, one can choose to couple into specific polarization, direction, and orbital angular momentum by judicious design and arrangement of meta-structures. Another advantage of such MSs lies in that the unidirectional emission above the metal surface is guaranteed, which increases the collection efficiency. In this direction, the group of S.~Bozhevolnyi and their collaborators have reported a series of experimental work by integrating nanodiamonds with color centers with dielectric nanostructures forming MS atop metal (e.g. gold, silver) films supporting the SPP propagation, leading to the directional generation of single photons with desired properties such as circular polarization~\cite{kan2020MetasurfaceEnabledGeneration}, radial polarization~\cite{komisar2021GenerationRadially}, circular polarization with orbital angular momentum~\cite{wu2022RoomtemperatureOnchip, liu2023OnchipGeneration}, linear polarization with orbital angular momentum~\cite{liu2023UltracompactSinglePhoton}. In their latest work, a multichannel signal-photon emission with control on both direction and polarization is demonstrated by embedding a nanodiamond with single Germanium vacancy center into a holographic MS which sits on a silver substrate with a thin SiO$_2$ spacer layer~\cite{komisar2023MultipleChannelling}. Efficient generation of two well-collimated (divergences $< 6.5^\circ$) single-photon beams at 602 nm propagating along different $15^\circ$ off-normal directions and featuring orthogonal linear polarizations is realized (Fig.\ref{fig2:source}a). The two-beam single photons show a second-order correlation function $g^{(2)}\sim0.1$ and an external quantum efficiency over 80\%. In another interesting experiment, S. Jia {\it et al.} demonstrated multichannel single-photon emissions from CdSe colloidal QDs with independent control on spin angular momentum and linear momentum (i.e. direction) by an anisotropic MS made by specially arranged nano-grooves etched into a metal film~\cite{jia2023MultichannelSinglePhoton}.

In most cases, precise positioning of the QE with respect to the metaphotonic structures is needed in the QE-meta-optics system in order to obtain the optimized performance. This is typically done by firstly locating the position of the QE before the nanofabrication of the meta-structures. Recently, Z. Xue {\it et al.} proposed a scalar-superposition MS supporting high-robust placement of the QE in tailoring the polarization of a QE~\cite{xue2022ScalarSuperpositionMetasurfaces}. The metasurface is formed by nano-scatters which are nano-holes etched on a metal substrate, as shown in Fig.\ref{fig2:source}b. The emitted single photons by the QE are firstly coupled to the SPP mode on the metal surface. The SPPs are then scattered to the far field by the scattering units, which are properly designed to support far-field scattering with the same polarization state. The total far-field scattered light will be the interference of all scattered fields and have the common polarization state of all scattering units. The most interesting part lies in that this polarized single photon emission is robust to the position of the QE on the metal surface. The requirement on the placement accuracy of the QE is released to three times of the wavelength in this work. Note that before this work, a similar idea of increasing the placement robustness of QE with an in-plane dipole was proposed and demonstrated by the simple thin-film meta-structures~\cite{checcucci2017BeamingLight}.

We would like to mention that current QEs can already support photon emission rate on the order of megahertz without Purcell acceleration, which is sufficient enough for many quantum applications. In this case, routing the emitted photon into desired channels is more critical. Metalens is an ideal meta-optics to collimate the single photons from a single QE by placing the QE in its focal plane. For example, T. Huang {\it et al.} fabricated nanopillars directly on the surface of diamond to function as a metalens and collimate the photons emitted from a vacancy center inside the diamond~\cite{huang2019MonolithicImmersion}. Direct fabrication on the high-index diamond eliminates the reflection loss at the diamond-air interface. The collection efficiency can be further improved by adding a back metal mirror, which also forms a mirror image of the QE~\cite{bao2020OndemandSpinstate}. By accurately integrating the QE and its mirror image on the two foci of a spin-splitting bifocal silicon metalens on the top, on-demand spin-state generation and splitting of single photons with direction control can be achieved. More recently, simultaneous multi-dimensional tailoring of direction, polarization, and orbital angular momentum of single photons from a defect in hexagonal boron nitride is demonstrated with a multifunctional metalens (Fig. \ref{fig1:general})~\cite{li2023ArbitrarilyStructured}. This  could unleash the full potential of single-photon QEs for their use as high-dimensional quantum sources for advanced quantum photonic applications. 

Although with great advancements on the rate enhancement and routing control of single photons from QEs, most of these works still rely on an external pump laser. Ultimate miniaturization of QE single photon sources would benefit from an integration with the pump laser. Towards this goal, X. Li {\it et al.} recently made a major step by deterministically-fabricating a planar circular Bragg grating as bright single photon source and an electrically-injected micropillar as a highly-directional pump microlaser on a single chip (Fig.\ref{fig2:source}c)~\cite{li2023BrightSemiconductor}. The circular Bragg grating and micropillar are individually optimized and heterogeneously integrated together by using a potentially scalable transfer printing process capable of fabricating a multitude of devices in a single run. The single QE was pumped by an on-chip micropillar laser under electrical injections, exhibiting high-performances in terms of the source brightness and single-photon purity thanks to the coupling of the QD to the cavity mode of the CBG. This work paves a way for realizing fully integrated metaphotonics engineered single photon sources.



\subsection{Two-photon sources}
Up to now, the most common way to generate two-photon source is based on the nonlinear SPDC process, where one pump photon goes through a second-order nonlinear material and spontaneously splits into two daughter photons called signal and idler. Conventionally, SPDC photon-pair sources rely on nonlinear crystals such as lithium niobate, BBO, KTP, which have a typical thickness on the scale of millimeters to centimeters. The stringent phase-matching condition limits the emission wavelengths and directions of the photon pairs to a certain range. A pioneering work from M.~Chekhova's group reveals that when the thickness of the nonlinear material is reduced to wavelength or subwavelength scale, the photon wavelength and angle are one order of magnitude broader than the that of thick crystals due to the relaxed phase matching condition~\cite{okoth2019MicroscaleGeneration, okoth2020IdealizedEinsteinPodolskyRosen}. This means strongly enhanced entanglement in the energy and momentum degrees of freedom, and promises to improve the time and spatial resolutions of many quantum photonic techniques beyond the current state of the art. Although the overall SPDC rate is much lower than that from thick nonlinear crystals, this work shows that the SPDC from subwavelength-thick nonlinear materials is measurable and potentially useful. 
The same group recently proposed an efficient way to remove the thermal background through time distillation, which increased the purity of the two-photon state from 0.002 to 0.99~\cite{sultanov2023TemporallyDistilled}. This method further increases the practicability of nanoscale and ultrathin SPDC sources.
Slightly before, A.~Sukhorukov {\it et al.} developed the quantum classical correspondence theory of SPDC, which stated that the quantum SPDC process can be explicitly described by the classical Green function~\cite{poddubny2016GenerationPhotonPlasmon}, or equivalently by its classical reverse process called sum frequency generation (SFG)~\cite{lenzini2018DirectCharacterization}. This theory not only provides a design and analysis tool for the metaphotonic SPDC source, but also proves that an efficient classical SFG source will serve as an efficient quantum SPDC source.
In the meantime, it is well developed that patterning thin films into nanostructures, i.e. forming nanoantennaes and MSs, can boost the classical nonlinear process at nanoscale by leveraging the optical resonances~\cite{li2017NonlinearPhotonic, krasnok2018NonlinearMetasurfaces, pertsch2020NonlinearOptics, gigli2022AlldielectricMetasurfaces}. 
Therefore, the aforementioned theory and experiment have laid the foundation to explore the ultrathin metaphotonic SPDC source.

As a start, single nanoantennaes were studied. Enhanced SPDC from an AlGaAs nanocylinder by Mie-type resonances at all three interacting wavelengths (i.e. pump, signal, and idler) was proposed and experimentally demonstrated by G. Marino {\it et al.}, generating photon pairs with a rate of 35 Hz~\cite{marino2019SpontaneousPhotonpairb}. The SPDC rate is significantly higher than conventional SPDC photon sources when normalized to the pump energy stored by the nanoantenna. This report of measurable SPDC from a single nanoantenna proves the feasibility of nanoscale SPDC source and opens the way for generating more complex photon states such as polarization correlation and entangled Bell states from a single nanoantenna~\cite{PhysRevA.103.043703, weissflog2024NonlinearNanoresonators} and multi-photon quantum states by multiplexing several antennas. Late, R.~Grange {\it et al.} reported SPDC experiments from single lithium niobate microcubes~\cite{duong2022SpontaneousParametric} and single GaAs nanowires~\cite{saerens2023BackgroundFreeNearInfrared}, further improving the SPDC rate by 40 times and showing the potential of realizing nanoscale SPDC sources at different material platforms.

To further increase the SPDC rate and directionality of the photon pairs, arranging the nanoantennaes into a periodic array, i.e. forming a MS, is a natural idea. The first nonlinear-MS SPDC experiment was reported by a collaborating team in Germany using a LiNbO$_3$ MS, which consists of nanoresonators in the shape of truncated pyramids~\cite{santiago-cruz2021PhotonPairs}. By leveraging the electric and magnetic Mie-like resonances at various wavelengths, the SPDC rate is enhanced up to two orders of magnitude within a narrow bandwidth, when compared with an unpatterned film of the same thickness and material. 
It also demonstrated the spectral control of the photon pairs by engineering the resonances and pump wavelength. This work opens the path toward the use of nonlinear MSs as versatile sources of photon pairs.
Later, more researches showed that the SPDC enhancement can be further boosted by leveraging other types of resonances with higher quality factor. For example, strong enhancement of SPDC through the high-quality-factor nonlocal resonances such as bound state in the continuum (BIC) was theoretically predicted in both AlGaAs and LiNbO$_3$ MSs~\cite{parry2021EnhancedGenerationb, jin2021EfficientSinglephoton, mazzanti2022EnhancedGeneration}, which predicted a SPDC enhancement over five orders of magnitude over the unpatterned film. 
In experiment, we demonstrated a LiNbO$_3$ MS with nonlocal guided mode resonances (GMR), which supported high and tunable quality factor and led to an 450-times enhancement of the SPDC over the thin film of LiNbO$_3$ of the same thickness (Fig. \ref{fig1:general})~\cite{zhang2022SpatiallyEntangled}. The proposed MS avoids nanofabrication on the nonlinear material and thus maximizes the nonlinear material volume and reduces noise from the fabrication induced material damage. The coincidence to accidental ratio (CAR) is up to 5000, which is 
larger than most of conventional crystal and waveguide SPDC sources. Furthermore, the non-classical spatial correlation of the generated photon pairs was experimentally verified in this work via Cauchy-Schwarz inequality violation experiment and later numerically quantified via Schmidt decomposition of the two-photon wavefunction~\cite{zhang2022PhotonPair}, which is promising for many free-space quantum applications such as quantum imaging as will be shown later.

\begin{figure}
    \includegraphics[width=\linewidth]{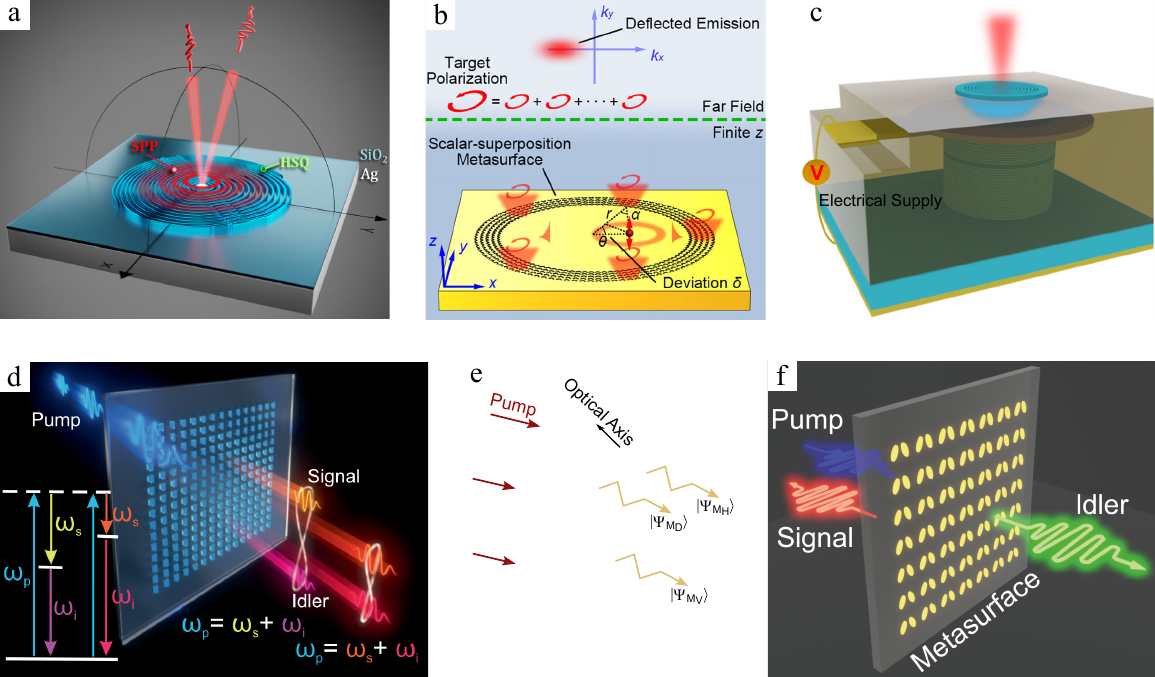}
    \caption{{\bf Meta-optics quantum-light sources based on quantum emitters and nonlinear metasurfaces.} (a) Collimated single-photon emission into two off-normal directions with orthogonal linear polarizations by integrating a QE with a MS atop a metal film~\cite{komisar2023MultipleChannelling}. Komisar \textit{et al.} Nat. Commun. 14, 6253 (2023); licensed under a Creative Commons Attribution (CC BY) license. 
    (b) Single-photon emission into a desired polarization by integrating a QE with a MS with robust placement of the QE on the MS~\cite{xue2022ScalarSuperpositionMetasurfaces}. Reproduced with permission from Xue \textit{et al.} Laser Photon. Rev. 16, 220179 (2022). Copyright 2022 John Wiley and Sons.
    (c) Directional single-photon emission by integrating a QE with a MS and an on-chip electrically-injected microlaser~\cite{li2023BrightSemiconductor}. 
    Li \textit{et al.} Light Sci. Appl. 12, 65 (2023); licensed under a Creative Commons Attribution (CC BY) license. 
    (d) Entangled photon pairs from a MS with multiple symmetry-protected quasi-BIC resonances~\cite{santiago-cruz2022ResonantMetasurfaces}. 
    From Santiago-Cruz \textit{et al.} Science 377, 991 (2022). Copyright 2022 AAAS.
    (e) Photon pairs from a multiplexed MS with polarization control beyond the fixed nonlinear susceptibility of the nonlinear material~\cite{ma2023PolarizationEngineering}. Reproduced with permission from Ma \textit{et al.} Nano Lett. 23, 8091 (2023). Copyright 2023 American Chemical Society.
    (f) Bidirectional photon pairs from a resonant MS~\cite{son2023PhotonPairs}. 
    Son \textit{et al.} Nanoscale 15, 2567 (2023); licensed under a Creative Commons Attribution 3.0 Unported Licence.}
    \label{fig2:source}
\end{figure}

With further development, MSs recently show a great potential to directly generate complex quantum states which are incapable for conventional SPDC sources, opening new opportunities for compact quantum information processing. One unique feature of MS over crystal lies in it is possible to fabricate several different MSs on a single substrate. Utilizing this feature, Santiago-Cruz {\it et al.} proposed a GaAs MS for direct generation of frequency cluster state by coherently pumping several MSs at multiple different wavelengths (Fig. \ref{fig2:source}d)~\cite{santiago-cruz2022ResonantMetasurfaces}. At the same time, the SPDC rate was strongly enhanced through the BIC resonances with quality factors up to 1000. 
In another work, we investigated the engineering of polarization state of the photon pairs from LiNbO$_3$ MS beyond the limit of the intrinsic nonlinear susceptibility tensor and formulated an efficient approach to generate arbitrary polarization-entangled qutrit states by simultaneous pumping of three closely patterned MSs with different orientations (Fig. \ref{fig2:source}e)~\cite{ma2023PolarizationEngineering}. The polarization qutrit state is optically tunable by the pump distributions on three MSs, which can be readily achieved by a spatial light modulator. 
Another exotic state comes from the freedom to choose the emission directions of the photons due to the relaxed phase matching condition. For example, the first MS SPDC experiment was performed in a configuration where both daughter photons were collected in the reflection direction~\cite{santiago-cruz2021PhotonPairs}, which is impossible in thick nonlinear crystals. As a matter of fact, it is experimentally verified that a single MS can simultaneously generate transmission-transmission, reflection-reflection, and transmission-reflection photon pairs (Fig. \ref{fig2:source}f)~\cite{son2023PhotonPairs, weissflog2024directionally}. The use of both directions of emission will fuel the development of more complicated architectures of nanoscale sources of quantum light. 

A summary on the performance of those experimentally demonstrated nonlinear-MS SPDC sources is shown in TABLE \ref{table:I}.
\begin{table}[h!]
\begin{center}
\begin{tabular}{ |c|c|c|c|c|c|c| } 
 \hline
 Reference & Material & Thickness & \makecell{Resonance \\ type} & \makecell{Measured rate$^{(a)}$ \\ @ pump power} & Enhancement$^{(a)}$ & CAR \\ 
 \hline
 [\cite{santiago-cruz2021PhotonPairs}] & LiNbO$_3$ & 680 nm & Mie & 5.4 Hz @ 70 mW & 20 & 361 \\ 
 \hline
 [\cite{zhang2022SpatiallyEntangled}] & LiNbO$_3$ & 304 nm & GMR & 1.8 Hz @ 85 mW & 450 & 5000 \\ 
 \hline
 [\cite{santiago-cruz2022ResonantMetasurfaces}] & GaAs & 500 nm & BIC & 0.08 Hz @ 9 mW & >1000 & $\sim$9.5$^{(b)}$ \\
 \hline 
 [\cite{ma2023PolarizationEngineering}] & LiNbO$_3$ & 300 nm & GMR & 0.83 Hz @ 85 mW & 210 & $\sim$1700$^{(b)}$ \\
 \hline
 [\cite{son2023PhotonPairs}] & GaP & 150 nm & BIC & 0.24 Hz @ 70 mW & 67 & $\sim$4.8$^{(c)}$ \\
 \hline
 [\cite{weissflog2024directionally}] & LiNbO$_3$ & 308 nm & GMR & 2.92 Hz @ 91 mW & NA & 7500 \\
 \hline
\end{tabular}
\caption{Summary of experimentally demonstrated SPDC sources from resonant metasurfaces. \\
$^{(a)}$Note that the measured rate and enhancement depend on the filtering bandwidth, collection angle and efficiency, detection efficiency of single-photon detectors, and the properties of optics in the setup. Therefore, the measured rate doesn’t represent the real internal generation rate from the metasurface. \\
$^{(b)}$Estimated from the second-order cross-correlation function by $\text{CAR} = g^{(2)}(0)-1$. \\
$^{(c)}$Estimated from the coincidence histogram in Fig. 2b of Ref.~[\cite{son2023PhotonPairs}].
}
\label{table:I}
\end{center}
\end{table}

Finally, it is worthy mentioning that quantum dots can also generate entangled photon pairs through the biexciton-exciton cascaded radiative process and its integration with broadband meta-optics such as circular Bragg grating can lead to two-photon sources with improved performance in terms of rate, efficiency, degree of entanglement, and indistinguishability~\cite{liu2019SolidstateSource, wang2019OnDemandSemiconductor, rota2022SourceEntangled}. Further development of the QE-meta-optics integration platform is promising for deterministic generation of two-photon sources.

\subsection{Multiphoton sources}
Multiphoton states with photon number over two are critical and desired in many quantum technologies to increase the scalability, capacity, sensitivity, and resolution~\cite{pan2012MultiphotonEntanglement}. The common way to create multiphoton states is through time-multiplexing of single photons from a QE or assembly of multiple photon-pair sources from nonlinear spontaneous photon-pair generation. Meta-optics enabled multiphoton source from QE has not been reported yet. While the idea of generating multiphoton states through multiplexing several nanoantenneas or MSs is straightforward, the practical challenge lies in that the rate of meta-optics enabled SPDC source is still too low to observe a multiphoton coincidence counting event.
Nevertheless, a groundbreaking experiment was performed by L. Li {\it et al.} to generate multiphoton states by integrating a $10\times10$ metalens array on the front surface of a nonlinear crystal (Fig. \ref{fig1:general})~\cite{li2020MetalensarrayBased}. Up to six photons generated from different metalenses with high indistinguishability was experimentally verified. This work provides a compact and practical platform for the development of advanced on-chip quantum photonic information processing.

\section{Metaphotonics for quantum light manipulation}
Light manipulation, transforming an initial state of light into a target state, is the most widely studied function of meta-optics since its born. On the other hand, almost all optical-manipulation elements for classical optics can find their use in quantum optics, such as beam splitters, waveplates, filters, and so on. Therefore, meta-optics enabled quantum light manipulation share the same advantages of meta-optics in classical light manipulation including ultracompact size, stability, controllability, multi-dimensionality, and multi-functionality. These have stimulated many meta-optics quantum optical devices which have potential to replace the convectional bulky counterparts with better performance or enable new light-manipulation functionalities beyond those can be achieved by conventional optical elements. In the following, we will summarize the most recent advancements in meta-optics for quantum light manipulation, including quantum interferences, entanglement manipulation, atom cooling and trapping, metasurface by atom array, metasurface for quantum computing, and some other emerging directions.  

\subsection{Quantum interferences}
Photon interferences play critical roles in many quantum photonic technologies ranging from quantum computing, communication, to quantum metrology and state characterization. Implementation of quantum interferences with MSs was pioneered by K. Wang {\it et al.}, who reported interference of multiphoton polarization-encoded states with photons coming from the same spatial modes on a one-input-six-output silicon MS (Fig. \ref{fig3:manipulate}a)~\cite{wang2018QuantumMetasurface}. The two-photon correlations between two distinct output ports at different time delays resemble those obtained in the conventional Hong-Ou-Mandel (HOM) experiment on a cube beam splitter~\cite{bouchard2021TwophotonInterference}, and expand their generalizations to lossy beam splitters. 
Comparing with the cube beam splitter, this MS enabled interferometer has much smaller size, flexibility of realizing different optical response by MS design, and 6 output ports with possibility of extending more. The later can enable 15 distinct two-photon correlations at the output. 
Furthermore, MS can have an non-unitary transmission matrix, being a new degree of freedom to engineer the photon interference behaviour. Q. Li {\it et al.} explored an non-unitary two-input-two-output MS for quantum interference (Fig. \ref{fig1:general})~\cite{li2021NonunitaryMetasurface}. 
They demonstrated dynamical and continuous control over the effective interaction of two single photons such that they show bosonic bunching, fermionic antibunching or arbitrarily intermediate behaviour, beyond their intrinsic bosonic nature. 
In the above work, the tuning was realized by mechanical rotation of the metasurface. One promising direction is to use reconfigurable MSs for tunable quantum interference~\cite{estakhri2021TunableQuantum}. 
It is possible to use MS to perform arbitrary U(2) transformations or even two-qubit U(4) operations by combining the spatial and polarization degrees of freedom~\cite{kang2021tailor}. 
In order to implement quantum interference for more photons, a multiport MS interferometer would be necessary. Recently, a promising method by employing the multiple diffraction orders of a MS grating was proposed for such a task~\cite{zhang2024SingleshotCharacterization}. The demonstrated multiport interferometer by a single MS supports an ultra-stable and tailored multiport transformation, which remained a challenge in the free-space configuration due to the requirement of deep subwavelength stability and its implementation has typically relied on active phase locking~\cite{Zhong:2020-1460:SCI} or integrated photonics circuits~\cite{Carolan:2015-711:SCI}.
These experiments prove that MSs not only can achieve an interference performance similar to the ones of traditional optical elements, but also reach new regions of interference patterns by tuning the transmission matrix. Therefore, MSs are promising candidates for integrated quantum interferometry and its related applications such as quantum state characterization~\cite{wang2018QuantumMetasurface, zhang2024SingleshotCharacterization} and quantum sensors~\cite{georgi2019MetasurfaceInterferometry}  (see more details later in the Detection section).
Lastly, MS has also been reported for remote quantum interferences~\cite{jha2015MetasurfaceEnabledRemote, liang2023continuous}.

\subsection{Entanglement manipulation}

Entanglement is a pure quantum feature underpinning many quantum enhanced technologies. As entanglement typically involve at least two modes in two degrees of freedom of the photons, MSs are very suitable to control entanglement due to their ability for multi-dimensional manipulation. A pioneering work from T. Stav {\it et al.} proposed a novel MS to create entanglement between spin and orbital angular momentum of a singe photon by using the geometric phase that arises from the photonic spin-orbit interaction (Fig. \ref{fig3:manipulate}b)~\cite{stav2018QuantumEntanglement}. Nonlocal correlations and entanglement between the spin and orbital angular momentum of two different photons is also created by the MS.
In the previously mentioned metalens array based multi-photon source, the same platform is also able to generate high-dimensional two-photon path entanglement with different phases and high fidelities~\cite{li2022HighdimensionalEntanglement}. 
MSs not only can generate entanglement between photons, but also it can realize disentanglement of two photons~\cite{georgi2019MetasurfaceInterferometry}, modification of the degree of entanglement~\cite{lung2020ComplexBirefringentDielectric}, entanglement between two qubits separated by macroscopic distances~\cite{jha2018MetasurfaceMediatedQuantum}, and entanglement distillation~\cite{asano2016DistillationPhoton, zhang2022AllopticalModulation}.
A plasmonic MS with all-optically tunable polarization dependent transmissions was demonstrated to continuously control the degree of entanglement between two photons from a non-maximally entangled one to that with fidelities higher than 98\%, enabling the function of entanglement distillation (Fig. \ref{fig1:general})~\cite{zhang2022AllopticalModulation}.
By placing two one-input-multi-output MSs at the paths of two polarization entangled photons, Y. Gao {\it et al.} reported the multichannel distribution and transformation of entangled photons (Fig. \ref{fig3:manipulate}e)~\cite{gao2022MultichannelDistribution}. In their experiments, $2 \times 2$ and $4 \times 4$ distributed entanglement states, including Bell states and superposition of Bell states, are demonstrated with high fidelity and strong polarization correlation. 

\subsection{Atom cooling and trapping}

A new application of MS enabled light manipulation is to construct the complex optical fields needed for cooling and trapping atoms in a very compact form. In atom trapping experiment, overlap of multiple circularly polarized beams coming from different directions is needed. This often requires multiple bulky beam splitters, waveplates and mirrors in conventional optics. 
L. Zhu {\it et al.} reported the use of a MS to replace the conventional bulky optical elements for creating cold atoms with a single incident laser beam (Fig. \ref{fig3:manipulate}d)~\cite{zhu2020DielectricMetasurface}. In their experiment, a single MS was applied for realizing both functions of beam splitting and polarization control. Atom numbers of $~10^7$ and temperatures of $~35 \mu K$ of relevance to quantum sensing were achieved in a compact and robust fashion. This temperature is comparable to what one would get in standard magneto optical trapping (MOT) systems. Recently, the same group fabricated a centimeter scale MS for MOT, which further increases the diffraction efficiency to 47\%, and successfully traps atoms with numbers up to $1.4\times 10^8$ and temperature down to $7 \mu K$~\cite{jin2023CentimeterScaleDielectric}. 
In parallel, T. Hsu {\it et al.} reported the use of MS lens to trap single atoms~\cite{hsu2022SingleAtomTrapping}. MS MOT paves a way toward fully integrated cold atom sources and quantum devices. Together with other intriguing proposals and experiments~\cite{mcgehee2021MagnetoopticalTrapping, zhu2023ConstructingMagnetooptical}, MS is becoming a powerful platform for optical twizzlers~\cite{shi2022OpticalManipulation}. One of the promising applications of MS MOT would be constructing a thin layer of cold atom array, which can function as a MS to manipulate both classical and quantum lights, as discussed below.

\subsection{Metasurfaces by atom arrays}

A new approach to construct optical MS is through a two-dimensional array of atoms or atom-like QEs, which play similar roles to the dielectric and plasmonic nanoantennaes in conventional MSs~\cite{zhou2017OpticalMetasurface, wang2017DesignMetasurface, facchinetti2018InteractionLight, bekenstein2020QuantumMetasurfaces, rui2020SubradiantOptical, alaee2020QuantumMetamaterials, ballantine2020OpticalMagnetism, ballantine2021CooperativeOptical, ballantine2021QuantumSinglePhoton, fernandez-fernandez2022TunableDirectional,levin2023ClusterStates}. For example, a planar atomic arrays with subwavelength spacing of atoms were proposed as MSs for collectively manipulating light~\cite{janne2012} and, in particular, they can act as a mirror to reflect light. J. Rui {\it et al.} performed the first experimental demonstration of this phenomena by using only a few hundred atoms (Fig. \ref{fig3:manipulate}e)~\cite{rui2020SubradiantOptical}. 
Importantly, the cooperative optical response of the atom array can be readily tuned by the atom density and ordering or an external pump light. 
They also observed the narrowing of the resonance transmission/reflection linewidth below the fundamental quantum limit of the single-atom Wigner-Weisskopf linewidth.  It is also possible to create entanglement between the atom MS and the scattered photons, constituting a new platform for simultaneous control of the quantum and spatiotemporal properties of light in free space, cavity-free parallel quantum operations on multiple photonic degrees of freedom, and the preparation of highly entangled photonic states suitable for quantum information~\cite{bekenstein2020QuantumMetasurfaces}. A recent study from Y. Levin {\it et al.} analyzed the  implementation of cluster states generation protocols by employing quantum MSs made out of sub-wavelength atomic arrays, including fundamental quantum logic gates useful for general quantum computation and communication purposes\cite{levin2023ClusterStates}.

\subsection{Metasurfaces for quantum computing}

\begin{figure}
    \includegraphics[width=\linewidth]{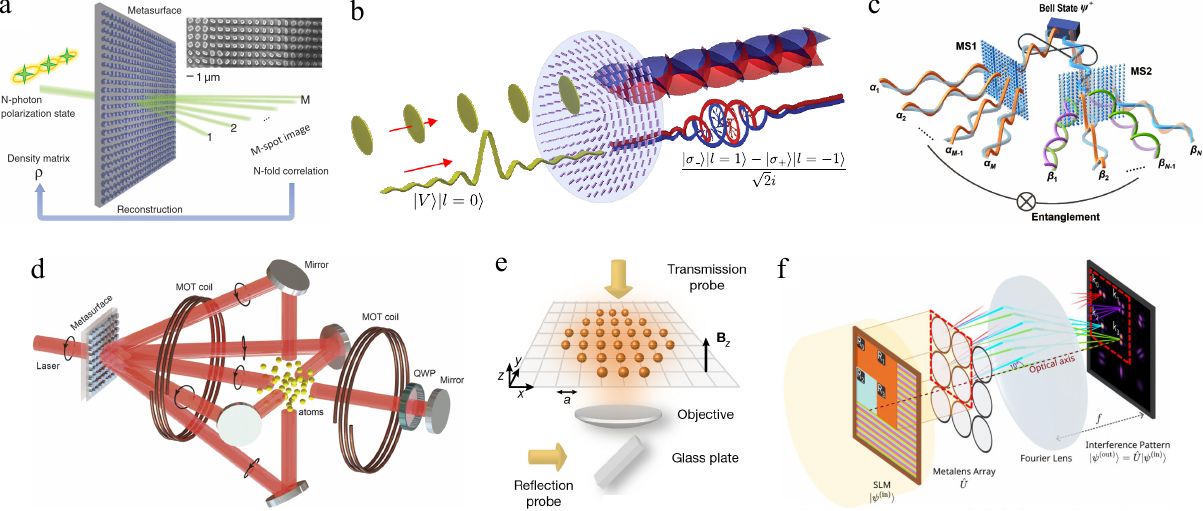}
    \caption{{\bf Metasurface-based light manipulation for quantum photonics.} 
    (a) Multiphoton interference and state reconstruction using a MS~\cite{wang2018QuantumMetasurface}. From Wang \textit{et al.} Science 361, 1104 (2018). Copyright 2018 AAAS.
    (b) Quantum entanglement of the spin and orbital angular momentum of photons using a MS~\cite{stav2018QuantumEntanglement}. From Stav \textit{et al.} Science 361, 1101 (2018). Copyright 2018 AAAS.
    (c) Multichannel distribution and transformation of polarization entangled photons using a MS~\cite{gao2022MultichannelDistribution}. Reproduced with permission from Gao \textit{et al.} Phys. Rev. Lett. 129, 023601 (2022). Copyright 2022 American Physical Society.
    (d) A centimeter-scale dielectric MS for the generation of cold atoms~\cite{jin2023CentimeterScaleDielectric}. Reproduced with permission from Jin \textit{et al.} Nano Lett. 23, 4008 (2023). Copyright 2023 American Chemical Society.
    (e) A subradiant optical mirror formed by a single structured atomic layer~\cite{rui2020SubradiantOptical}. Reproduced with permission from Rui \textit{et al.} Nature 583, 369 (2020). Copyright 2020 Springer Nature.
    (f) MS for programmable quantum algorithms with quantum and classical light~\cite{tanuwijaya2024metasurface}. Tanuwijaya \textit{et al.} Nanophotonics (2024); licensed under a Creative Commons Attribution 4.0 International License.
    }
    \label{fig3:manipulate}
\end{figure}

Quantum computing promises to outperform the classical computing in certain tasks and has attracted much attention. The basic building blocks of quantum computing include a physical platform to implement a quantum bit, logic gates, and algorithms. Recently, MSs have been proposed and demonstrated for all these essential parts of quantum computing in a compact and multifunctional manner.
L. Chen {\it et al.} proposed a concept of analog quantum bit to emulate a qubit and used a Pancharatnam-Berry (PB) phase MS as a physical platform to implement such a concept~\cite{chen2023AnalogQuantum}. A PB phase MS was experimentally fabricated to validate that the proposed analog qubit can be used to emulate the quantum bit in terms of both mathematical and geometrical representations. 
X. Ding {\it et al.} reported a novel MS-based all-optical diffractive neural network to implement quantum logic gate operations (Fig. \ref{fig1:general})~\cite{ding2023MetasurfacebasedOptical}. In comparison to previous works of quantum computing systems based on bulky optical elements, the proposed optical quantum logic gate is only composed of a single-layer MS with a compact size, which provides significant superiority in integration. Furthermore, it could be extended to more complex quantum architectures such as CNOT gates, rotating operators, multi-quantum bit operation, and others via increasing the layers with optimized channels and multiplexing. 
In another work, R. Tanuwijaya {\it et al.} proposed and experimentally demonstrated a programmable MS capable of performing quantum algorithms at the single photon level (Fig. \ref{fig3:manipulate}f)~\cite{tanuwijaya2024metasurface}. By selectively exciting subsets of metalenses and interpreting the interference patterns at specific output directions, two programmable quantum algorithms, i.e. the Grover's algorithm and the quantum Fourier transform, were demonstrated onto the same metalens array on a MS. 
These works open the door for applications of MSs in integrated and large-scale quantum computing and algorithms.



\section{Metaphotonics for quantum light detection}

Metaphotonics has also been widely studied for the detection and characterization of quantum photonic states. It should be noted that the light-detection metaphotonics is strongly associated with the light-manipulation metaphotonics in the above section. Most meta-optics, e.g. MSs, are still playing the role of light manipulation, while the detection of photons is based on single photon detectors (SPDs) positioned after the meta-optics. It is the quantum state manipulations or transformations implemented by the meta-optics that enable or enhance the desired detection applications such as sensing and imaging. Therefore, the optical functions of the MSs in this section are similar to those in the previous section or in some cases the same MS enables both manipulation and detection.

\subsection{Quantum sensing}

Quantum sensors can support a sensitivity beyond the classical shot noise limit. Metaphotonic quantum sensing has a rich history from plasmonic quantum sensors~\cite{lee2021QuantumPlasmonic} to MS quantum sensors~\cite{xu2023MetasurfacesOptical}. M. Dowran {\it et al.} demonstrated a quantum enhanced plasmonic sensor to enhance the sensitivity in measuring the local changes of the refractive index in air~\cite{dowran2018QuantumenhancedPlasmonica}. By using a plasmonic MS consisting of a triangle nanohole array drilled in a 100nm-thick silver film and bright entangled twin beams as the source, they measured a sensitivity on the order of $10^{-10} RIU/\sqrt{Hz}$, which is nearly 5 orders of magnitude better than previous proof-of-principle implementations of quantum-enhanced plasmonic sensors.  
L. Kim {\it et al.} used a plasmonic quantum MS to confine the incident infrared probe light in a micrometer-thick nitrogen vacancy layer beneath the diamond surface through the plasmonic lattice resonance and thus enhanced the infrared absorption for the readout of the nitrogen vacancy singlet transition for a magnetometer (Fig. \ref{fig4:detection}a)~\cite{kim2021AbsorptionBasedDiamond}. 
By optimizing the spin readout, this plasmonic quantum sensing metasurface can enable a near-spin-projection-noise-limited sensitivity below $1 nT /\sqrt{Hz}$ per $\mu m^2$ of sensing area.
As mentioned before, P. Georgi {\it et al.} developed a dielectric MS to generate a two-photon path-entangled NOON state~\cite{georgi2019MetasurfaceInterferometry}. Based on the same MS, they then built an interferometer to probe the phase changes of one path and obtained an enhanced sensing visibility for the entangled state at zero time delay over the disentangled one at a large time delay (Fig. \ref{fig1:general}).


\subsection{Quantum state characterization}

\begin{figure}
    \includegraphics[width=\linewidth]{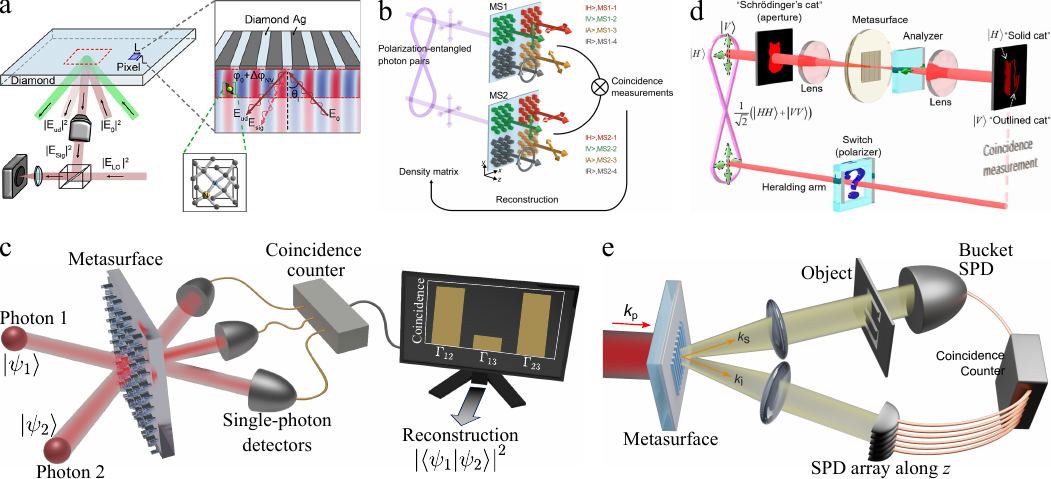}
    \caption{{\bf Metasurface-enabled quantum light detection.} 
    (a) Absorption-based diamond spin microscopy on a MS~\cite{kim2021AbsorptionBasedDiamond}. Reproduced with permission from Kim \textit{et al.} ACS Photon. 8, 3218 (2021). Copyright 2021 American Chemical Society.
    (b) Quantum tomography of polarization-entangled states with a MS~\cite{wang2022ImplementQuantum}. Reproduced from Wang \textit{et al.} Appl. Phys. Lett. 121, 081703 (2022), with the permission of AIP Publishing
    (c) Single-shot characterization of photon distinguishability with a MS~\cite{zhang2024SingleshotCharacterization}. Reproduced with permission from Zhang \textit{et al.} arXiv:2401.01485 (2024).
    (d) Quantum edge detection with a MS~\cite{zhou2020MetasurfaceEnabled}. Zhou \textit{et al.} Sci. Adv. 6, eabc4385 (2020); licensed under a Creative Commons Attribution (CC BY) license.
    (e) Quantum imaging using entangled photon pairs from a nonlinear MS~\cite{zhang2023QuantumImaging}. Reproduced with permission from Zhang \textit{et al.} Asia Communications and Photonics Conference/2023
     International Photonics and Optoelectronics Meetings (2023). Copyright IEEE.
    }
    \label{fig4:detection}
\end{figure}

Quantum state characterization is typically accomplished by performing a sequence of identical measurements in a series of different bases, a process called quantum state tomography~\cite{altepeter2005PhotonicState}. The change of projection into different bases is realized by tuning of several bulky optical elements. For example, the measurement of the two-photon polarization state requires 16 projection measurements by rotating the angles of four waveplates positioned before two polarizers~\cite{james2001MeasurementQubits}. 
MSs, due to their ability to manipulate multiple degrees of freedom and multiple bases in the same time, can significantly simplify the setup and process needed for the quantum state characterization. 
For example, the silicon MS proposed by K. Wang {\it et al.} for quantum interference, as shown in Fig. \ref{fig3:manipulate}a, was utilized for a robust reconstruction of amplitude, phase, coherence, and entanglement of multiphoton polarization-encoded states by simultaneously imaging multiple projections of quantum states~\cite{wang2018QuantumMetasurface}. In their experiments, two-photon states are reconstructed through nonlocal photon correlation measurements with polarization insensitive click detectors positioned after the MS, and the scalability to higher photon numbers is established theoretically.
Later, Z. Wang {\it et al.} proposed and demonstrated a two-MS scheme for the measurement of two-photon polarization state, where each MS has four districts functioning as a polarizer for four different polarization states (Fig. \ref{fig4:detection}b)~\cite{wang2022ImplementQuantum}. The needed 16 projection measurements were done by spatially translating the MS into different districts. As a result, reconstruction of the four Bell states was achieved with fidelity over 93.45\%. 
A similar two-MS approach was proposed to characterize the polarization Bell state without the need of spatial translation of the MSs~\cite{gao2023MetasurfaceComplete}.
Other than the polarization state, M. Wang {\it et al.} recently reported the characterization of OAM state of single photons by a dielectric MS (Fig. \ref{fig1:general})~\cite{wang2023CharacterizationOrbital}. By replacing the conventional bulky optical components in OAM measurements such as spiral phase plate, q-plate and spatial light modulator by a single MS, they reconstructed the density matrix of an arbitrary OAM state with high fidelity and measured the Schmidt number of the OAM entanglement. 
Recently, we proposed and demonstrated a two-input-three-output dielectric MS to realize the single-shot characterization of indistinguishability between two photons in several different degrees of freedom (Fig. \ref{fig4:detection}c)~\cite{zhang2024SingleshotCharacterization}. Topology optimization is employed to design a silicon MS with multiple targets, i.e. polarization independence, high transmission, and high tolerance to measurement noise. Based on the fabricated MS, we experimentally quantified the indistinguishability of two photons from a nonlinear crystal with fidelity over 98.4\%, without any reconfigurable and phase-locking elements.
MS has also been used for quantum weak measurements~\cite{chen2017DielectricMetasurfaces}, randomized measurement of photonic qubits with mitigated estimation error~\cite{ren2023error}, and efficient characterization of multiphoton entanglement with fewer measurements, higher accuracy, and robustness against optical loss~\cite{an2023efficient}.

\subsection{Quantum imaging and image processing}

MSs and metalens have been extensively studied for imaging and image processing using classical light source~\cite{pan2022DielectricMetalens, he2022ComputingMetasurfaces}. The combination of MS and quantum light source is leading to new opportunities in the field of quantum imaging~\cite{altuzarra2019ImagingPolarizationsensitivea, zhou2020MetasurfaceEnabled, vega2021MetasurfaceAssistedQuantum, yung2022PolarizationCoincidence, yung2022JonesmatrixImaging, liu2023MetasurfacesEnableda, zhang2023QuantumImaging}. 
J. Zhou {\it et al.} proposed and experimentally demonstrated a switchable optical edge detection by a high-efficiency dielectric MS and a polarization-entangled photon source (Fig. \ref{fig4:detection}d)~\cite{zhou2020MetasurfaceEnabled}. By selecting a proper polarization state in the heralding arm of the entangled photon source, either normal image or edge image is obtained. Importantly, compared to the case by using classical light sources, the quantum edge detection scheme shows a high signal-to-noise ratio at the same photon flux level. Recently, a similar concept has been demonstrated for the nonlocal weak-measurement microscopy~\cite{PhysRevLett.132.043601}.

T. Yung {\it et al.} combined the polarization-sensitive capability of MSs with HOM-type interference to generate images with tailor-made two-photon interference and coincidence signatures (Fig. \ref{fig1:general})~\cite{yung2022PolarizationCoincidence}.
As mentioned before, MS based SPDC source can support large angle photon emission with strong spatial entanglement, which can facilitate the field of view and resolution in quantum imaging. Recently, we performed the first quantum imaging experiment using MS based quantum light source with spatial entanglement (Fig. \ref{fig4:detection}e)~\cite{zhang2023QuantumImaging}. By combining quantum ghost and scanning imaging protocol, we realized two-dimensional quantum imaging using only a one-dimensional detector array.

\section{Conclusion and outlook}

Metaphotonics provides novel opportunities to design and fabricate optical devices for manipulating and routing non-classical light. It is underpinned by a smart design and the physics of metamaterials originating from the study of optically-induced electric and magnetic resonances. Quantum metasystems can find many applications including, among others, the development of unbreakable encryption, as well as they can open the door to new possibilities for quantum information systems on a chip for integrated quantum photonics~\cite{meng2021OpticalMetawaveguides, wang2022MetasurfaceIntegrated}, employing powerful tools of inverse design and machine learning developed for quantum metasurfaces~\cite{kudyshev2021MachineLearning}. We note that current experiments all rely on the integration of conventional bulky optical elements with metasurfaces which play one of roles in the generation, manipulation, and detection of quantum lights. In the future, we expect the development of metaphotonic quantum optical systems by integrating multiple metasurfaces for quantum light generation, manipulation, and detection.

Nonlinear metasurfaces are used as special metadevices for conversion of light frequency and wave mixing~\cite{lee2017PhotonPairGeneration}.
For quantum applications, nonlinear metasurfaces are employed for the efficient generation of photon pairs, and novel materials are being explored for SPDC, including multilayer stacked two-dimensional materials~\cite{guo2023UltrathinQuantum, weissflog2023TunableTransition,Braun2024spontaneousparametric}. In addition, metasurface-based SPDC sources can be developed for exotic wavelength ranges such as UV, mid-IR and THz, with applications in quantum imaging with undetected photons~\cite{lemos2014QuantumImaginga}. Furthermore, nonlinear metasurfaces can be employed for multiphoton parametric down-conversion and high-harmonic generation (HHG)~\cite{gorlach2020quantum,gorlach2023high}. HHG can be used as a source of coherent broadband radiation in the form of pulses with duration reaching attosecond timescales, and the emitted high harmonics can be squeezed or entangled. Quantum effects can modify the spectrum and photon correlations showing when individual frequency components become squeezed.
An attractive direction is integration of such metasurfaces with electrically injected quantum emitters or integration of nonlinear metasurfaces with electrically injected pump light.

Structuring light in multiple degrees of freedom, from spatial to temporal, holds great promise for advancing modern photonics~\cite{forbes2021StructuredLighta,bliokh2023RoadmapStructured}. Structuring light as single photons and entangled states allows accessing high-dimensional Hilbert spaces for fundamental tests of quantum mechanics and advanced quantum information processing~\cite{forbes2023}. Metasurfaces have been developed for the generation, manipulation, and detection of structured classical light, but they can offer a versatile platform for structuring the quantum states of light toward high-dimensional photonic quantum processing.

More recent developments in quantum metasurfaces aim exploring the new physics that can lead to breakthrough applications in quantum technologies.
This includes, for example, the use of metasurfaces for levitation~\cite{bao2023dipolar, holland2023demand} and a novel direction of spatiotemporal quantum metasurfaces~\cite{kort-kamp2021SpaceTimeQuantum}. In addition,  active and tunable metasurfaces with externally-driven change of their properties to control quantum light and quantum properties such as single-photon emission, manipulation, and non-classical detection, will become crucially important~\cite{zhang2022AllopticalModulation, uriri2018ActiveControl}. Coherent photon absorption by metasurfaces is also worth further exploring~\cite{roger2015CoherentPerfect, altuzarra2017CoherentPerfect, lyons2019CoherentMetamaterial}.

The further steps will include the development of new approaches and tools to bring quantum metasurfaces to atomic and solid-state based systems involving atom-atom, atom-photon and photon-photon entanglements, for example by employing atomic planar arrays where coherent scattering of incident light beams can be highly collimated in the forward and backward direction~\cite{janne2023}. The atomic planar arrays share features with fabricated metasurfaces, but a specific feature of atomic arrays is the possibility for the state manipulation via internal levels for photon storage, switching, gates, etc. Solid-based quantum metasurfaces can also be realized by deterministically preparing arrays of silicon-vacancy centers, which has been proved recently as promising quantum emitters for photon generation~\cite{knall2022}.

In summary, quantum metaphotonics has emerged as a cutting-edge development of metamaterial driven concepts to generate, manipulate, and detect quantum states of light. It can be employed for the miniaturization of current bulky quantum optical elements as well as design of on-chip quantum systems for quantum technologies. Over the past few years, this field has witnessed a surge of intriguing theoretical ideas, groundbreaking experiments, and novel applications, that we summarized in this paper to encourage the further development of this exciting research field. 

\begin{acknowledgments}
J.Z. acknowledges the support from Innovation Program for Quantum Science and Technology (No. 2021ZD0302300) and Songshan Lake Materials Laboratory (No. XMYS20230020), and thanks Andrey Sukhorukov, Jinyong Ma, and Dragomir Neshev for fruitful discussions and collaboration. The work of Y.K. was supported by Australian Research Council (Grants Nos. DP200101168 and DP210101292) and the International Technology Center Indo-Pacific (ITC IPAC) via Army Research Office (contract FA520923C0023). 
We thank many of our colleagues for productive collaboration, and more specifically  Janne Ruostekoski, Alexander Solntsev, Sergey Bozhevolnyi, Jie Zhao, and Kai Wang for useful and constructive comments on the manuscript.  
\end{acknowledgments}

\vspace{15 pt}
The authors have no conflicts to disclose.


%
%

%



\newpage

\bibliography{References}

\begin{thebibliography}{143}%
\makeatletter
\providecommand \@ifxundefined [1]{%
 \@ifx{#1\undefined}
}%
\providecommand \@ifnum [1]{%
 \ifnum #1\expandafter \@firstoftwo
 \else \expandafter \@secondoftwo
 \fi
}%
\providecommand \@ifx [1]{%
 \ifx #1\expandafter \@firstoftwo
 \else \expandafter \@secondoftwo
 \fi
}%
\providecommand \natexlab [1]{#1}%
\providecommand \enquote  [1]{``#1''}%
\providecommand \bibnamefont  [1]{#1}%
\providecommand \bibfnamefont [1]{#1}%
\providecommand \citenamefont [1]{#1}%
\providecommand \href@noop [0]{\@secondoftwo}%
\providecommand \href [0]{\begingroup \@sanitize@url \@href}%
\providecommand \@href[1]{\@@startlink{#1}\@@href}%
\providecommand \@@href[1]{\endgroup#1\@@endlink}%
\providecommand \@sanitize@url [0]{\catcode `\\12\catcode `\$12\catcode `\&12\catcode `\#12\catcode `\^12\catcode `\_12\catcode `\%12\relax}%
\providecommand \@@startlink[1]{}%
\providecommand \@@endlink[0]{}%
\providecommand \url  [0]{\begingroup\@sanitize@url \@url }%
\providecommand \@url [1]{\endgroup\@href {#1}{\urlprefix }}%
\providecommand \urlprefix  [0]{URL }%
\providecommand \Eprint [0]{\href }%
\providecommand \doibase [0]{http://dx.doi.org/}%
\providecommand \selectlanguage [0]{\@gobble}%
\providecommand \bibinfo  [0]{\@secondoftwo}%
\providecommand \bibfield  [0]{\@secondoftwo}%
\providecommand \translation [1]{[#1]}%
\providecommand \BibitemOpen [0]{}%
\providecommand \bibitemStop [0]{}%
\providecommand \bibitemNoStop [0]{.\EOS\space}%
\providecommand \EOS [0]{\spacefactor3000\relax}%
\providecommand \BibitemShut  [1]{\csname bibitem#1\endcsname}%
\let\auto@bib@innerbib\@empty
\bibitem [{\citenamefont {Qiu}\ \emph {et~al.}(2021)\citenamefont {Qiu}, \citenamefont {Zhang}, \citenamefont {Hu},\ and\ \citenamefont {Kivshar}}]{yuri2021}%
  \BibitemOpen
  \bibfield  {author} {\bibinfo {author} {\bibfnamefont {C.}~\bibnamefont {Qiu}}, \bibinfo {author} {\bibfnamefont {T.}~\bibnamefont {Zhang}}, \bibinfo {author} {\bibfnamefont {G.}~\bibnamefont {Hu}}, \ and\ \bibinfo {author} {\bibfnamefont {Y.}~\bibnamefont {Kivshar}},\ }\bibfield  {title} {\enquote {\bibinfo {title} {Quo vadis, metasurfaces?}}\ }\href {\doibase 10.1103/PhysRevA.64.052312} {\bibfield  {journal} {\bibinfo  {journal} {Nano Letters}\ }\textbf {\bibinfo {volume} {21}},\ \bibinfo {pages} {5461–5474} (\bibinfo {year} {2021})}\BibitemShut {NoStop}%
\bibitem [{\citenamefont {Fan}\ \emph {et~al.}(2022)\citenamefont {Fan}, \citenamefont {Liang}, \citenamefont {Li}, \citenamefont {Tsai},\ and\ \citenamefont {Zhang}}]{fan2022EmergingTrend}%
  \BibitemOpen
  \bibfield  {author} {\bibinfo {author} {\bibfnamefont {Y.}~\bibnamefont {Fan}}, \bibinfo {author} {\bibfnamefont {H.}~\bibnamefont {Liang}}, \bibinfo {author} {\bibfnamefont {J.}~\bibnamefont {Li}}, \bibinfo {author} {\bibfnamefont {D.~P.}\ \bibnamefont {Tsai}}, \ and\ \bibinfo {author} {\bibfnamefont {S.}~\bibnamefont {Zhang}},\ }\bibfield  {title} {\enquote {\bibinfo {title} {Emerging trend in unconventional metasurfaces: From nonlinear, non-hermitian to nonclassical metasurfaces},}\ }\href {\doibase 10.1021/acsphotonics.2c00816} {\bibfield  {journal} {\bibinfo  {journal} {ACS Photonics}\ }\textbf {\bibinfo {volume} {9}},\ \bibinfo {pages} {2872--2890} (\bibinfo {year} {2022})}\BibitemShut {NoStop}%
\bibitem [{\citenamefont {Wang}, \citenamefont {Chekhova},\ and\ \citenamefont {Kivshar}(2022)}]{wang2022MetasurfacesQuantum}%
  \BibitemOpen
  \bibfield  {author} {\bibinfo {author} {\bibfnamefont {K.}~\bibnamefont {Wang}}, \bibinfo {author} {\bibfnamefont {M.}~\bibnamefont {Chekhova}}, \ and\ \bibinfo {author} {\bibfnamefont {Y.}~\bibnamefont {Kivshar}},\ }\bibfield  {title} {\enquote {\bibinfo {title} {Metasurfaces for quantum technologies},}\ }\href {\doibase 10.1063/PT.3.5062} {\bibfield  {journal} {\bibinfo  {journal} {Physics Today}\ }\textbf {\bibinfo {volume} {75}},\ \bibinfo {pages} {38--44} (\bibinfo {year} {2022})}\BibitemShut {NoStop}%
\bibitem [{\citenamefont {Solntsev}, \citenamefont {Agarwal},\ and\ \citenamefont {Kivshar}(2021)}]{solntsev2021MetasurfacesQuantum}%
  \BibitemOpen
  \bibfield  {author} {\bibinfo {author} {\bibfnamefont {A.~S.}\ \bibnamefont {Solntsev}}, \bibinfo {author} {\bibfnamefont {G.~S.}\ \bibnamefont {Agarwal}}, \ and\ \bibinfo {author} {\bibfnamefont {Y.~S.}\ \bibnamefont {Kivshar}},\ }\bibfield  {title} {\enquote {\bibinfo {title} {Metasurfaces for quantum photonics},}\ }\href {\doibase 10.1038/s41566-021-00793-z} {\bibfield  {journal} {\bibinfo  {journal} {Nature Photonics}\ }\textbf {\bibinfo {volume} {15}},\ \bibinfo {pages} {327--336} (\bibinfo {year} {2021})}\BibitemShut {NoStop}%
\bibitem [{\citenamefont {Liu}\ \emph {et~al.}(2021)\citenamefont {Liu}, \citenamefont {Shi}, \citenamefont {Chen}, \citenamefont {Wang}, \citenamefont {Wang},\ and\ \citenamefont {Zhu}}]{liu2021QuantumPhotonics}%
  \BibitemOpen
  \bibfield  {author} {\bibinfo {author} {\bibfnamefont {J.}~\bibnamefont {Liu}}, \bibinfo {author} {\bibfnamefont {M.}~\bibnamefont {Shi}}, \bibinfo {author} {\bibfnamefont {Z.}~\bibnamefont {Chen}}, \bibinfo {author} {\bibfnamefont {S.}~\bibnamefont {Wang}}, \bibinfo {author} {\bibfnamefont {Z.}~\bibnamefont {Wang}}, \ and\ \bibinfo {author} {\bibfnamefont {S.}~\bibnamefont {Zhu}},\ }\bibfield  {title} {\enquote {\bibinfo {title} {Quantum photonics based on metasurfaces},}\ }\href {\doibase 10.29026/oea.2021.200092} {\bibfield  {journal} {\bibinfo  {journal} {Opto-Electronic Advances}\ }\textbf {\bibinfo {volume} {4}},\ \bibinfo {pages} {200092} (\bibinfo {year} {2021})}\BibitemShut {NoStop}%
\bibitem [{\citenamefont {Sharapova}, \citenamefont {Kruk},\ and\ \citenamefont {Solntsev}(2023)}]{sharapova2023NonlinearDielectric}%
  \BibitemOpen
  \bibfield  {author} {\bibinfo {author} {\bibfnamefont {P.~R.}\ \bibnamefont {Sharapova}}, \bibinfo {author} {\bibfnamefont {S.~S.}\ \bibnamefont {Kruk}}, \ and\ \bibinfo {author} {\bibfnamefont {A.~S.}\ \bibnamefont {Solntsev}},\ }\bibfield  {title} {\enquote {\bibinfo {title} {Nonlinear dielectric nanoresonators and metasurfaces: Toward efficient generation of entangled photons},}\ }\href {\doibase 10.1002/lpor.202200408} {\bibfield  {journal} {\bibinfo  {journal} {Laser \& Photonics Reviews}\ }\textbf {\bibinfo {volume} {17}},\ \bibinfo {pages} {2200408} (\bibinfo {year} {2023})}\BibitemShut {NoStop}%
\bibitem [{\citenamefont {Kan}\ and\ \citenamefont {Bozhevolnyi}(2023)}]{kan2023AdvancesMetaphotonics}%
  \BibitemOpen
  \bibfield  {author} {\bibinfo {author} {\bibfnamefont {Y.}~\bibnamefont {Kan}}\ and\ \bibinfo {author} {\bibfnamefont {S.~I.}\ \bibnamefont {Bozhevolnyi}},\ }\bibfield  {title} {\enquote {\bibinfo {title} {Advances in metaphotonics empowered single photon emission},}\ }\href {\doibase 10.1002/adom.202202759} {\bibfield  {journal} {\bibinfo  {journal} {Advanced Optical Materials}\ }\textbf {\bibinfo {volume} {11}},\ \bibinfo {pages} {2202759} (\bibinfo {year} {2023})}\BibitemShut {NoStop}%
\bibitem [{\citenamefont {Ding}\ and\ \citenamefont {Bozhevolnyi}(2023)}]{ding2023AdvancesQuantum}%
  \BibitemOpen
  \bibfield  {author} {\bibinfo {author} {\bibfnamefont {F.}~\bibnamefont {Ding}}\ and\ \bibinfo {author} {\bibfnamefont {S.~I.}\ \bibnamefont {Bozhevolnyi}},\ }\bibfield  {title} {\enquote {\bibinfo {title} {Advances in quantum meta-optics},}\ }\href {\doibase 10.1016/j.mattod.2023.07.021} {\bibfield  {journal} {\bibinfo  {journal} {Materials Today}\ }\textbf {\bibinfo {volume} {71}},\ \bibinfo {pages} {63--72} (\bibinfo {year} {2023})}\BibitemShut {NoStop}%
\bibitem [{\citenamefont {Xu}\ \emph {et~al.}(2023)\citenamefont {Xu}, \citenamefont {Su}, \citenamefont {Chai},\ and\ \citenamefont {Li}}]{xu2023MetasurfacesOptical}%
  \BibitemOpen
  \bibfield  {author} {\bibinfo {author} {\bibfnamefont {Y.}~\bibnamefont {Xu}}, \bibinfo {author} {\bibfnamefont {X.}~\bibnamefont {Su}}, \bibinfo {author} {\bibfnamefont {Z.}~\bibnamefont {Chai}}, \ and\ \bibinfo {author} {\bibfnamefont {J.}~\bibnamefont {Li}},\ }\bibfield  {title} {\enquote {\bibinfo {title} {Metasurfaces toward optical manipulation technologies for quantum precision measurement},}\ }\href {\doibase 10.1002/lpor.202300355} {\bibfield  {journal} {\bibinfo  {journal} {Laser \& Photonics Reviews}\ }\textbf {\bibinfo {volume} {n/a}},\ \bibinfo {pages} {2300355} (\bibinfo {year} {2023})}\BibitemShut {NoStop}%
\bibitem [{\citenamefont {Parry}\ \emph {et~al.}(2024{\natexlab{a}})\citenamefont {Parry}, \citenamefont {Mazzanti}, \citenamefont {Poddubny}, \citenamefont {Valle}, \citenamefont {Neshev},\ and\ \citenamefont {Sukhorukov}}]{parry2024ChapterNonlineara}%
  \BibitemOpen
  \bibfield  {author} {\bibinfo {author} {\bibfnamefont {M.}~\bibnamefont {Parry}}, \bibinfo {author} {\bibfnamefont {A.}~\bibnamefont {Mazzanti}}, \bibinfo {author} {\bibfnamefont {A.~N.}\ \bibnamefont {Poddubny}}, \bibinfo {author} {\bibfnamefont {G.~D.}\ \bibnamefont {Valle}}, \bibinfo {author} {\bibfnamefont {D.~N.}\ \bibnamefont {Neshev}}, \ and\ \bibinfo {author} {\bibfnamefont {A.~A.}\ \bibnamefont {Sukhorukov}},\ }\bibfield  {title} {\enquote {\bibinfo {title} {Chapter 8 - nonlinear metasurfaces for the generation of quantum photon-pair states},}\ }in\ \href {\doibase 10.1016/B978-0-323-90614-2.00013-4} {\emph {\bibinfo {booktitle} {Fundamentals and Applications of Nonlinear Nanophotonics}}},\ \bibinfo {series and number} {Nanophotonics},\ \bibinfo {editor} {edited by\ \bibinfo {editor} {\bibfnamefont {N.~C.}\ \bibnamefont {Panoiu}}}\ (\bibinfo  {publisher} {Elsevier},\ \bibinfo {year} {2024})\ pp.\ \bibinfo {pages} {271--287}\BibitemShut {NoStop}%
\bibitem [{\citenamefont {Li}\ \emph {et~al.}(2023{\natexlab{a}})\citenamefont {Li}, \citenamefont {Jang}, \citenamefont {Badloe}, \citenamefont {Yang}, \citenamefont {Kim}, \citenamefont {Kim}, \citenamefont {Nguyen}, \citenamefont {Maier}, \citenamefont {Rho}, \citenamefont {Ren},\ and\ \citenamefont {Aharonovich}}]{li2023ArbitrarilyStructured}%
  \BibitemOpen
  \bibfield  {author} {\bibinfo {author} {\bibfnamefont {C.}~\bibnamefont {Li}}, \bibinfo {author} {\bibfnamefont {J.}~\bibnamefont {Jang}}, \bibinfo {author} {\bibfnamefont {T.}~\bibnamefont {Badloe}}, \bibinfo {author} {\bibfnamefont {T.}~\bibnamefont {Yang}}, \bibinfo {author} {\bibfnamefont {J.}~\bibnamefont {Kim}}, \bibinfo {author} {\bibfnamefont {J.}~\bibnamefont {Kim}}, \bibinfo {author} {\bibfnamefont {M.}~\bibnamefont {Nguyen}}, \bibinfo {author} {\bibfnamefont {S.~A.}\ \bibnamefont {Maier}}, \bibinfo {author} {\bibfnamefont {J.}~\bibnamefont {Rho}}, \bibinfo {author} {\bibfnamefont {H.}~\bibnamefont {Ren}}, \ and\ \bibinfo {author} {\bibfnamefont {I.}~\bibnamefont {Aharonovich}},\ }\bibfield  {title} {\enquote {\bibinfo {title} {Arbitrarily structured quantum emission with a multifunctional metalens},}\ }\href {\doibase 10.1186/s43593-023-00052-4} {\bibfield  {journal} {\bibinfo  {journal} {eLight}\ }\textbf {\bibinfo {volume} {3}},\ \bibinfo {pages} {19} (\bibinfo {year}
  {2023}{\natexlab{a}})}\BibitemShut {NoStop}%
\bibitem [{\citenamefont {Zhang}\ \emph {et~al.}(2022{\natexlab{a}})\citenamefont {Zhang}, \citenamefont {Ma}, \citenamefont {Parry}, \citenamefont {Cai}, \citenamefont {{Camacho-Morales}}, \citenamefont {Xu}, \citenamefont {Neshev},\ and\ \citenamefont {Sukhorukov}}]{zhang2022SpatiallyEntangled}%
  \BibitemOpen
  \bibfield  {author} {\bibinfo {author} {\bibfnamefont {J.}~\bibnamefont {Zhang}}, \bibinfo {author} {\bibfnamefont {J.}~\bibnamefont {Ma}}, \bibinfo {author} {\bibfnamefont {M.}~\bibnamefont {Parry}}, \bibinfo {author} {\bibfnamefont {M.}~\bibnamefont {Cai}}, \bibinfo {author} {\bibfnamefont {R.}~\bibnamefont {{Camacho-Morales}}}, \bibinfo {author} {\bibfnamefont {L.}~\bibnamefont {Xu}}, \bibinfo {author} {\bibfnamefont {D.~N.}\ \bibnamefont {Neshev}}, \ and\ \bibinfo {author} {\bibfnamefont {A.~A.}\ \bibnamefont {Sukhorukov}},\ }\bibfield  {title} {\enquote {\bibinfo {title} {Spatially entangled photon pairs from lithium niobate nonlocal metasurfaces},}\ }\href {\doibase 10.1126/sciadv.abq4240} {\bibfield  {journal} {\bibinfo  {journal} {Science Advances}\ }\textbf {\bibinfo {volume} {8}},\ \bibinfo {pages} {eabq4240} (\bibinfo {year} {2022}{\natexlab{a}})}\BibitemShut {NoStop}%
\bibitem [{\citenamefont {Li}\ \emph {et~al.}(2020)\citenamefont {Li}, \citenamefont {Liu}, \citenamefont {Ren}, \citenamefont {Wang}, \citenamefont {Su}, \citenamefont {Chen}, \citenamefont {Chu}, \citenamefont {Kuo}, \citenamefont {Liu}, \citenamefont {Zang}, \citenamefont {Guo}, \citenamefont {Zhang}, \citenamefont {Wang}, \citenamefont {Zhu},\ and\ \citenamefont {Tsai}}]{li2020MetalensarrayBased}%
  \BibitemOpen
  \bibfield  {author} {\bibinfo {author} {\bibfnamefont {L.}~\bibnamefont {Li}}, \bibinfo {author} {\bibfnamefont {Z.}~\bibnamefont {Liu}}, \bibinfo {author} {\bibfnamefont {X.}~\bibnamefont {Ren}}, \bibinfo {author} {\bibfnamefont {S.}~\bibnamefont {Wang}}, \bibinfo {author} {\bibfnamefont {V.-C.}\ \bibnamefont {Su}}, \bibinfo {author} {\bibfnamefont {M.-K.}\ \bibnamefont {Chen}}, \bibinfo {author} {\bibfnamefont {C.~H.}\ \bibnamefont {Chu}}, \bibinfo {author} {\bibfnamefont {H.~Y.}\ \bibnamefont {Kuo}}, \bibinfo {author} {\bibfnamefont {B.}~\bibnamefont {Liu}}, \bibinfo {author} {\bibfnamefont {W.}~\bibnamefont {Zang}}, \bibinfo {author} {\bibfnamefont {G.}~\bibnamefont {Guo}}, \bibinfo {author} {\bibfnamefont {L.}~\bibnamefont {Zhang}}, \bibinfo {author} {\bibfnamefont {Z.}~\bibnamefont {Wang}}, \bibinfo {author} {\bibfnamefont {S.}~\bibnamefont {Zhu}}, \ and\ \bibinfo {author} {\bibfnamefont {D.~P.}\ \bibnamefont {Tsai}},\ }\bibfield  {title} {\enquote {\bibinfo {title} {Metalens-array{\textendash}based
  high-dimensional and multiphoton quantum source},}\ }\href {\doibase 10.1126/science.aba9779} {\bibfield  {journal} {\bibinfo  {journal} {Science}\ }\textbf {\bibinfo {volume} {368}},\ \bibinfo {pages} {1487--1490} (\bibinfo {year} {2020})}\BibitemShut {NoStop}%
\bibitem [{\citenamefont {Li}\ \emph {et~al.}(2021)\citenamefont {Li}, \citenamefont {Bao}, \citenamefont {Nie}, \citenamefont {Xia}, \citenamefont {Xue}, \citenamefont {Wang}, \citenamefont {Yang},\ and\ \citenamefont {Zhang}}]{li2021NonunitaryMetasurface}%
  \BibitemOpen
  \bibfield  {author} {\bibinfo {author} {\bibfnamefont {Q.}~\bibnamefont {Li}}, \bibinfo {author} {\bibfnamefont {W.}~\bibnamefont {Bao}}, \bibinfo {author} {\bibfnamefont {Z.}~\bibnamefont {Nie}}, \bibinfo {author} {\bibfnamefont {Y.}~\bibnamefont {Xia}}, \bibinfo {author} {\bibfnamefont {Y.}~\bibnamefont {Xue}}, \bibinfo {author} {\bibfnamefont {Y.}~\bibnamefont {Wang}}, \bibinfo {author} {\bibfnamefont {S.}~\bibnamefont {Yang}}, \ and\ \bibinfo {author} {\bibfnamefont {X.}~\bibnamefont {Zhang}},\ }\bibfield  {title} {\enquote {\bibinfo {title} {A non-unitary metasurface enables continuous control of quantum photon{\textendash}photon interactions from bosonic to fermionic},}\ }\href {\doibase 10.1038/s41566-021-00762-6} {\bibfield  {journal} {\bibinfo  {journal} {Nature Photonics}\ }\textbf {\bibinfo {volume} {15}},\ \bibinfo {pages} {267--271} (\bibinfo {year} {2021})}\BibitemShut {NoStop}%
\bibitem [{\citenamefont {Zhang}\ \emph {et~al.}(2022{\natexlab{b}})\citenamefont {Zhang}, \citenamefont {Chen}, \citenamefont {Gong}, \citenamefont {Wu}, \citenamefont {Cai}, \citenamefont {Ren}, \citenamefont {Ren}, \citenamefont {Zhang}, \citenamefont {Guo},\ and\ \citenamefont {Xu}}]{zhang2022AllopticalModulation}%
  \BibitemOpen
  \bibfield  {author} {\bibinfo {author} {\bibfnamefont {D.}~\bibnamefont {Zhang}}, \bibinfo {author} {\bibfnamefont {Y.}~\bibnamefont {Chen}}, \bibinfo {author} {\bibfnamefont {S.}~\bibnamefont {Gong}}, \bibinfo {author} {\bibfnamefont {W.}~\bibnamefont {Wu}}, \bibinfo {author} {\bibfnamefont {W.}~\bibnamefont {Cai}}, \bibinfo {author} {\bibfnamefont {M.}~\bibnamefont {Ren}}, \bibinfo {author} {\bibfnamefont {X.}~\bibnamefont {Ren}}, \bibinfo {author} {\bibfnamefont {S.}~\bibnamefont {Zhang}}, \bibinfo {author} {\bibfnamefont {G.}~\bibnamefont {Guo}}, \ and\ \bibinfo {author} {\bibfnamefont {J.}~\bibnamefont {Xu}},\ }\bibfield  {title} {\enquote {\bibinfo {title} {All-optical modulation of quantum states by nonlinear metasurface},}\ }\href {\doibase 10.1038/s41377-022-00744-5} {\bibfield  {journal} {\bibinfo  {journal} {Light: Science \& Applications}\ }\textbf {\bibinfo {volume} {11}},\ \bibinfo {pages} {58} (\bibinfo {year} {2022}{\natexlab{b}})}\BibitemShut {NoStop}%
\bibitem [{\citenamefont {Ding}\ \emph {et~al.}(2023)\citenamefont {Ding}, \citenamefont {Zhao}, \citenamefont {Xie}, \citenamefont {Cai}, \citenamefont {Meng}, \citenamefont {Wang}, \citenamefont {Wu}, \citenamefont {Liu}, \citenamefont {Burokur},\ and\ \citenamefont {Hu}}]{ding2023MetasurfacebasedOptical}%
  \BibitemOpen
  \bibfield  {author} {\bibinfo {author} {\bibfnamefont {X.}~\bibnamefont {Ding}}, \bibinfo {author} {\bibfnamefont {Z.}~\bibnamefont {Zhao}}, \bibinfo {author} {\bibfnamefont {P.}~\bibnamefont {Xie}}, \bibinfo {author} {\bibfnamefont {D.}~\bibnamefont {Cai}}, \bibinfo {author} {\bibfnamefont {F.}~\bibnamefont {Meng}}, \bibinfo {author} {\bibfnamefont {C.}~\bibnamefont {Wang}}, \bibinfo {author} {\bibfnamefont {Q.}~\bibnamefont {Wu}}, \bibinfo {author} {\bibfnamefont {J.}~\bibnamefont {Liu}}, \bibinfo {author} {\bibfnamefont {S.~N.}\ \bibnamefont {Burokur}}, \ and\ \bibinfo {author} {\bibfnamefont {G.}~\bibnamefont {Hu}},\ }\bibfield  {title} {\enquote {\bibinfo {title} {Metasurface-based optical logic operators driven by diffractive neural networks},}\ }\href {\doibase 10.1002/adma.202308993} {\bibfield  {journal} {\bibinfo  {journal} {Advanced Materials}\ }\textbf {\bibinfo {volume} {36}},\ \bibinfo {pages} {2308993} (\bibinfo {year} {2023})}\BibitemShut {NoStop}%
\bibitem [{\citenamefont {Georgi}\ \emph {et~al.}(2019)\citenamefont {Georgi}, \citenamefont {Massaro}, \citenamefont {Luo}, \citenamefont {Sain}, \citenamefont {Montaut}, \citenamefont {Herrmann}, \citenamefont {Weiss}, \citenamefont {Li}, \citenamefont {Silberhorn},\ and\ \citenamefont {Zentgraf}}]{georgi2019MetasurfaceInterferometry}%
  \BibitemOpen
  \bibfield  {author} {\bibinfo {author} {\bibfnamefont {P.}~\bibnamefont {Georgi}}, \bibinfo {author} {\bibfnamefont {M.}~\bibnamefont {Massaro}}, \bibinfo {author} {\bibfnamefont {K.-H.}\ \bibnamefont {Luo}}, \bibinfo {author} {\bibfnamefont {B.}~\bibnamefont {Sain}}, \bibinfo {author} {\bibfnamefont {N.}~\bibnamefont {Montaut}}, \bibinfo {author} {\bibfnamefont {H.}~\bibnamefont {Herrmann}}, \bibinfo {author} {\bibfnamefont {T.}~\bibnamefont {Weiss}}, \bibinfo {author} {\bibfnamefont {G.}~\bibnamefont {Li}}, \bibinfo {author} {\bibfnamefont {C.}~\bibnamefont {Silberhorn}}, \ and\ \bibinfo {author} {\bibfnamefont {T.}~\bibnamefont {Zentgraf}},\ }\bibfield  {title} {\enquote {\bibinfo {title} {Metasurface interferometry toward quantum sensors},}\ }\href {\doibase 10.1038/s41377-019-0182-6} {\bibfield  {journal} {\bibinfo  {journal} {Light: Science \& Applications}\ }\textbf {\bibinfo {volume} {8}},\ \bibinfo {pages} {70} (\bibinfo {year} {2019})}\BibitemShut {NoStop}%
\bibitem [{\citenamefont {Wang}\ \emph {et~al.}(2023)\citenamefont {Wang}, \citenamefont {Chen}, \citenamefont {Choi}, \citenamefont {Huang}, \citenamefont {Wang}, \citenamefont {Tu}, \citenamefont {Cheng}, \citenamefont {Tian}, \citenamefont {Li}, \citenamefont {Chen},\ and\ \citenamefont {Wang}}]{wang2023CharacterizationOrbital}%
  \BibitemOpen
  \bibfield  {author} {\bibinfo {author} {\bibfnamefont {M.}~\bibnamefont {Wang}}, \bibinfo {author} {\bibfnamefont {L.}~\bibnamefont {Chen}}, \bibinfo {author} {\bibfnamefont {D.-Y.}\ \bibnamefont {Choi}}, \bibinfo {author} {\bibfnamefont {S.}~\bibnamefont {Huang}}, \bibinfo {author} {\bibfnamefont {Q.}~\bibnamefont {Wang}}, \bibinfo {author} {\bibfnamefont {C.}~\bibnamefont {Tu}}, \bibinfo {author} {\bibfnamefont {H.}~\bibnamefont {Cheng}}, \bibinfo {author} {\bibfnamefont {J.}~\bibnamefont {Tian}}, \bibinfo {author} {\bibfnamefont {Y.}~\bibnamefont {Li}}, \bibinfo {author} {\bibfnamefont {S.}~\bibnamefont {Chen}}, \ and\ \bibinfo {author} {\bibfnamefont {H.-T.}\ \bibnamefont {Wang}},\ }\bibfield  {title} {\enquote {\bibinfo {title} {Characterization of orbital angular momentum quantum states empowered by metasurfaces},}\ }\href {\doibase 10.1021/acs.nanolett.3c00554} {\bibfield  {journal} {\bibinfo  {journal} {Nano Letters}\ }\textbf {\bibinfo {volume} {23}},\ \bibinfo {pages} {3921--3928} (\bibinfo {year}
  {2023})}\BibitemShut {NoStop}%
\bibitem [{\citenamefont {Yung}\ \emph {et~al.}(2022{\natexlab{a}})\citenamefont {Yung}, \citenamefont {Xi}, \citenamefont {Liang}, \citenamefont {Lau}, \citenamefont {Wong}, \citenamefont {Tanuwijaya}, \citenamefont {Zhong}, \citenamefont {Liu}, \citenamefont {Tam},\ and\ \citenamefont {Li}}]{yung2022PolarizationCoincidence}%
  \BibitemOpen
  \bibfield  {author} {\bibinfo {author} {\bibfnamefont {T.~K.}\ \bibnamefont {Yung}}, \bibinfo {author} {\bibfnamefont {J.}~\bibnamefont {Xi}}, \bibinfo {author} {\bibfnamefont {H.}~\bibnamefont {Liang}}, \bibinfo {author} {\bibfnamefont {K.~M.}\ \bibnamefont {Lau}}, \bibinfo {author} {\bibfnamefont {W.~C.}\ \bibnamefont {Wong}}, \bibinfo {author} {\bibfnamefont {R.~S.}\ \bibnamefont {Tanuwijaya}}, \bibinfo {author} {\bibfnamefont {F.}~\bibnamefont {Zhong}}, \bibinfo {author} {\bibfnamefont {H.}~\bibnamefont {Liu}}, \bibinfo {author} {\bibfnamefont {W.~Y.}\ \bibnamefont {Tam}}, \ and\ \bibinfo {author} {\bibfnamefont {J.}~\bibnamefont {Li}},\ }\bibfield  {title} {\enquote {\bibinfo {title} {Polarization coincidence images from metasurfaces with hom-type interference},}\ }\href {\doibase 10.1016/j.isci.2022.104155} {\bibfield  {journal} {\bibinfo  {journal} {iScience}\ }\textbf {\bibinfo {volume} {25}},\ \bibinfo {pages} {104155} (\bibinfo {year} {2022}{\natexlab{a}})}\BibitemShut {NoStop}%
\bibitem [{\citenamefont {Lodahl}, \citenamefont {Mahmoodian},\ and\ \citenamefont {Stobbe}(2015)}]{lodahl2015InterfacingSingle}%
  \BibitemOpen
  \bibfield  {author} {\bibinfo {author} {\bibfnamefont {P.}~\bibnamefont {Lodahl}}, \bibinfo {author} {\bibfnamefont {S.}~\bibnamefont {Mahmoodian}}, \ and\ \bibinfo {author} {\bibfnamefont {S.}~\bibnamefont {Stobbe}},\ }\bibfield  {title} {\enquote {\bibinfo {title} {Interfacing single photons and single quantum dots with photonic nanostructures},}\ }\href {\doibase 10.1103/RevModPhys.87.347} {\bibfield  {journal} {\bibinfo  {journal} {Reviews of Modern Physics}\ }\textbf {\bibinfo {volume} {87}},\ \bibinfo {pages} {347--400} (\bibinfo {year} {2015})}\BibitemShut {NoStop}%
\bibitem [{\citenamefont {Pelton}(2015)}]{pelton2015ModifiedSpontaneousa}%
  \BibitemOpen
  \bibfield  {author} {\bibinfo {author} {\bibfnamefont {M.}~\bibnamefont {Pelton}},\ }\bibfield  {title} {\enquote {\bibinfo {title} {Modified spontaneous emission in nanophotonic structures},}\ }\href {\doibase 10.1038/nphoton.2015.103} {\bibfield  {journal} {\bibinfo  {journal} {Nature Photonics}\ }\textbf {\bibinfo {volume} {9}},\ \bibinfo {pages} {427--435} (\bibinfo {year} {2015})}\BibitemShut {NoStop}%
\bibitem [{\citenamefont {Qian}\ \emph {et~al.}(2021)\citenamefont {Qian}, \citenamefont {Shan}, \citenamefont {Zhang}, \citenamefont {Liu}, \citenamefont {Ma}, \citenamefont {Gong},\ and\ \citenamefont {Gu}}]{qian2021SpontaneousEmission}%
  \BibitemOpen
  \bibfield  {author} {\bibinfo {author} {\bibfnamefont {Z.}~\bibnamefont {Qian}}, \bibinfo {author} {\bibfnamefont {L.}~\bibnamefont {Shan}}, \bibinfo {author} {\bibfnamefont {X.}~\bibnamefont {Zhang}}, \bibinfo {author} {\bibfnamefont {Q.}~\bibnamefont {Liu}}, \bibinfo {author} {\bibfnamefont {Y.}~\bibnamefont {Ma}}, \bibinfo {author} {\bibfnamefont {Q.}~\bibnamefont {Gong}}, \ and\ \bibinfo {author} {\bibfnamefont {Y.}~\bibnamefont {Gu}},\ }\bibfield  {title} {\enquote {\bibinfo {title} {Spontaneous emission in micro- or nanophotonic structures},}\ }\href {\doibase 10.1186/s43074-021-00043-z} {\bibfield  {journal} {\bibinfo  {journal} {PhotoniX}\ }\textbf {\bibinfo {volume} {2}},\ \bibinfo {pages} {21} (\bibinfo {year} {2021})}\BibitemShut {NoStop}%
\bibitem [{\citenamefont {Barnes}, \citenamefont {Horsley},\ and\ \citenamefont {Vos}(2020)}]{barnes2020ClassicalAntennas}%
  \BibitemOpen
  \bibfield  {author} {\bibinfo {author} {\bibfnamefont {W.~L.}\ \bibnamefont {Barnes}}, \bibinfo {author} {\bibfnamefont {S.~A.~R.}\ \bibnamefont {Horsley}}, \ and\ \bibinfo {author} {\bibfnamefont {W.~L.}\ \bibnamefont {Vos}},\ }\bibfield  {title} {\enquote {\bibinfo {title} {Classical antennas, quantum emitters, and densities of optical states},}\ }\href {\doibase 10.1088/2040-8986/ab7b01} {\bibfield  {journal} {\bibinfo  {journal} {Journal of Optics}\ }\textbf {\bibinfo {volume} {22}},\ \bibinfo {pages} {073501} (\bibinfo {year} {2020})}\BibitemShut {NoStop}%
\bibitem [{\citenamefont {Poddubny}, \citenamefont {Iorsh},\ and\ \citenamefont {Sukhorukov}(2016)}]{poddubny2016GenerationPhotonPlasmon}%
  \BibitemOpen
  \bibfield  {author} {\bibinfo {author} {\bibfnamefont {A.~N.}\ \bibnamefont {Poddubny}}, \bibinfo {author} {\bibfnamefont {I.~V.}\ \bibnamefont {Iorsh}}, \ and\ \bibinfo {author} {\bibfnamefont {A.~A.}\ \bibnamefont {Sukhorukov}},\ }\bibfield  {title} {\enquote {\bibinfo {title} {Generation of photon-plasmon quantum states in nonlinear hyperbolic metamaterials},}\ }\href {\doibase 10.1103/PhysRevLett.117.123901} {\bibfield  {journal} {\bibinfo  {journal} {Physical Review Letters}\ }\textbf {\bibinfo {volume} {117}},\ \bibinfo {pages} {123901} (\bibinfo {year} {2016})}\BibitemShut {NoStop}%
\bibitem [{\citenamefont {Parry}\ \emph {et~al.}(2024{\natexlab{b}})\citenamefont {Parry}, \citenamefont {Mazzanti}, \citenamefont {Poddubny}, \citenamefont {Valle}, \citenamefont {Neshev},\ and\ \citenamefont {Sukhorukov}}]{parry2024ChapterNonlinear}%
  \BibitemOpen
  \bibfield  {author} {\bibinfo {author} {\bibfnamefont {M.}~\bibnamefont {Parry}}, \bibinfo {author} {\bibfnamefont {A.}~\bibnamefont {Mazzanti}}, \bibinfo {author} {\bibfnamefont {A.~N.}\ \bibnamefont {Poddubny}}, \bibinfo {author} {\bibfnamefont {G.~D.}\ \bibnamefont {Valle}}, \bibinfo {author} {\bibfnamefont {D.~N.}\ \bibnamefont {Neshev}}, \ and\ \bibinfo {author} {\bibfnamefont {A.~A.}\ \bibnamefont {Sukhorukov}},\ }\bibfield  {title} {\enquote {\bibinfo {title} {Chapter 8 - nonlinear metasurfaces for the generation of quantum photon-pair states},}\ }in\ \href {\doibase 10.1016/B978-0-323-90614-2.00013-4} {\emph {\bibinfo {booktitle} {Fundamentals and Applications of Nonlinear Nanophotonics}}},\ \bibinfo {series and number} {Nanophotonics},\ \bibinfo {editor} {edited by\ \bibinfo {editor} {\bibfnamefont {N.~C.}\ \bibnamefont {Panoiu}}}\ (\bibinfo  {publisher} {Elsevier},\ \bibinfo {year} {2024})\ pp.\ \bibinfo {pages} {271--287}\BibitemShut {NoStop}%
\bibitem [{\citenamefont {Lalanne}\ \emph {et~al.}(2018)\citenamefont {Lalanne}, \citenamefont {Yan}, \citenamefont {Vynck}, \citenamefont {Sauvan},\ and\ \citenamefont {Hugonin}}]{lalanne2018LightInteraction}%
  \BibitemOpen
  \bibfield  {author} {\bibinfo {author} {\bibfnamefont {P.}~\bibnamefont {Lalanne}}, \bibinfo {author} {\bibfnamefont {W.}~\bibnamefont {Yan}}, \bibinfo {author} {\bibfnamefont {K.}~\bibnamefont {Vynck}}, \bibinfo {author} {\bibfnamefont {C.}~\bibnamefont {Sauvan}}, \ and\ \bibinfo {author} {\bibfnamefont {J.-P.}\ \bibnamefont {Hugonin}},\ }\bibfield  {title} {\enquote {\bibinfo {title} {Light interaction with photonic and plasmonic resonances},}\ }\href {\doibase 10.1002/lpor.201700113} {\bibfield  {journal} {\bibinfo  {journal} {Laser \& Photonics Reviews}\ }\textbf {\bibinfo {volume} {12}},\ \bibinfo {pages} {1700113} (\bibinfo {year} {2018})}\BibitemShut {NoStop}%
\bibitem [{\citenamefont {Weissflog}\ \emph {et~al.}(2021)\citenamefont {Weissflog}, \citenamefont {Saravi}, \citenamefont {Gigli}, \citenamefont {Marino}, \citenamefont {Dezert}, \citenamefont {Vinel}, \citenamefont {Borne}, \citenamefont {Leo}, \citenamefont {Pertsch},\ and\ \citenamefont {Setzpfandt}}]{weissflog2021ModellingPhotonpair}%
  \BibitemOpen
  \bibfield  {author} {\bibinfo {author} {\bibfnamefont {M.~A.}\ \bibnamefont {Weissflog}}, \bibinfo {author} {\bibfnamefont {S.}~\bibnamefont {Saravi}}, \bibinfo {author} {\bibfnamefont {C.}~\bibnamefont {Gigli}}, \bibinfo {author} {\bibfnamefont {G.}~\bibnamefont {Marino}}, \bibinfo {author} {\bibfnamefont {R.}~\bibnamefont {Dezert}}, \bibinfo {author} {\bibfnamefont {V.}~\bibnamefont {Vinel}}, \bibinfo {author} {\bibfnamefont {A.}~\bibnamefont {Borne}}, \bibinfo {author} {\bibfnamefont {G.}~\bibnamefont {Leo}}, \bibinfo {author} {\bibfnamefont {T.}~\bibnamefont {Pertsch}}, \ and\ \bibinfo {author} {\bibfnamefont {F.}~\bibnamefont {Setzpfandt}},\ }\bibfield  {title} {\enquote {\bibinfo {title} {Modelling {{Photon-pair Generation}} in {{Nanoresonators Using Quasinormal Mode Expansions}}},}\ }in\ \href {\doibase 10.1364/FIO.2021.FTh6C.5} {\emph {\bibinfo {booktitle} {Frontiers in {{Optics}} + {{Laser Science}} 2021 (2021), Paper {{FTh6C}}.5}}}\ (\bibinfo  {publisher} {{Optica Publishing Group}},\ \bibinfo
  {year} {2021})\ p.\ \bibinfo {pages} {FTh6C.5}\BibitemShut {NoStop}%
\bibitem [{\citenamefont {Weissflog}\ \emph {et~al.}(2024{\natexlab{a}})\citenamefont {Weissflog}, \citenamefont {Dezert}, \citenamefont {Vinel}, \citenamefont {Gigli}, \citenamefont {Leo}, \citenamefont {Pertsch}, \citenamefont {Setzpfandt}, \citenamefont {Borne},\ and\ \citenamefont {Saravi}}]{weissflog2024NonlinearNanoresonators}%
  \BibitemOpen
  \bibfield  {author} {\bibinfo {author} {\bibfnamefont {M.~A.}\ \bibnamefont {Weissflog}}, \bibinfo {author} {\bibfnamefont {R.}~\bibnamefont {Dezert}}, \bibinfo {author} {\bibfnamefont {V.}~\bibnamefont {Vinel}}, \bibinfo {author} {\bibfnamefont {C.}~\bibnamefont {Gigli}}, \bibinfo {author} {\bibfnamefont {G.}~\bibnamefont {Leo}}, \bibinfo {author} {\bibfnamefont {T.}~\bibnamefont {Pertsch}}, \bibinfo {author} {\bibfnamefont {F.}~\bibnamefont {Setzpfandt}}, \bibinfo {author} {\bibfnamefont {A.}~\bibnamefont {Borne}}, \ and\ \bibinfo {author} {\bibfnamefont {S.}~\bibnamefont {Saravi}},\ }\bibfield  {title} {\enquote {\bibinfo {title} {Nonlinear nanoresonators for bell state generation},}\ }\href {\doibase 10.1063/5.0172240} {\bibfield  {journal} {\bibinfo  {journal} {Applied Physics Reviews}\ }\textbf {\bibinfo {volume} {11}},\ \bibinfo {pages} {011403} (\bibinfo {year} {2024}{\natexlab{a}})}\BibitemShut {NoStop}%
\bibitem [{\citenamefont {Aharonovich}, \citenamefont {Englund},\ and\ \citenamefont {Toth}(2016)}]{aharonovich2016SolidstateSinglephoton}%
  \BibitemOpen
  \bibfield  {author} {\bibinfo {author} {\bibfnamefont {I.}~\bibnamefont {Aharonovich}}, \bibinfo {author} {\bibfnamefont {D.}~\bibnamefont {Englund}}, \ and\ \bibinfo {author} {\bibfnamefont {M.}~\bibnamefont {Toth}},\ }\bibfield  {title} {\enquote {\bibinfo {title} {Solid-state single-photon emitters},}\ }\href {\doibase 10.1038/nphoton.2016.186} {\bibfield  {journal} {\bibinfo  {journal} {Nature Photonics}\ }\textbf {\bibinfo {volume} {10}},\ \bibinfo {pages} {631--641} (\bibinfo {year} {2016})}\BibitemShut {NoStop}%
\bibitem [{\citenamefont {Ristori}(2023)}]{ristori2023SinglePhoton}%
  \BibitemOpen
  \bibfield  {author} {\bibinfo {author} {\bibfnamefont {A.}~\bibnamefont {Ristori}},\ }\emph {\bibinfo {title} {Single Photon Emitters}},\ \href {https://hdl.handle.net/2158/1304674} {Ph.D. thesis},\ \bibinfo  {school} {University of Florence} (\bibinfo {year} {2023})\BibitemShut {NoStop}%
\bibitem [{\citenamefont {Kan}\ \emph {et~al.}(2020)\citenamefont {Kan}, \citenamefont {Andersen}, \citenamefont {Ding}, \citenamefont {Kumar}, \citenamefont {Zhao},\ and\ \citenamefont {Bozhevolnyi}}]{kan2020MetasurfaceEnabledGeneration}%
  \BibitemOpen
  \bibfield  {author} {\bibinfo {author} {\bibfnamefont {Y.}~\bibnamefont {Kan}}, \bibinfo {author} {\bibfnamefont {S.~K.~H.}\ \bibnamefont {Andersen}}, \bibinfo {author} {\bibfnamefont {F.}~\bibnamefont {Ding}}, \bibinfo {author} {\bibfnamefont {S.}~\bibnamefont {Kumar}}, \bibinfo {author} {\bibfnamefont {C.}~\bibnamefont {Zhao}}, \ and\ \bibinfo {author} {\bibfnamefont {S.~I.}\ \bibnamefont {Bozhevolnyi}},\ }\bibfield  {title} {\enquote {\bibinfo {title} {Metasurface-enabled generation of circularly polarized single photons},}\ }\href {\doibase 10.1002/adma.201907832} {\bibfield  {journal} {\bibinfo  {journal} {Advanced Materials}\ }\textbf {\bibinfo {volume} {32}},\ \bibinfo {pages} {1907832} (\bibinfo {year} {2020})}\BibitemShut {NoStop}%
\bibitem [{\citenamefont {Komisar}\ \emph {et~al.}(2021)\citenamefont {Komisar}, \citenamefont {Kumar}, \citenamefont {Kan}, \citenamefont {Wu},\ and\ \citenamefont {Bozhevolnyi}}]{komisar2021GenerationRadially}%
  \BibitemOpen
  \bibfield  {author} {\bibinfo {author} {\bibfnamefont {D.}~\bibnamefont {Komisar}}, \bibinfo {author} {\bibfnamefont {S.}~\bibnamefont {Kumar}}, \bibinfo {author} {\bibfnamefont {Y.}~\bibnamefont {Kan}}, \bibinfo {author} {\bibfnamefont {C.}~\bibnamefont {Wu}}, \ and\ \bibinfo {author} {\bibfnamefont {S.~I.}\ \bibnamefont {Bozhevolnyi}},\ }\bibfield  {title} {\enquote {\bibinfo {title} {Generation of radially polarized single photons with plasmonic bullseye antennas},}\ }\href {\doibase 10.1021/acsphotonics.1c00459} {\bibfield  {journal} {\bibinfo  {journal} {ACS Photonics}\ }\textbf {\bibinfo {volume} {8}},\ \bibinfo {pages} {2190--2196} (\bibinfo {year} {2021})}\BibitemShut {NoStop}%
\bibitem [{\citenamefont {Wu}\ \emph {et~al.}(2022)\citenamefont {Wu}, \citenamefont {Kumar}, \citenamefont {Kan}, \citenamefont {Komisar}, \citenamefont {Wang}, \citenamefont {Bozhevolnyi},\ and\ \citenamefont {Ding}}]{wu2022RoomtemperatureOnchip}%
  \BibitemOpen
  \bibfield  {author} {\bibinfo {author} {\bibfnamefont {C.}~\bibnamefont {Wu}}, \bibinfo {author} {\bibfnamefont {S.}~\bibnamefont {Kumar}}, \bibinfo {author} {\bibfnamefont {Y.}~\bibnamefont {Kan}}, \bibinfo {author} {\bibfnamefont {D.}~\bibnamefont {Komisar}}, \bibinfo {author} {\bibfnamefont {Z.}~\bibnamefont {Wang}}, \bibinfo {author} {\bibfnamefont {S.~I.}\ \bibnamefont {Bozhevolnyi}}, \ and\ \bibinfo {author} {\bibfnamefont {F.}~\bibnamefont {Ding}},\ }\bibfield  {title} {\enquote {\bibinfo {title} {Room-temperature on-chip orbital angular momentum single-photon sources},}\ }\href {\doibase 10.1126/sciadv.abk3075} {\bibfield  {journal} {\bibinfo  {journal} {Science Advances}\ }\textbf {\bibinfo {volume} {8}},\ \bibinfo {pages} {eabk3075} (\bibinfo {year} {2022})}\BibitemShut {NoStop}%
\bibitem [{\citenamefont {Liu}\ \emph {et~al.}(2023{\natexlab{a}})\citenamefont {Liu}, \citenamefont {Kan}, \citenamefont {Kumar}, \citenamefont {Komisar}, \citenamefont {Zhao},\ and\ \citenamefont {Bozhevolnyi}}]{liu2023OnchipGeneration}%
  \BibitemOpen
  \bibfield  {author} {\bibinfo {author} {\bibfnamefont {X.}~\bibnamefont {Liu}}, \bibinfo {author} {\bibfnamefont {Y.}~\bibnamefont {Kan}}, \bibinfo {author} {\bibfnamefont {S.}~\bibnamefont {Kumar}}, \bibinfo {author} {\bibfnamefont {D.}~\bibnamefont {Komisar}}, \bibinfo {author} {\bibfnamefont {C.}~\bibnamefont {Zhao}}, \ and\ \bibinfo {author} {\bibfnamefont {S.~I.}\ \bibnamefont {Bozhevolnyi}},\ }\bibfield  {title} {\enquote {\bibinfo {title} {On-chip generation of single-photon circularly polarized single-mode vortex beams},}\ }\href {\doibase 10.1126/sciadv.adh0725} {\bibfield  {journal} {\bibinfo  {journal} {Science Advances}\ }\textbf {\bibinfo {volume} {9}},\ \bibinfo {pages} {eadh0725} (\bibinfo {year} {2023}{\natexlab{a}})}\BibitemShut {NoStop}%
\bibitem [{\citenamefont {Liu}\ \emph {et~al.}(2023{\natexlab{b}})\citenamefont {Liu}, \citenamefont {Kan}, \citenamefont {Kumar}, \citenamefont {Kulikova}, \citenamefont {Davydov}, \citenamefont {Agafonov}, \citenamefont {Zhao},\ and\ \citenamefont {Bozhevolnyi}}]{liu2023UltracompactSinglePhoton}%
  \BibitemOpen
  \bibfield  {author} {\bibinfo {author} {\bibfnamefont {X.}~\bibnamefont {Liu}}, \bibinfo {author} {\bibfnamefont {Y.}~\bibnamefont {Kan}}, \bibinfo {author} {\bibfnamefont {S.}~\bibnamefont {Kumar}}, \bibinfo {author} {\bibfnamefont {L.~F.}\ \bibnamefont {Kulikova}}, \bibinfo {author} {\bibfnamefont {V.~A.}\ \bibnamefont {Davydov}}, \bibinfo {author} {\bibfnamefont {V.~N.}\ \bibnamefont {Agafonov}}, \bibinfo {author} {\bibfnamefont {C.}~\bibnamefont {Zhao}}, \ and\ \bibinfo {author} {\bibfnamefont {S.~I.}\ \bibnamefont {Bozhevolnyi}},\ }\bibfield  {title} {\enquote {\bibinfo {title} {Ultracompact single-photon sources of linearly polarized vortex beams},}\ }\href {\doibase 10.1002/adma.202304495} {\bibfield  {journal} {\bibinfo  {journal} {Advanced Materials}\ }\textbf {\bibinfo {volume} {n/a}},\ \bibinfo {pages} {2304495} (\bibinfo {year} {2023}{\natexlab{b}})}\BibitemShut {NoStop}%
\bibitem [{\citenamefont {Komisar}\ \emph {et~al.}(2023)\citenamefont {Komisar}, \citenamefont {Kumar}, \citenamefont {Kan}, \citenamefont {Meng}, \citenamefont {Kulikova}, \citenamefont {Davydov}, \citenamefont {Agafonov},\ and\ \citenamefont {Bozhevolnyi}}]{komisar2023MultipleChannelling}%
  \BibitemOpen
  \bibfield  {author} {\bibinfo {author} {\bibfnamefont {D.}~\bibnamefont {Komisar}}, \bibinfo {author} {\bibfnamefont {S.}~\bibnamefont {Kumar}}, \bibinfo {author} {\bibfnamefont {Y.}~\bibnamefont {Kan}}, \bibinfo {author} {\bibfnamefont {C.}~\bibnamefont {Meng}}, \bibinfo {author} {\bibfnamefont {L.~F.}\ \bibnamefont {Kulikova}}, \bibinfo {author} {\bibfnamefont {V.~A.}\ \bibnamefont {Davydov}}, \bibinfo {author} {\bibfnamefont {V.~N.}\ \bibnamefont {Agafonov}}, \ and\ \bibinfo {author} {\bibfnamefont {S.~I.}\ \bibnamefont {Bozhevolnyi}},\ }\bibfield  {title} {\enquote {\bibinfo {title} {Multiple channelling single-photon emission with scattering holography designed metasurfaces},}\ }\href {\doibase 10.1038/s41467-023-42046-3} {\bibfield  {journal} {\bibinfo  {journal} {Nature Communications}\ }\textbf {\bibinfo {volume} {14}},\ \bibinfo {pages} {6253} (\bibinfo {year} {2023})}\BibitemShut {NoStop}%
\bibitem [{\citenamefont {Jia}\ \emph {et~al.}(2023)\citenamefont {Jia}, \citenamefont {Li}, \citenamefont {Xue}, \citenamefont {Chen}, \citenamefont {Li}, \citenamefont {Gong},\ and\ \citenamefont {Chen}}]{jia2023MultichannelSinglePhoton}%
  \BibitemOpen
  \bibfield  {author} {\bibinfo {author} {\bibfnamefont {S.}~\bibnamefont {Jia}}, \bibinfo {author} {\bibfnamefont {Y.}~\bibnamefont {Li}}, \bibinfo {author} {\bibfnamefont {Z.}~\bibnamefont {Xue}}, \bibinfo {author} {\bibfnamefont {K.}~\bibnamefont {Chen}}, \bibinfo {author} {\bibfnamefont {Z.}~\bibnamefont {Li}}, \bibinfo {author} {\bibfnamefont {Q.}~\bibnamefont {Gong}}, \ and\ \bibinfo {author} {\bibfnamefont {J.}~\bibnamefont {Chen}},\ }\bibfield  {title} {\enquote {\bibinfo {title} {Multichannel single-photon emissions with on-demand momentums by using anisotropic quantum metasurfaces},}\ }\href {\doibase 10.1002/adma.202212244} {\bibfield  {journal} {\bibinfo  {journal} {Advanced Materials}\ }\textbf {\bibinfo {volume} {35}},\ \bibinfo {pages} {2212244} (\bibinfo {year} {2023})}\BibitemShut {NoStop}%
\bibitem [{\citenamefont {Xue}\ \emph {et~al.}(2022)\citenamefont {Xue}, \citenamefont {Jia}, \citenamefont {Li}, \citenamefont {Li}, \citenamefont {Gong},\ and\ \citenamefont {Chen}}]{xue2022ScalarSuperpositionMetasurfaces}%
  \BibitemOpen
  \bibfield  {author} {\bibinfo {author} {\bibfnamefont {Z.}~\bibnamefont {Xue}}, \bibinfo {author} {\bibfnamefont {S.}~\bibnamefont {Jia}}, \bibinfo {author} {\bibfnamefont {X.}~\bibnamefont {Li}}, \bibinfo {author} {\bibfnamefont {Z.}~\bibnamefont {Li}}, \bibinfo {author} {\bibfnamefont {Q.}~\bibnamefont {Gong}}, \ and\ \bibinfo {author} {\bibfnamefont {J.}~\bibnamefont {Chen}},\ }\bibfield  {title} {\enquote {\bibinfo {title} {Scalar-superposition metasurfaces with robust placement of quantum emitters for tailoring single-photon emission polarization},}\ }\href {\doibase 10.1002/lpor.202200179} {\bibfield  {journal} {\bibinfo  {journal} {Laser \& Photonics Reviews}\ }\textbf {\bibinfo {volume} {16}},\ \bibinfo {pages} {2200179} (\bibinfo {year} {2022})}\BibitemShut {NoStop}%
\bibitem [{\citenamefont {Checcucci}\ \emph {et~al.}(2017)\citenamefont {Checcucci}, \citenamefont {Lombardi}, \citenamefont {Rizvi}, \citenamefont {Sgrignuoli}, \citenamefont {Gruhler}, \citenamefont {Dieleman}, \citenamefont {S~Cataliotti}, \citenamefont {Pernice}, \citenamefont {Agio},\ and\ \citenamefont {Toninelli}}]{checcucci2017BeamingLight}%
  \BibitemOpen
  \bibfield  {author} {\bibinfo {author} {\bibfnamefont {S.}~\bibnamefont {Checcucci}}, \bibinfo {author} {\bibfnamefont {P.}~\bibnamefont {Lombardi}}, \bibinfo {author} {\bibfnamefont {S.}~\bibnamefont {Rizvi}}, \bibinfo {author} {\bibfnamefont {F.}~\bibnamefont {Sgrignuoli}}, \bibinfo {author} {\bibfnamefont {N.}~\bibnamefont {Gruhler}}, \bibinfo {author} {\bibfnamefont {F.~B.}\ \bibnamefont {Dieleman}}, \bibinfo {author} {\bibfnamefont {F.}~\bibnamefont {S~Cataliotti}}, \bibinfo {author} {\bibfnamefont {W.~H.}\ \bibnamefont {Pernice}}, \bibinfo {author} {\bibfnamefont {M.}~\bibnamefont {Agio}}, \ and\ \bibinfo {author} {\bibfnamefont {C.}~\bibnamefont {Toninelli}},\ }\bibfield  {title} {\enquote {\bibinfo {title} {Beaming light from a quantum emitter with a planar optical antenna},}\ }\href {\doibase 10.1038/lsa.2016.245} {\bibfield  {journal} {\bibinfo  {journal} {Light: Science \& Applications}\ }\textbf {\bibinfo {volume} {6}},\ \bibinfo {pages} {e16245} (\bibinfo {year} {2017})}\BibitemShut {NoStop}%
\bibitem [{\citenamefont {Huang}\ \emph {et~al.}(2019)\citenamefont {Huang}, \citenamefont {Grote}, \citenamefont {Mann}, \citenamefont {Hopper}, \citenamefont {Exarhos}, \citenamefont {Lopez}, \citenamefont {Klein}, \citenamefont {Garnett},\ and\ \citenamefont {Bassett}}]{huang2019MonolithicImmersion}%
  \BibitemOpen
  \bibfield  {author} {\bibinfo {author} {\bibfnamefont {T.-Y.}\ \bibnamefont {Huang}}, \bibinfo {author} {\bibfnamefont {R.~R.}\ \bibnamefont {Grote}}, \bibinfo {author} {\bibfnamefont {S.~A.}\ \bibnamefont {Mann}}, \bibinfo {author} {\bibfnamefont {D.~A.}\ \bibnamefont {Hopper}}, \bibinfo {author} {\bibfnamefont {A.~L.}\ \bibnamefont {Exarhos}}, \bibinfo {author} {\bibfnamefont {G.~G.}\ \bibnamefont {Lopez}}, \bibinfo {author} {\bibfnamefont {A.~R.}\ \bibnamefont {Klein}}, \bibinfo {author} {\bibfnamefont {E.~C.}\ \bibnamefont {Garnett}}, \ and\ \bibinfo {author} {\bibfnamefont {L.~C.}\ \bibnamefont {Bassett}},\ }\bibfield  {title} {\enquote {\bibinfo {title} {A monolithic immersion metalens for imaging solid-state quantum emitters},}\ }\href {\doibase 10.1038/s41467-019-10238-5} {\bibfield  {journal} {\bibinfo  {journal} {Nature Communications}\ }\textbf {\bibinfo {volume} {10}},\ \bibinfo {pages} {2392} (\bibinfo {year} {2019})}\BibitemShut {NoStop}%
\bibitem [{\citenamefont {Bao}\ \emph {et~al.}(2020)\citenamefont {Bao}, \citenamefont {Lin}, \citenamefont {Su}, \citenamefont {Zhou}, \citenamefont {Song}, \citenamefont {Li},\ and\ \citenamefont {Wang}}]{bao2020OndemandSpinstate}%
  \BibitemOpen
  \bibfield  {author} {\bibinfo {author} {\bibfnamefont {Y.}~\bibnamefont {Bao}}, \bibinfo {author} {\bibfnamefont {Q.}~\bibnamefont {Lin}}, \bibinfo {author} {\bibfnamefont {R.}~\bibnamefont {Su}}, \bibinfo {author} {\bibfnamefont {Z.-K.}\ \bibnamefont {Zhou}}, \bibinfo {author} {\bibfnamefont {J.}~\bibnamefont {Song}}, \bibinfo {author} {\bibfnamefont {J.}~\bibnamefont {Li}}, \ and\ \bibinfo {author} {\bibfnamefont {X.-H.}\ \bibnamefont {Wang}},\ }\bibfield  {title} {\enquote {\bibinfo {title} {On-demand spin-state manipulation of single-photon emission from quantum dot integrated with metasurface},}\ }\href {\doibase 10.1126/sciadv.aba8761} {\bibfield  {journal} {\bibinfo  {journal} {Science Advances}\ }\textbf {\bibinfo {volume} {6}},\ \bibinfo {pages} {eaba8761} (\bibinfo {year} {2020})}\BibitemShut {NoStop}%
\bibitem [{\citenamefont {Li}\ \emph {et~al.}(2023{\natexlab{b}})\citenamefont {Li}, \citenamefont {Liu}, \citenamefont {Wei}, \citenamefont {Ma}, \citenamefont {Song}, \citenamefont {Yu}, \citenamefont {Su}, \citenamefont {Geng}, \citenamefont {Ni}, \citenamefont {Liu}, \citenamefont {Su}, \citenamefont {Niu}, \citenamefont {Chen},\ and\ \citenamefont {Liu}}]{li2023BrightSemiconductor}%
  \BibitemOpen
  \bibfield  {author} {\bibinfo {author} {\bibfnamefont {X.}~\bibnamefont {Li}}, \bibinfo {author} {\bibfnamefont {S.}~\bibnamefont {Liu}}, \bibinfo {author} {\bibfnamefont {Y.}~\bibnamefont {Wei}}, \bibinfo {author} {\bibfnamefont {J.}~\bibnamefont {Ma}}, \bibinfo {author} {\bibfnamefont {C.}~\bibnamefont {Song}}, \bibinfo {author} {\bibfnamefont {Y.}~\bibnamefont {Yu}}, \bibinfo {author} {\bibfnamefont {R.}~\bibnamefont {Su}}, \bibinfo {author} {\bibfnamefont {W.}~\bibnamefont {Geng}}, \bibinfo {author} {\bibfnamefont {H.}~\bibnamefont {Ni}}, \bibinfo {author} {\bibfnamefont {H.}~\bibnamefont {Liu}}, \bibinfo {author} {\bibfnamefont {X.}~\bibnamefont {Su}}, \bibinfo {author} {\bibfnamefont {Z.}~\bibnamefont {Niu}}, \bibinfo {author} {\bibfnamefont {Y.-l.}\ \bibnamefont {Chen}}, \ and\ \bibinfo {author} {\bibfnamefont {J.}~\bibnamefont {Liu}},\ }\bibfield  {title} {\enquote {\bibinfo {title} {Bright semiconductor single-photon sources pumped by heterogeneously integrated micropillar lasers with electrical
  injections},}\ }\href {\doibase 10.1038/s41377-023-01110-9} {\bibfield  {journal} {\bibinfo  {journal} {Light: Science \& Applications}\ }\textbf {\bibinfo {volume} {12}},\ \bibinfo {pages} {65} (\bibinfo {year} {2023}{\natexlab{b}})}\BibitemShut {NoStop}%
\bibitem [{\citenamefont {Okoth}\ \emph {et~al.}(2019)\citenamefont {Okoth}, \citenamefont {Cavanna}, \citenamefont {{Santiago-Cruz}},\ and\ \citenamefont {Chekhova}}]{okoth2019MicroscaleGeneration}%
  \BibitemOpen
  \bibfield  {author} {\bibinfo {author} {\bibfnamefont {C.}~\bibnamefont {Okoth}}, \bibinfo {author} {\bibfnamefont {A.}~\bibnamefont {Cavanna}}, \bibinfo {author} {\bibfnamefont {T.}~\bibnamefont {{Santiago-Cruz}}}, \ and\ \bibinfo {author} {\bibfnamefont {M.~V.}\ \bibnamefont {Chekhova}},\ }\bibfield  {title} {\enquote {\bibinfo {title} {Microscale generation of entangled photons without momentum conservation},}\ }\href {\doibase 10.1103/PhysRevLett.123.263602} {\bibfield  {journal} {\bibinfo  {journal} {Physical Review Letters}\ }\textbf {\bibinfo {volume} {123}},\ \bibinfo {pages} {263602} (\bibinfo {year} {2019})}\BibitemShut {NoStop}%
\bibitem [{\citenamefont {Okoth}\ \emph {et~al.}(2020)\citenamefont {Okoth}, \citenamefont {Kovlakov}, \citenamefont {B{\"o}nsel}, \citenamefont {Cavanna}, \citenamefont {Straupe}, \citenamefont {Kulik},\ and\ \citenamefont {Chekhova}}]{okoth2020IdealizedEinsteinPodolskyRosen}%
  \BibitemOpen
  \bibfield  {author} {\bibinfo {author} {\bibfnamefont {C.}~\bibnamefont {Okoth}}, \bibinfo {author} {\bibfnamefont {E.}~\bibnamefont {Kovlakov}}, \bibinfo {author} {\bibfnamefont {F.}~\bibnamefont {B{\"o}nsel}}, \bibinfo {author} {\bibfnamefont {A.}~\bibnamefont {Cavanna}}, \bibinfo {author} {\bibfnamefont {S.}~\bibnamefont {Straupe}}, \bibinfo {author} {\bibfnamefont {S.~P.}\ \bibnamefont {Kulik}}, \ and\ \bibinfo {author} {\bibfnamefont {M.~V.}\ \bibnamefont {Chekhova}},\ }\bibfield  {title} {\enquote {\bibinfo {title} {Idealized einstein-podolsky-rosen states from non--phase-matched parametric down-conversion},}\ }\href {\doibase 10.1103/PhysRevA.101.011801} {\bibfield  {journal} {\bibinfo  {journal} {Physical Review A}\ }\textbf {\bibinfo {volume} {101}},\ \bibinfo {pages} {011801} (\bibinfo {year} {2020})}\BibitemShut {NoStop}%
\bibitem [{\citenamefont {Sultanov}\ and\ \citenamefont {Chekhova}(2023)}]{sultanov2023TemporallyDistilled}%
  \BibitemOpen
  \bibfield  {author} {\bibinfo {author} {\bibfnamefont {V.}~\bibnamefont {Sultanov}}\ and\ \bibinfo {author} {\bibfnamefont {M.}~\bibnamefont {Chekhova}},\ }\bibfield  {title} {\enquote {\bibinfo {title} {Temporally distilled high-dimensional biphotonic states from thin sources},}\ }\href {\doibase 10.1021/acsphotonics.3c01169} {\bibfield  {journal} {\bibinfo  {journal} {ACS Photonics}\ }\textbf {\bibinfo {volume} {11}},\ \bibinfo {pages} {2--6} (\bibinfo {year} {2023})}\BibitemShut {NoStop}%
\bibitem [{\citenamefont {Lenzini}\ \emph {et~al.}(2018)\citenamefont {Lenzini}, \citenamefont {Poddubny}, \citenamefont {Titchener}, \citenamefont {Fisher}, \citenamefont {Boes}, \citenamefont {Kasture}, \citenamefont {Haylock}, \citenamefont {Villa}, \citenamefont {Mitchell}, \citenamefont {Solntsev}, \citenamefont {Sukhorukov},\ and\ \citenamefont {Lobino}}]{lenzini2018DirectCharacterization}%
  \BibitemOpen
  \bibfield  {author} {\bibinfo {author} {\bibfnamefont {F.}~\bibnamefont {Lenzini}}, \bibinfo {author} {\bibfnamefont {A.~N.}\ \bibnamefont {Poddubny}}, \bibinfo {author} {\bibfnamefont {J.}~\bibnamefont {Titchener}}, \bibinfo {author} {\bibfnamefont {P.}~\bibnamefont {Fisher}}, \bibinfo {author} {\bibfnamefont {A.}~\bibnamefont {Boes}}, \bibinfo {author} {\bibfnamefont {S.}~\bibnamefont {Kasture}}, \bibinfo {author} {\bibfnamefont {B.}~\bibnamefont {Haylock}}, \bibinfo {author} {\bibfnamefont {M.}~\bibnamefont {Villa}}, \bibinfo {author} {\bibfnamefont {A.}~\bibnamefont {Mitchell}}, \bibinfo {author} {\bibfnamefont {A.~S.}\ \bibnamefont {Solntsev}}, \bibinfo {author} {\bibfnamefont {A.~A.}\ \bibnamefont {Sukhorukov}}, \ and\ \bibinfo {author} {\bibfnamefont {M.}~\bibnamefont {Lobino}},\ }\bibfield  {title} {\enquote {\bibinfo {title} {Direct characterization of a nonlinear photonic circuit’s wave function with laser light},}\ }\href {\doibase 10.1038/lsa.2017.143} {\bibfield  {journal} {\bibinfo
  {journal} {Light: Science \& Applications}\ }\textbf {\bibinfo {volume} {7}},\ \bibinfo {pages} {17143} (\bibinfo {year} {2018})}\BibitemShut {NoStop}%
\bibitem [{\citenamefont {Li}, \citenamefont {Zhang},\ and\ \citenamefont {Zentgraf}(2017)}]{li2017NonlinearPhotonic}%
  \BibitemOpen
  \bibfield  {author} {\bibinfo {author} {\bibfnamefont {G.}~\bibnamefont {Li}}, \bibinfo {author} {\bibfnamefont {S.}~\bibnamefont {Zhang}}, \ and\ \bibinfo {author} {\bibfnamefont {T.}~\bibnamefont {Zentgraf}},\ }\bibfield  {title} {\enquote {\bibinfo {title} {Nonlinear photonic metasurfaces},}\ }\href {\doibase 10.1038/natrevmats.2017.10} {\bibfield  {journal} {\bibinfo  {journal} {Nature Reviews Materials}\ }\textbf {\bibinfo {volume} {2}},\ \bibinfo {pages} {1--14} (\bibinfo {year} {2017})}\BibitemShut {NoStop}%
\bibitem [{\citenamefont {Krasnok}, \citenamefont {Tymchenko},\ and\ \citenamefont {Alù}(2018)}]{krasnok2018NonlinearMetasurfaces}%
  \BibitemOpen
  \bibfield  {author} {\bibinfo {author} {\bibfnamefont {A.}~\bibnamefont {Krasnok}}, \bibinfo {author} {\bibfnamefont {M.}~\bibnamefont {Tymchenko}}, \ and\ \bibinfo {author} {\bibfnamefont {A.}~\bibnamefont {Alù}},\ }\bibfield  {title} {\enquote {\bibinfo {title} {Nonlinear metasurfaces: A paradigm shift in nonlinear optics},}\ }\href {\doibase 10.1016/j.mattod.2017.06.007} {\bibfield  {journal} {\bibinfo  {journal} {Materials Today}\ }\textbf {\bibinfo {volume} {21}},\ \bibinfo {pages} {8--21} (\bibinfo {year} {2018})}\BibitemShut {NoStop}%
\bibitem [{\citenamefont {Pertsch}\ and\ \citenamefont {Kivshar}(2020)}]{pertsch2020NonlinearOptics}%
  \BibitemOpen
  \bibfield  {author} {\bibinfo {author} {\bibfnamefont {T.}~\bibnamefont {Pertsch}}\ and\ \bibinfo {author} {\bibfnamefont {Y.}~\bibnamefont {Kivshar}},\ }\bibfield  {title} {\enquote {\bibinfo {title} {Nonlinear optics with resonant metasurfaces},}\ }\href {\doibase 10.1557/mrs.2020.65} {\bibfield  {journal} {\bibinfo  {journal} {MRS Bulletin}\ }\textbf {\bibinfo {volume} {45}},\ \bibinfo {pages} {210--220} (\bibinfo {year} {2020})}\BibitemShut {NoStop}%
\bibitem [{\citenamefont {Gigli}\ and\ \citenamefont {Leo}(2022)}]{gigli2022AlldielectricMetasurfaces}%
  \BibitemOpen
  \bibfield  {author} {\bibinfo {author} {\bibfnamefont {C.}~\bibnamefont {Gigli}}\ and\ \bibinfo {author} {\bibfnamefont {G.}~\bibnamefont {Leo}},\ }\bibfield  {title} {\enquote {\bibinfo {title} {All-dielectric {X}{\textsuperscript{(2)}} metasurfaces: Recent progress},}\ }\href {\doibase 10.29026/oea.2022.210093} {\bibfield  {journal} {\bibinfo  {journal} {Opto-Electronic Advances}\ }\textbf {\bibinfo {volume} {5}},\ \bibinfo {pages} {210093--14} (\bibinfo {year} {2022})}\BibitemShut {NoStop}%
\bibitem [{\citenamefont {Marino}\ \emph {et~al.}(2019)\citenamefont {Marino}, \citenamefont {Solntsev}, \citenamefont {Xu}, \citenamefont {Gili}, \citenamefont {Carletti}, \citenamefont {Poddubny}, \citenamefont {Rahmani}, \citenamefont {Smirnova}, \citenamefont {Chen}, \citenamefont {Lema{\^i}tre}, \citenamefont {Zhang}, \citenamefont {Zayats}, \citenamefont {De~Angelis}, \citenamefont {Leo}, \citenamefont {Sukhorukov},\ and\ \citenamefont {Neshev}}]{marino2019SpontaneousPhotonpairb}%
  \BibitemOpen
  \bibfield  {author} {\bibinfo {author} {\bibfnamefont {G.}~\bibnamefont {Marino}}, \bibinfo {author} {\bibfnamefont {A.~S.}\ \bibnamefont {Solntsev}}, \bibinfo {author} {\bibfnamefont {L.}~\bibnamefont {Xu}}, \bibinfo {author} {\bibfnamefont {V.~F.}\ \bibnamefont {Gili}}, \bibinfo {author} {\bibfnamefont {L.}~\bibnamefont {Carletti}}, \bibinfo {author} {\bibfnamefont {A.~N.}\ \bibnamefont {Poddubny}}, \bibinfo {author} {\bibfnamefont {M.}~\bibnamefont {Rahmani}}, \bibinfo {author} {\bibfnamefont {D.~A.}\ \bibnamefont {Smirnova}}, \bibinfo {author} {\bibfnamefont {H.}~\bibnamefont {Chen}}, \bibinfo {author} {\bibfnamefont {A.}~\bibnamefont {Lema{\^i}tre}}, \bibinfo {author} {\bibfnamefont {G.}~\bibnamefont {Zhang}}, \bibinfo {author} {\bibfnamefont {A.~V.}\ \bibnamefont {Zayats}}, \bibinfo {author} {\bibfnamefont {C.}~\bibnamefont {De~Angelis}}, \bibinfo {author} {\bibfnamefont {G.}~\bibnamefont {Leo}}, \bibinfo {author} {\bibfnamefont {A.~A.}\ \bibnamefont {Sukhorukov}}, \ and\ \bibinfo {author}
  {\bibfnamefont {D.~N.}\ \bibnamefont {Neshev}},\ }\bibfield  {title} {\enquote {\bibinfo {title} {Spontaneous photon-pair generation from a dielectric nanoantenna},}\ }\href {\doibase 10.1364/OPTICA.6.001416} {\bibfield  {journal} {\bibinfo  {journal} {Optica}\ }\textbf {\bibinfo {volume} {6}},\ \bibinfo {pages} {1416} (\bibinfo {year} {2019})}\BibitemShut {NoStop}%
\bibitem [{\citenamefont {Nikolaeva}\ \emph {et~al.}(2021)\citenamefont {Nikolaeva}, \citenamefont {Frizyuk}, \citenamefont {Olekhno}, \citenamefont {Solntsev},\ and\ \citenamefont {Petrov}}]{PhysRevA.103.043703}%
  \BibitemOpen
  \bibfield  {author} {\bibinfo {author} {\bibfnamefont {A.}~\bibnamefont {Nikolaeva}}, \bibinfo {author} {\bibfnamefont {K.}~\bibnamefont {Frizyuk}}, \bibinfo {author} {\bibfnamefont {N.}~\bibnamefont {Olekhno}}, \bibinfo {author} {\bibfnamefont {A.}~\bibnamefont {Solntsev}}, \ and\ \bibinfo {author} {\bibfnamefont {M.}~\bibnamefont {Petrov}},\ }\bibfield  {title} {\enquote {\bibinfo {title} {Directional emission of down-converted photons from a dielectric nanoresonator},}\ }\href {\doibase 10.1103/PhysRevA.103.043703} {\bibfield  {journal} {\bibinfo  {journal} {Phys. Rev. A}\ }\textbf {\bibinfo {volume} {103}},\ \bibinfo {pages} {043703} (\bibinfo {year} {2021})}\BibitemShut {NoStop}%
\bibitem [{\citenamefont {Duong}\ \emph {et~al.}(2022)\citenamefont {Duong}, \citenamefont {Saerens}, \citenamefont {Timpu}, \citenamefont {Buscaglia}, \citenamefont {Buscaglia}, \citenamefont {Morandi}, \citenamefont {Müller}, \citenamefont {Maeder}, \citenamefont {Kaufmann}, \citenamefont {Solntsev},\ and\ \citenamefont {Grange}}]{duong2022SpontaneousParametric}%
  \BibitemOpen
  \bibfield  {author} {\bibinfo {author} {\bibfnamefont {N.~M.~H.}\ \bibnamefont {Duong}}, \bibinfo {author} {\bibfnamefont {G.}~\bibnamefont {Saerens}}, \bibinfo {author} {\bibfnamefont {F.}~\bibnamefont {Timpu}}, \bibinfo {author} {\bibfnamefont {M.~T.}\ \bibnamefont {Buscaglia}}, \bibinfo {author} {\bibfnamefont {V.}~\bibnamefont {Buscaglia}}, \bibinfo {author} {\bibfnamefont {A.}~\bibnamefont {Morandi}}, \bibinfo {author} {\bibfnamefont {J.~S.}\ \bibnamefont {Müller}}, \bibinfo {author} {\bibfnamefont {A.}~\bibnamefont {Maeder}}, \bibinfo {author} {\bibfnamefont {F.}~\bibnamefont {Kaufmann}}, \bibinfo {author} {\bibfnamefont {A.~S.}\ \bibnamefont {Solntsev}}, \ and\ \bibinfo {author} {\bibfnamefont {R.}~\bibnamefont {Grange}},\ }\bibfield  {title} {\enquote {\bibinfo {title} {Spontaneous parametric down-conversion in bottom-up grown lithium niobate microcubes},}\ }\href {\doibase 10.1364/OME.462981} {\bibfield  {journal} {\bibinfo  {journal} {Optical Materials Express}\ }\textbf {\bibinfo {volume}
  {12}},\ \bibinfo {pages} {3696--3704} (\bibinfo {year} {2022})}\BibitemShut {NoStop}%
\bibitem [{\citenamefont {Saerens}\ \emph {et~al.}(2023)\citenamefont {Saerens}, \citenamefont {Dursap}, \citenamefont {Hesner}, \citenamefont {Duong}, \citenamefont {Solntsev}, \citenamefont {Morandi}, \citenamefont {Maeder}, \citenamefont {Karvounis}, \citenamefont {Regreny}, \citenamefont {Chapman}, \citenamefont {Danescu}, \citenamefont {Chauvin}, \citenamefont {Penuelas},\ and\ \citenamefont {Grange}}]{saerens2023BackgroundFreeNearInfrared}%
  \BibitemOpen
  \bibfield  {author} {\bibinfo {author} {\bibfnamefont {G.}~\bibnamefont {Saerens}}, \bibinfo {author} {\bibfnamefont {T.}~\bibnamefont {Dursap}}, \bibinfo {author} {\bibfnamefont {I.}~\bibnamefont {Hesner}}, \bibinfo {author} {\bibfnamefont {N.~M.~H.}\ \bibnamefont {Duong}}, \bibinfo {author} {\bibfnamefont {A.~S.}\ \bibnamefont {Solntsev}}, \bibinfo {author} {\bibfnamefont {A.}~\bibnamefont {Morandi}}, \bibinfo {author} {\bibfnamefont {A.}~\bibnamefont {Maeder}}, \bibinfo {author} {\bibfnamefont {A.}~\bibnamefont {Karvounis}}, \bibinfo {author} {\bibfnamefont {P.}~\bibnamefont {Regreny}}, \bibinfo {author} {\bibfnamefont {R.~J.}\ \bibnamefont {Chapman}}, \bibinfo {author} {\bibfnamefont {A.}~\bibnamefont {Danescu}}, \bibinfo {author} {\bibfnamefont {N.}~\bibnamefont {Chauvin}}, \bibinfo {author} {\bibfnamefont {J.}~\bibnamefont {Penuelas}}, \ and\ \bibinfo {author} {\bibfnamefont {R.}~\bibnamefont {Grange}},\ }\bibfield  {title} {\enquote {\bibinfo {title} {Background-free near-infrared biphoton emission
  from single gaas nanowires},}\ }\href {\doibase 10.1021/acs.nanolett.3c00026} {\bibfield  {journal} {\bibinfo  {journal} {Nano Letters}\ }\textbf {\bibinfo {volume} {23}},\ \bibinfo {pages} {3245--3250} (\bibinfo {year} {2023})}\BibitemShut {NoStop}%
\bibitem [{\citenamefont {{Santiago-Cruz}}\ \emph {et~al.}(2021)\citenamefont {{Santiago-Cruz}}, \citenamefont {Fedotova}, \citenamefont {Sultanov}, \citenamefont {Weissflog}, \citenamefont {Arslan}, \citenamefont {Younesi}, \citenamefont {Pertsch}, \citenamefont {Staude}, \citenamefont {Setzpfandt},\ and\ \citenamefont {Chekhova}}]{santiago-cruz2021PhotonPairs}%
  \BibitemOpen
  \bibfield  {author} {\bibinfo {author} {\bibfnamefont {T.}~\bibnamefont {{Santiago-Cruz}}}, \bibinfo {author} {\bibfnamefont {A.}~\bibnamefont {Fedotova}}, \bibinfo {author} {\bibfnamefont {V.}~\bibnamefont {Sultanov}}, \bibinfo {author} {\bibfnamefont {M.~A.}\ \bibnamefont {Weissflog}}, \bibinfo {author} {\bibfnamefont {D.}~\bibnamefont {Arslan}}, \bibinfo {author} {\bibfnamefont {M.}~\bibnamefont {Younesi}}, \bibinfo {author} {\bibfnamefont {T.}~\bibnamefont {Pertsch}}, \bibinfo {author} {\bibfnamefont {I.}~\bibnamefont {Staude}}, \bibinfo {author} {\bibfnamefont {F.}~\bibnamefont {Setzpfandt}}, \ and\ \bibinfo {author} {\bibfnamefont {M.}~\bibnamefont {Chekhova}},\ }\bibfield  {title} {\enquote {\bibinfo {title} {Photon pairs from resonant metasurfaces},}\ }\href {\doibase 10.1021/acs.nanolett.1c01125} {\bibfield  {journal} {\bibinfo  {journal} {Nano Letters}\ }\textbf {\bibinfo {volume} {21}},\ \bibinfo {pages} {4423--4429} (\bibinfo {year} {2021})}\BibitemShut {NoStop}%
\bibitem [{\citenamefont {Parry}\ \emph {et~al.}(2021)\citenamefont {Parry}, \citenamefont {Mazzanti}, \citenamefont {Poddubny}, \citenamefont {Valle}, \citenamefont {Neshev},\ and\ \citenamefont {Sukhorukov}}]{parry2021EnhancedGenerationb}%
  \BibitemOpen
  \bibfield  {author} {\bibinfo {author} {\bibfnamefont {M.}~\bibnamefont {Parry}}, \bibinfo {author} {\bibfnamefont {A.}~\bibnamefont {Mazzanti}}, \bibinfo {author} {\bibfnamefont {A.~N.}\ \bibnamefont {Poddubny}}, \bibinfo {author} {\bibfnamefont {G.~D.}\ \bibnamefont {Valle}}, \bibinfo {author} {\bibfnamefont {D.~N.}\ \bibnamefont {Neshev}}, \ and\ \bibinfo {author} {\bibfnamefont {A.~A.}\ \bibnamefont {Sukhorukov}},\ }\bibfield  {title} {\enquote {\bibinfo {title} {Enhanced generation of nondegenerate photon pairs in nonlinear metasurfaces},}\ }\href {\doibase 10.1117/1.AP.3.5.055001} {\bibfield  {journal} {\bibinfo  {journal} {Advanced Photonics}\ }\textbf {\bibinfo {volume} {3}},\ \bibinfo {pages} {055001} (\bibinfo {year} {2021})}\BibitemShut {NoStop}%
\bibitem [{\citenamefont {Jin}, \citenamefont {Mishra},\ and\ \citenamefont {Argyropoulos}(2021)}]{jin2021EfficientSinglephoton}%
  \BibitemOpen
  \bibfield  {author} {\bibinfo {author} {\bibfnamefont {B.}~\bibnamefont {Jin}}, \bibinfo {author} {\bibfnamefont {D.}~\bibnamefont {Mishra}}, \ and\ \bibinfo {author} {\bibfnamefont {C.}~\bibnamefont {Argyropoulos}},\ }\bibfield  {title} {\enquote {\bibinfo {title} {Efficient single-photon pair generation by spontaneous parametric down-conversion in nonlinear plasmonic metasurfaces},}\ }\href {\doibase 10.1039/D1NR05379E} {\bibfield  {journal} {\bibinfo  {journal} {Nanoscale}\ }\textbf {\bibinfo {volume} {13}},\ \bibinfo {pages} {19903--19914} (\bibinfo {year} {2021})}\BibitemShut {NoStop}%
\bibitem [{\citenamefont {Mazzanti}\ \emph {et~al.}(2022)\citenamefont {Mazzanti}, \citenamefont {Parry}, \citenamefont {Poddubny}, \citenamefont {Valle}, \citenamefont {Neshev},\ and\ \citenamefont {Sukhorukov}}]{mazzanti2022EnhancedGeneration}%
  \BibitemOpen
  \bibfield  {author} {\bibinfo {author} {\bibfnamefont {A.}~\bibnamefont {Mazzanti}}, \bibinfo {author} {\bibfnamefont {M.}~\bibnamefont {Parry}}, \bibinfo {author} {\bibfnamefont {A.}~\bibnamefont {Poddubny}}, \bibinfo {author} {\bibfnamefont {G.~D.}\ \bibnamefont {Valle}}, \bibinfo {author} {\bibfnamefont {D.}~\bibnamefont {Neshev}}, \ and\ \bibinfo {author} {\bibfnamefont {A.~A.}\ \bibnamefont {Sukhorukov}},\ }\bibfield  {title} {\enquote {\bibinfo {title} {Enhanced generation of angle correlated photon-pairs in nonlinear metasurfaces},}\ }\href {\doibase 10.1088/1367-2630/ac599e} {\bibfield  {journal} {\bibinfo  {journal} {New Journal of Physics}\ }\textbf {\bibinfo {volume} {24}},\ \bibinfo {pages} {035006} (\bibinfo {year} {2022})}\BibitemShut {NoStop}%
\bibitem [{\citenamefont {Zhang}\ \emph {et~al.}(2022{\natexlab{c}})\citenamefont {Zhang}, \citenamefont {Ma}, \citenamefont {Neshev},\ and\ \citenamefont {Sukhorukov}}]{zhang2022PhotonPair}%
  \BibitemOpen
  \bibfield  {author} {\bibinfo {author} {\bibfnamefont {J.}~\bibnamefont {Zhang}}, \bibinfo {author} {\bibfnamefont {J.}~\bibnamefont {Ma}}, \bibinfo {author} {\bibfnamefont {D.~N.}\ \bibnamefont {Neshev}}, \ and\ \bibinfo {author} {\bibfnamefont {A.~A.}\ \bibnamefont {Sukhorukov}},\ }\bibfield  {title} {\enquote {\bibinfo {title} {Photon pair generation from lithium niobate metasurface with tunable spatial entanglement [{{Invited}}]},}\ }\href {\doibase 10.3788/COL202321.010005} {\bibfield  {journal} {\bibinfo  {journal} {Chinese Optics Letters}\ }\textbf {\bibinfo {volume} {21}},\ \bibinfo {pages} {010005} (\bibinfo {year} {2022}{\natexlab{c}})}\BibitemShut {NoStop}%
\bibitem [{\citenamefont {{Santiago-Cruz}}\ \emph {et~al.}(2022)\citenamefont {{Santiago-Cruz}}, \citenamefont {Gennaro}, \citenamefont {Mitrofanov}, \citenamefont {Addamane}, \citenamefont {Reno}, \citenamefont {Brener},\ and\ \citenamefont {Chekhova}}]{santiago-cruz2022ResonantMetasurfaces}%
  \BibitemOpen
  \bibfield  {author} {\bibinfo {author} {\bibfnamefont {T.}~\bibnamefont {{Santiago-Cruz}}}, \bibinfo {author} {\bibfnamefont {S.~D.}\ \bibnamefont {Gennaro}}, \bibinfo {author} {\bibfnamefont {O.}~\bibnamefont {Mitrofanov}}, \bibinfo {author} {\bibfnamefont {S.}~\bibnamefont {Addamane}}, \bibinfo {author} {\bibfnamefont {J.}~\bibnamefont {Reno}}, \bibinfo {author} {\bibfnamefont {I.}~\bibnamefont {Brener}}, \ and\ \bibinfo {author} {\bibfnamefont {M.~V.}\ \bibnamefont {Chekhova}},\ }\bibfield  {title} {\enquote {\bibinfo {title} {Resonant metasurfaces for generating complex quantum states},}\ }\href {\doibase 10.1126/science.abq8684} {\bibfield  {journal} {\bibinfo  {journal} {Science}\ }\textbf {\bibinfo {volume} {377}},\ \bibinfo {pages} {991--995} (\bibinfo {year} {2022})}\BibitemShut {NoStop}%
\bibitem [{\citenamefont {Ma}\ \emph {et~al.}(2023)\citenamefont {Ma}, \citenamefont {Zhang}, \citenamefont {Jiang}, \citenamefont {Fan}, \citenamefont {Parry}, \citenamefont {Neshev},\ and\ \citenamefont {Sukhorukov}}]{ma2023PolarizationEngineering}%
  \BibitemOpen
  \bibfield  {author} {\bibinfo {author} {\bibfnamefont {J.}~\bibnamefont {Ma}}, \bibinfo {author} {\bibfnamefont {J.}~\bibnamefont {Zhang}}, \bibinfo {author} {\bibfnamefont {Y.}~\bibnamefont {Jiang}}, \bibinfo {author} {\bibfnamefont {T.}~\bibnamefont {Fan}}, \bibinfo {author} {\bibfnamefont {M.}~\bibnamefont {Parry}}, \bibinfo {author} {\bibfnamefont {D.~N.}\ \bibnamefont {Neshev}}, \ and\ \bibinfo {author} {\bibfnamefont {A.~A.}\ \bibnamefont {Sukhorukov}},\ }\bibfield  {title} {\enquote {\bibinfo {title} {Polarization engineering of entangled photons from a lithium niobate nonlinear metasurface},}\ }\href {\doibase 10.1021/acs.nanolett.3c02055} {\bibfield  {journal} {\bibinfo  {journal} {Nano Letters}\ }\textbf {\bibinfo {volume} {23}},\ \bibinfo {pages} {8091--8098} (\bibinfo {year} {2023})}\BibitemShut {NoStop}%
\bibitem [{\citenamefont {Son}\ \emph {et~al.}(2023)\citenamefont {Son}, \citenamefont {Sultanov}, \citenamefont {{Santiago-Cruz}}, \citenamefont {Anthur}, \citenamefont {Zhang}, \citenamefont {{Paniagua-Dominguez}}, \citenamefont {Krivitskiy}, \citenamefont {Kuznetsov},\ and\ \citenamefont {Chekhova}}]{son2023PhotonPairs}%
  \BibitemOpen
  \bibfield  {author} {\bibinfo {author} {\bibfnamefont {C.}~\bibnamefont {Son}}, \bibinfo {author} {\bibfnamefont {V.}~\bibnamefont {Sultanov}}, \bibinfo {author} {\bibfnamefont {T.}~\bibnamefont {{Santiago-Cruz}}}, \bibinfo {author} {\bibfnamefont {A.~P.}\ \bibnamefont {Anthur}}, \bibinfo {author} {\bibfnamefont {H.}~\bibnamefont {Zhang}}, \bibinfo {author} {\bibfnamefont {R.}~\bibnamefont {{Paniagua-Dominguez}}}, \bibinfo {author} {\bibfnamefont {L.}~\bibnamefont {Krivitskiy}}, \bibinfo {author} {\bibfnamefont {A.~I.}\ \bibnamefont {Kuznetsov}}, \ and\ \bibinfo {author} {\bibfnamefont {M.}~\bibnamefont {Chekhova}},\ }\bibfield  {title} {\enquote {\bibinfo {title} {Photon pairs bi-directionally emitted from a resonant metasurface},}\ }\href {\doibase 10.1039/D2NR05499J} {\bibfield  {journal} {\bibinfo  {journal} {Nanoscale}\ }\textbf {\bibinfo {volume} {15}},\ \bibinfo {pages} {2567--2572} (\bibinfo {year} {2023})}\BibitemShut {NoStop}%
\bibitem [{\citenamefont {Weissflog}\ \emph {et~al.}(2024{\natexlab{b}})\citenamefont {Weissflog}, \citenamefont {Ma}, \citenamefont {Zhang}, \citenamefont {Fan}, \citenamefont {Pertsch}, \citenamefont {Neshev}, \citenamefont {Saravi}, \citenamefont {Setzpfandt},\ and\ \citenamefont {Sukhorukov}}]{weissflog2024directionally}%
  \BibitemOpen
  \bibfield  {author} {\bibinfo {author} {\bibfnamefont {M.~A.}\ \bibnamefont {Weissflog}}, \bibinfo {author} {\bibfnamefont {J.}~\bibnamefont {Ma}}, \bibinfo {author} {\bibfnamefont {J.}~\bibnamefont {Zhang}}, \bibinfo {author} {\bibfnamefont {T.}~\bibnamefont {Fan}}, \bibinfo {author} {\bibfnamefont {T.}~\bibnamefont {Pertsch}}, \bibinfo {author} {\bibfnamefont {D.~N.}\ \bibnamefont {Neshev}}, \bibinfo {author} {\bibfnamefont {S.}~\bibnamefont {Saravi}}, \bibinfo {author} {\bibfnamefont {F.}~\bibnamefont {Setzpfandt}}, \ and\ \bibinfo {author} {\bibfnamefont {A.~A.}\ \bibnamefont {Sukhorukov}},\ }\href@noop {} {\enquote {\bibinfo {title} {Directionally tunable co- and counter-propagating photon pairs from a nonlinear metasurface},}\ } (\bibinfo {year} {2024}{\natexlab{b}}),\ \Eprint {http://arxiv.org/abs/2403.07651} {arXiv:2403.07651 [physics.optics]} \BibitemShut {NoStop}%
\bibitem [{\citenamefont {Liu}\ \emph {et~al.}(2019)\citenamefont {Liu}, \citenamefont {Su}, \citenamefont {Wei}, \citenamefont {Yao}, \citenamefont {da~Silva}, \citenamefont {Yu}, \citenamefont {{Iles-Smith}}, \citenamefont {Srinivasan}, \citenamefont {Rastelli}, \citenamefont {Li},\ and\ \citenamefont {Wang}}]{liu2019SolidstateSource}%
  \BibitemOpen
  \bibfield  {author} {\bibinfo {author} {\bibfnamefont {J.}~\bibnamefont {Liu}}, \bibinfo {author} {\bibfnamefont {R.}~\bibnamefont {Su}}, \bibinfo {author} {\bibfnamefont {Y.}~\bibnamefont {Wei}}, \bibinfo {author} {\bibfnamefont {B.}~\bibnamefont {Yao}}, \bibinfo {author} {\bibfnamefont {S.~F.~C.}\ \bibnamefont {da~Silva}}, \bibinfo {author} {\bibfnamefont {Y.}~\bibnamefont {Yu}}, \bibinfo {author} {\bibfnamefont {J.}~\bibnamefont {{Iles-Smith}}}, \bibinfo {author} {\bibfnamefont {K.}~\bibnamefont {Srinivasan}}, \bibinfo {author} {\bibfnamefont {A.}~\bibnamefont {Rastelli}}, \bibinfo {author} {\bibfnamefont {J.}~\bibnamefont {Li}}, \ and\ \bibinfo {author} {\bibfnamefont {X.}~\bibnamefont {Wang}},\ }\bibfield  {title} {\enquote {\bibinfo {title} {A solid-state source of strongly entangled photon pairs with high brightness and indistinguishability},}\ }\href {\doibase 10.1038/s41565-019-0435-9} {\bibfield  {journal} {\bibinfo  {journal} {Nature Nanotechnology}\ }\textbf {\bibinfo {volume} {14}},\ \bibinfo
  {pages} {586--593} (\bibinfo {year} {2019})}\BibitemShut {NoStop}%
\bibitem [{\citenamefont {Wang}\ \emph {et~al.}(2019)\citenamefont {Wang}, \citenamefont {Hu}, \citenamefont {Chung}, \citenamefont {Qin}, \citenamefont {Yang}, \citenamefont {Li}, \citenamefont {Liu}, \citenamefont {Zhong}, \citenamefont {He}, \citenamefont {Ding}, \citenamefont {Deng}, \citenamefont {Dai}, \citenamefont {Huo}, \citenamefont {H{\"o}fling}, \citenamefont {Lu},\ and\ \citenamefont {Pan}}]{wang2019OnDemandSemiconductor}%
  \BibitemOpen
  \bibfield  {author} {\bibinfo {author} {\bibfnamefont {H.}~\bibnamefont {Wang}}, \bibinfo {author} {\bibfnamefont {H.}~\bibnamefont {Hu}}, \bibinfo {author} {\bibfnamefont {T.-H.}\ \bibnamefont {Chung}}, \bibinfo {author} {\bibfnamefont {J.}~\bibnamefont {Qin}}, \bibinfo {author} {\bibfnamefont {X.}~\bibnamefont {Yang}}, \bibinfo {author} {\bibfnamefont {J.-P.}\ \bibnamefont {Li}}, \bibinfo {author} {\bibfnamefont {R.-Z.}\ \bibnamefont {Liu}}, \bibinfo {author} {\bibfnamefont {H.-S.}\ \bibnamefont {Zhong}}, \bibinfo {author} {\bibfnamefont {Y.-M.}\ \bibnamefont {He}}, \bibinfo {author} {\bibfnamefont {X.}~\bibnamefont {Ding}}, \bibinfo {author} {\bibfnamefont {Y.-H.}\ \bibnamefont {Deng}}, \bibinfo {author} {\bibfnamefont {Q.}~\bibnamefont {Dai}}, \bibinfo {author} {\bibfnamefont {Y.-H.}\ \bibnamefont {Huo}}, \bibinfo {author} {\bibfnamefont {S.}~\bibnamefont {H{\"o}fling}}, \bibinfo {author} {\bibfnamefont {C.-Y.}\ \bibnamefont {Lu}}, \ and\ \bibinfo {author} {\bibfnamefont {J.-W.}\ \bibnamefont {Pan}},\
  }\bibfield  {title} {\enquote {\bibinfo {title} {On-demand semiconductor source of entangled photons which simultaneously has high fidelity, efficiency, and indistinguishability},}\ }\href {\doibase 10.1103/PhysRevLett.122.113602} {\bibfield  {journal} {\bibinfo  {journal} {Physical Review Letters}\ }\textbf {\bibinfo {volume} {122}},\ \bibinfo {pages} {113602} (\bibinfo {year} {2019})}\BibitemShut {NoStop}%
\bibitem [{\citenamefont {Rota}\ \emph {et~al.}(2022)\citenamefont {Rota}, \citenamefont {Krieger}, \citenamefont {Buchinger}, \citenamefont {Beccaceci}, \citenamefont {Neuwirth}, \citenamefont {Huet}, \citenamefont {Horov{\'a}}, \citenamefont {Lovicu}, \citenamefont {Ronco}, \citenamefont {{da Silva}}, \citenamefont {Pettinari}, \citenamefont {{Mocza{\l}a-Dusanowska}}, \citenamefont {Kohlberger}, \citenamefont {Manna}, \citenamefont {Stroj}, \citenamefont {Freund}, \citenamefont {Yuan}, \citenamefont {Schneider}, \citenamefont {Je{\v z}ek}, \citenamefont {H{\"o}fling}, \citenamefont {Basset}, \citenamefont {{Huber-Loyola}}, \citenamefont {Rastelli},\ and\ \citenamefont {Trotta}}]{rota2022SourceEntangled}%
  \BibitemOpen
  \bibfield  {author} {\bibinfo {author} {\bibfnamefont {M.~B.}\ \bibnamefont {Rota}}, \bibinfo {author} {\bibfnamefont {T.~M.}\ \bibnamefont {Krieger}}, \bibinfo {author} {\bibfnamefont {Q.}~\bibnamefont {Buchinger}}, \bibinfo {author} {\bibfnamefont {M.}~\bibnamefont {Beccaceci}}, \bibinfo {author} {\bibfnamefont {J.}~\bibnamefont {Neuwirth}}, \bibinfo {author} {\bibfnamefont {H.}~\bibnamefont {Huet}}, \bibinfo {author} {\bibfnamefont {N.}~\bibnamefont {Horov{\'a}}}, \bibinfo {author} {\bibfnamefont {G.}~\bibnamefont {Lovicu}}, \bibinfo {author} {\bibfnamefont {G.}~\bibnamefont {Ronco}}, \bibinfo {author} {\bibfnamefont {S.~F.~C.}\ \bibnamefont {{da Silva}}}, \bibinfo {author} {\bibfnamefont {G.}~\bibnamefont {Pettinari}}, \bibinfo {author} {\bibfnamefont {M.}~\bibnamefont {{Mocza{\l}a-Dusanowska}}}, \bibinfo {author} {\bibfnamefont {C.}~\bibnamefont {Kohlberger}}, \bibinfo {author} {\bibfnamefont {S.}~\bibnamefont {Manna}}, \bibinfo {author} {\bibfnamefont {S.}~\bibnamefont {Stroj}}, \bibinfo {author}
  {\bibfnamefont {J.}~\bibnamefont {Freund}}, \bibinfo {author} {\bibfnamefont {X.}~\bibnamefont {Yuan}}, \bibinfo {author} {\bibfnamefont {C.}~\bibnamefont {Schneider}}, \bibinfo {author} {\bibfnamefont {M.}~\bibnamefont {Je{\v z}ek}}, \bibinfo {author} {\bibfnamefont {S.}~\bibnamefont {H{\"o}fling}}, \bibinfo {author} {\bibfnamefont {F.~B.}\ \bibnamefont {Basset}}, \bibinfo {author} {\bibfnamefont {T.}~\bibnamefont {{Huber-Loyola}}}, \bibinfo {author} {\bibfnamefont {A.}~\bibnamefont {Rastelli}}, \ and\ \bibinfo {author} {\bibfnamefont {R.}~\bibnamefont {Trotta}},\ }\href {\doibase 10.48550/arXiv.2212.12506} {\enquote {\bibinfo {title} {A source of entangled photons based on a cavity-enhanced and strain-tuned {G}a{A}s quantum dot},}\ } (\bibinfo {year} {2022}),\ \Eprint {http://arxiv.org/abs/2212.12506} {arxiv:2212.12506 [quant-ph]} \BibitemShut {NoStop}%
\bibitem [{\citenamefont {Pan}\ \emph {et~al.}(2012)\citenamefont {Pan}, \citenamefont {Chen}, \citenamefont {Lu}, \citenamefont {Weinfurter}, \citenamefont {Zeilinger},\ and\ \citenamefont {Żukowski}}]{pan2012MultiphotonEntanglement}%
  \BibitemOpen
  \bibfield  {author} {\bibinfo {author} {\bibfnamefont {J.-W.}\ \bibnamefont {Pan}}, \bibinfo {author} {\bibfnamefont {Z.-B.}\ \bibnamefont {Chen}}, \bibinfo {author} {\bibfnamefont {C.-Y.}\ \bibnamefont {Lu}}, \bibinfo {author} {\bibfnamefont {H.}~\bibnamefont {Weinfurter}}, \bibinfo {author} {\bibfnamefont {A.}~\bibnamefont {Zeilinger}}, \ and\ \bibinfo {author} {\bibfnamefont {M.}~\bibnamefont {Żukowski}},\ }\bibfield  {title} {\enquote {\bibinfo {title} {Multiphoton entanglement and interferometry},}\ }\href {\doibase 10.1103/RevModPhys.84.777} {\bibfield  {journal} {\bibinfo  {journal} {Reviews of Modern Physics}\ }\textbf {\bibinfo {volume} {84}},\ \bibinfo {pages} {777--838} (\bibinfo {year} {2012})}\BibitemShut {NoStop}%
\bibitem [{\citenamefont {Wang}\ \emph {et~al.}(2018)\citenamefont {Wang}, \citenamefont {Titchener}, \citenamefont {Kruk}, \citenamefont {Xu}, \citenamefont {Chung}, \citenamefont {Parry}, \citenamefont {Kravchenko}, \citenamefont {Chen}, \citenamefont {Solntsev}, \citenamefont {Kivshar}, \citenamefont {Neshev},\ and\ \citenamefont {Sukhorukov}}]{wang2018QuantumMetasurface}%
  \BibitemOpen
  \bibfield  {author} {\bibinfo {author} {\bibfnamefont {K.}~\bibnamefont {Wang}}, \bibinfo {author} {\bibfnamefont {J.~G.}\ \bibnamefont {Titchener}}, \bibinfo {author} {\bibfnamefont {S.~S.}\ \bibnamefont {Kruk}}, \bibinfo {author} {\bibfnamefont {L.}~\bibnamefont {Xu}}, \bibinfo {author} {\bibfnamefont {H.-P.}\ \bibnamefont {Chung}}, \bibinfo {author} {\bibfnamefont {M.}~\bibnamefont {Parry}}, \bibinfo {author} {\bibfnamefont {I.~I.}\ \bibnamefont {Kravchenko}}, \bibinfo {author} {\bibfnamefont {Y.-H.}\ \bibnamefont {Chen}}, \bibinfo {author} {\bibfnamefont {A.~S.}\ \bibnamefont {Solntsev}}, \bibinfo {author} {\bibfnamefont {Y.~S.}\ \bibnamefont {Kivshar}}, \bibinfo {author} {\bibfnamefont {D.~N.}\ \bibnamefont {Neshev}}, \ and\ \bibinfo {author} {\bibfnamefont {A.~A.}\ \bibnamefont {Sukhorukov}},\ }\bibfield  {title} {\enquote {\bibinfo {title} {Quantum metasurface for multiphoton interference and state reconstruction},}\ }\href {\doibase 10.1126/science.aat8196} {\bibfield  {journal} {\bibinfo  {journal}
  {Science}\ }\textbf {\bibinfo {volume} {361}},\ \bibinfo {pages} {1104--1108} (\bibinfo {year} {2018})}\BibitemShut {NoStop}%
\bibitem [{\citenamefont {Bouchard}\ \emph {et~al.}(2021)\citenamefont {Bouchard}, \citenamefont {Sit}, \citenamefont {Zhang}, \citenamefont {Fickler}, \citenamefont {Miatto}, \citenamefont {Yao}, \citenamefont {Sciarrino},\ and\ \citenamefont {Karimi}}]{bouchard2021TwophotonInterference}%
  \BibitemOpen
  \bibfield  {author} {\bibinfo {author} {\bibfnamefont {F.}~\bibnamefont {Bouchard}}, \bibinfo {author} {\bibfnamefont {A.}~\bibnamefont {Sit}}, \bibinfo {author} {\bibfnamefont {Y.}~\bibnamefont {Zhang}}, \bibinfo {author} {\bibfnamefont {R.}~\bibnamefont {Fickler}}, \bibinfo {author} {\bibfnamefont {F.~M.}\ \bibnamefont {Miatto}}, \bibinfo {author} {\bibfnamefont {Y.}~\bibnamefont {Yao}}, \bibinfo {author} {\bibfnamefont {F.}~\bibnamefont {Sciarrino}}, \ and\ \bibinfo {author} {\bibfnamefont {E.}~\bibnamefont {Karimi}},\ }\bibfield  {title} {\enquote {\bibinfo {title} {Two-photon interference: The hong–ou–mandel effect},}\ }\href {\doibase 10.1088/1361-6633/abcd7a} {\bibfield  {journal} {\bibinfo  {journal} {Reports on Progress in Physics}\ }\textbf {\bibinfo {volume} {84}},\ \bibinfo {pages} {012402} (\bibinfo {year} {2021})}\BibitemShut {NoStop}%
\bibitem [{\citenamefont {Estakhri}\ \emph {et~al.}(2021)\citenamefont {Estakhri}, \citenamefont {Estakhri}, \citenamefont {Norris},\ and\ \citenamefont {Norris}}]{estakhri2021TunableQuantum}%
  \BibitemOpen
  \bibfield  {author} {\bibinfo {author} {\bibfnamefont {N.~M.}\ \bibnamefont {Estakhri}}, \bibinfo {author} {\bibfnamefont {N.~M.}\ \bibnamefont {Estakhri}}, \bibinfo {author} {\bibfnamefont {T.~B.}\ \bibnamefont {Norris}}, \ and\ \bibinfo {author} {\bibfnamefont {T.~B.}\ \bibnamefont {Norris}},\ }\bibfield  {title} {\enquote {\bibinfo {title} {Tunable quantum two-photon interference with reconfigurable metasurfaces using phase-change materials},}\ }\href {\doibase 10.1364/OE.419892} {\bibfield  {journal} {\bibinfo  {journal} {Optics Express}\ }\textbf {\bibinfo {volume} {29}},\ \bibinfo {pages} {14245--14259} (\bibinfo {year} {2021})}\BibitemShut {NoStop}%
\bibitem [{\citenamefont {Kang}\ \emph {et~al.}(2021)\citenamefont {Kang}, \citenamefont {Lau}, \citenamefont {Yung}, \citenamefont {Du}, \citenamefont {Tam},\ and\ \citenamefont {Li}}]{kang2021tailor}%
  \BibitemOpen
  \bibfield  {author} {\bibinfo {author} {\bibfnamefont {M.}~\bibnamefont {Kang}}, \bibinfo {author} {\bibfnamefont {K.~M.}\ \bibnamefont {Lau}}, \bibinfo {author} {\bibfnamefont {T.~K.}\ \bibnamefont {Yung}}, \bibinfo {author} {\bibfnamefont {S.}~\bibnamefont {Du}}, \bibinfo {author} {\bibfnamefont {W.~Y.}\ \bibnamefont {Tam}}, \ and\ \bibinfo {author} {\bibfnamefont {J.}~\bibnamefont {Li}},\ }\bibfield  {title} {\enquote {\bibinfo {title} {Tailor-made unitary operations using dielectric metasurfaces},}\ }\href@noop {} {\bibfield  {journal} {\bibinfo  {journal} {Optics Express}\ }\textbf {\bibinfo {volume} {29}},\ \bibinfo {pages} {5677--5686} (\bibinfo {year} {2021})}\BibitemShut {NoStop}%
\bibitem [{\citenamefont {Zhang}\ \emph {et~al.}(2024)\citenamefont {Zhang}, \citenamefont {Ma}, \citenamefont {Li}, \citenamefont {Lung},\ and\ \citenamefont {Sukhorukov}}]{zhang2024SingleshotCharacterization}%
  \BibitemOpen
  \bibfield  {author} {\bibinfo {author} {\bibfnamefont {J.}~\bibnamefont {Zhang}}, \bibinfo {author} {\bibfnamefont {J.}~\bibnamefont {Ma}}, \bibinfo {author} {\bibfnamefont {N.}~\bibnamefont {Li}}, \bibinfo {author} {\bibfnamefont {S.}~\bibnamefont {Lung}}, \ and\ \bibinfo {author} {\bibfnamefont {A.~A.}\ \bibnamefont {Sukhorukov}},\ }\href {https://arxiv.org/abs/2401.01485v1} {\enquote {\bibinfo {title} {Single-shot characterization of photon indistinguishability with dielectric metasurfaces},}\ } (\bibinfo {year} {2024}),\ \Eprint {http://arxiv.org/abs/2401.01485} {arXiv:2401.01485 [physics.optics]} \BibitemShut {NoStop}%
\bibitem [{\citenamefont {Zhong}\ \emph {et~al.}(2020)\citenamefont {Zhong}, \citenamefont {Wang}, \citenamefont {Deng}, \citenamefont {Chen}, \citenamefont {Peng}, \citenamefont {Luo}, \citenamefont {Qin}, \citenamefont {Wu}, \citenamefont {Ding}, \citenamefont {Hu}, \citenamefont {Hu}, \citenamefont {Yang}, \citenamefont {Zhang}, \citenamefont {Li}, \citenamefont {Li}, \citenamefont {Jiang}, \citenamefont {Gan}, \citenamefont {Yang}, \citenamefont {You}, \citenamefont {Wang}, \citenamefont {Li}, \citenamefont {Liu}, \citenamefont {Lu},\ and\ \citenamefont {Pan}}]{Zhong:2020-1460:SCI}%
  \BibitemOpen
  \bibfield  {author} {\bibinfo {author} {\bibfnamefont {H.~S.}\ \bibnamefont {Zhong}}, \bibinfo {author} {\bibfnamefont {H.}~\bibnamefont {Wang}}, \bibinfo {author} {\bibfnamefont {Y.~H.}\ \bibnamefont {Deng}}, \bibinfo {author} {\bibfnamefont {M.~C.}\ \bibnamefont {Chen}}, \bibinfo {author} {\bibfnamefont {L.~C.}\ \bibnamefont {Peng}}, \bibinfo {author} {\bibfnamefont {Y.~H.}\ \bibnamefont {Luo}}, \bibinfo {author} {\bibfnamefont {J.}~\bibnamefont {Qin}}, \bibinfo {author} {\bibfnamefont {D.}~\bibnamefont {Wu}}, \bibinfo {author} {\bibfnamefont {X.}~\bibnamefont {Ding}}, \bibinfo {author} {\bibfnamefont {Y.}~\bibnamefont {Hu}}, \bibinfo {author} {\bibfnamefont {P.}~\bibnamefont {Hu}}, \bibinfo {author} {\bibfnamefont {X.~Y.}\ \bibnamefont {Yang}}, \bibinfo {author} {\bibfnamefont {W.~J.}\ \bibnamefont {Zhang}}, \bibinfo {author} {\bibfnamefont {H.}~\bibnamefont {Li}}, \bibinfo {author} {\bibfnamefont {Y.~X.}\ \bibnamefont {Li}}, \bibinfo {author} {\bibfnamefont {X.}~\bibnamefont {Jiang}}, \bibinfo {author}
  {\bibfnamefont {L.}~\bibnamefont {Gan}}, \bibinfo {author} {\bibfnamefont {G.~W.}\ \bibnamefont {Yang}}, \bibinfo {author} {\bibfnamefont {L.~X.}\ \bibnamefont {You}}, \bibinfo {author} {\bibfnamefont {Z.}~\bibnamefont {Wang}}, \bibinfo {author} {\bibfnamefont {L.}~\bibnamefont {Li}}, \bibinfo {author} {\bibfnamefont {N.~L.}\ \bibnamefont {Liu}}, \bibinfo {author} {\bibfnamefont {C.~Y.}\ \bibnamefont {Lu}}, \ and\ \bibinfo {author} {\bibfnamefont {J.~W.}\ \bibnamefont {Pan}},\ }\bibfield  {title} {\enquote {\bibinfo {title} {Quantum computational advantage using photons},}\ }\href {\doibase 10.1126/science.abe8770} {\bibfield  {journal} {\bibinfo  {journal} {Science}\ }\textbf {\bibinfo {volume} {370}},\ \bibinfo {pages} {1460--1463} (\bibinfo {year} {2020})}\BibitemShut {NoStop}%
\bibitem [{\citenamefont {Carolan}\ \emph {et~al.}(2015)\citenamefont {Carolan}, \citenamefont {Harrold}, \citenamefont {Sparrow}, \citenamefont {Martin-Lopez}, \citenamefont {Russell}, \citenamefont {Silverstone}, \citenamefont {Shadbolt}, \citenamefont {Matsuda}, \citenamefont {Oguma}, \citenamefont {Itoh}, \citenamefont {Marshall}, \citenamefont {Thompson}, \citenamefont {Matthews}, \citenamefont {Hashimoto}, \citenamefont {O'Brien},\ and\ \citenamefont {Laing}}]{Carolan:2015-711:SCI}%
  \BibitemOpen
  \bibfield  {author} {\bibinfo {author} {\bibfnamefont {J.}~\bibnamefont {Carolan}}, \bibinfo {author} {\bibfnamefont {C.}~\bibnamefont {Harrold}}, \bibinfo {author} {\bibfnamefont {C.}~\bibnamefont {Sparrow}}, \bibinfo {author} {\bibfnamefont {E.}~\bibnamefont {Martin-Lopez}}, \bibinfo {author} {\bibfnamefont {N.~J.}\ \bibnamefont {Russell}}, \bibinfo {author} {\bibfnamefont {J.~W.}\ \bibnamefont {Silverstone}}, \bibinfo {author} {\bibfnamefont {P.~J.}\ \bibnamefont {Shadbolt}}, \bibinfo {author} {\bibfnamefont {N.}~\bibnamefont {Matsuda}}, \bibinfo {author} {\bibfnamefont {M.}~\bibnamefont {Oguma}}, \bibinfo {author} {\bibfnamefont {M.}~\bibnamefont {Itoh}}, \bibinfo {author} {\bibfnamefont {G.~D.}\ \bibnamefont {Marshall}}, \bibinfo {author} {\bibfnamefont {M.~G.}\ \bibnamefont {Thompson}}, \bibinfo {author} {\bibfnamefont {J.~C.~F.}\ \bibnamefont {Matthews}}, \bibinfo {author} {\bibfnamefont {T.}~\bibnamefont {Hashimoto}}, \bibinfo {author} {\bibfnamefont {J.~L.}\ \bibnamefont {O'Brien}}, \ and\ \bibinfo
  {author} {\bibfnamefont {A.}~\bibnamefont {Laing}},\ }\bibfield  {title} {\enquote {\bibinfo {title} {Universal linear optics},}\ }\href {\doibase 10.1126/science.aab3642} {\bibfield  {journal} {\bibinfo  {journal} {Science}\ }\textbf {\bibinfo {volume} {349}},\ \bibinfo {pages} {711--716} (\bibinfo {year} {2015})}\BibitemShut {NoStop}%
\bibitem [{\citenamefont {Jha}\ \emph {et~al.}(2015)\citenamefont {Jha}, \citenamefont {Ni}, \citenamefont {Wu}, \citenamefont {Wang},\ and\ \citenamefont {Zhang}}]{jha2015MetasurfaceEnabledRemote}%
  \BibitemOpen
  \bibfield  {author} {\bibinfo {author} {\bibfnamefont {P.~K.}\ \bibnamefont {Jha}}, \bibinfo {author} {\bibfnamefont {X.}~\bibnamefont {Ni}}, \bibinfo {author} {\bibfnamefont {C.}~\bibnamefont {Wu}}, \bibinfo {author} {\bibfnamefont {Y.}~\bibnamefont {Wang}}, \ and\ \bibinfo {author} {\bibfnamefont {X.}~\bibnamefont {Zhang}},\ }\bibfield  {title} {\enquote {\bibinfo {title} {Metasurface-enabled remote quantum interference},}\ }\href {\doibase 10.1103/PhysRevLett.115.025501} {\bibfield  {journal} {\bibinfo  {journal} {Physical Review Letters}\ }\textbf {\bibinfo {volume} {115}},\ \bibinfo {pages} {025501} (\bibinfo {year} {2015})}\BibitemShut {NoStop}%
\bibitem [{\citenamefont {Liang}\ \emph {et~al.}(2023)\citenamefont {Liang}, \citenamefont {Ahmed}, \citenamefont {Tam}, \citenamefont {Chen},\ and\ \citenamefont {Li}}]{liang2023continuous}%
  \BibitemOpen
  \bibfield  {author} {\bibinfo {author} {\bibfnamefont {H.}~\bibnamefont {Liang}}, \bibinfo {author} {\bibfnamefont {H.}~\bibnamefont {Ahmed}}, \bibinfo {author} {\bibfnamefont {W.~Y.}\ \bibnamefont {Tam}}, \bibinfo {author} {\bibfnamefont {X.}~\bibnamefont {Chen}}, \ and\ \bibinfo {author} {\bibfnamefont {J.}~\bibnamefont {Li}},\ }\bibfield  {title} {\enquote {\bibinfo {title} {Continuous heralding control of vortex beams using quantum metasurface},}\ }\href@noop {} {\bibfield  {journal} {\bibinfo  {journal} {Communications Physics}\ }\textbf {\bibinfo {volume} {6}},\ \bibinfo {pages} {140} (\bibinfo {year} {2023})}\BibitemShut {NoStop}%
\bibitem [{\citenamefont {Stav}\ \emph {et~al.}(2018)\citenamefont {Stav}, \citenamefont {Faerman}, \citenamefont {Maguid}, \citenamefont {Oren}, \citenamefont {Kleiner}, \citenamefont {Hasman},\ and\ \citenamefont {Segev}}]{stav2018QuantumEntanglement}%
  \BibitemOpen
  \bibfield  {author} {\bibinfo {author} {\bibfnamefont {T.}~\bibnamefont {Stav}}, \bibinfo {author} {\bibfnamefont {A.}~\bibnamefont {Faerman}}, \bibinfo {author} {\bibfnamefont {E.}~\bibnamefont {Maguid}}, \bibinfo {author} {\bibfnamefont {D.}~\bibnamefont {Oren}}, \bibinfo {author} {\bibfnamefont {V.}~\bibnamefont {Kleiner}}, \bibinfo {author} {\bibfnamefont {E.}~\bibnamefont {Hasman}}, \ and\ \bibinfo {author} {\bibfnamefont {M.}~\bibnamefont {Segev}},\ }\bibfield  {title} {\enquote {\bibinfo {title} {Quantum entanglement of the spin and orbital angular momentum of photons using metamaterials},}\ }\href {\doibase 10.1126/science.aat9042} {\bibfield  {journal} {\bibinfo  {journal} {Science}\ }\textbf {\bibinfo {volume} {361}},\ \bibinfo {pages} {1101--1104} (\bibinfo {year} {2018})}\BibitemShut {NoStop}%
\bibitem [{\citenamefont {Li}\ \emph {et~al.}(2022)\citenamefont {Li}, \citenamefont {Zhu}, \citenamefont {Lin}, \citenamefont {Huo}, \citenamefont {Xia}, \citenamefont {Liu}, \citenamefont {Ruan}, \citenamefont {Tang}, \citenamefont {Cai}, \citenamefont {Wu}, \citenamefont {Meng}, \citenamefont {Zhang}, \citenamefont {Chen}, \citenamefont {Xu}, \citenamefont {Xia}, \citenamefont {Zhang},\ and\ \citenamefont {Lu}}]{li2022HighdimensionalEntanglement}%
  \BibitemOpen
  \bibfield  {author} {\bibinfo {author} {\bibfnamefont {Z.-X.}\ \bibnamefont {Li}}, \bibinfo {author} {\bibfnamefont {D.}~\bibnamefont {Zhu}}, \bibinfo {author} {\bibfnamefont {P.-C.}\ \bibnamefont {Lin}}, \bibinfo {author} {\bibfnamefont {P.-C.}\ \bibnamefont {Huo}}, \bibinfo {author} {\bibfnamefont {H.-K.}\ \bibnamefont {Xia}}, \bibinfo {author} {\bibfnamefont {M.-Z.}\ \bibnamefont {Liu}}, \bibinfo {author} {\bibfnamefont {Y.-P.}\ \bibnamefont {Ruan}}, \bibinfo {author} {\bibfnamefont {J.-S.}\ \bibnamefont {Tang}}, \bibinfo {author} {\bibfnamefont {M.}~\bibnamefont {Cai}}, \bibinfo {author} {\bibfnamefont {H.-D.}\ \bibnamefont {Wu}}, \bibinfo {author} {\bibfnamefont {C.-Y.}\ \bibnamefont {Meng}}, \bibinfo {author} {\bibfnamefont {H.}~\bibnamefont {Zhang}}, \bibinfo {author} {\bibfnamefont {P.}~\bibnamefont {Chen}}, \bibinfo {author} {\bibfnamefont {T.}~\bibnamefont {Xu}}, \bibinfo {author} {\bibfnamefont {K.-Y.}\ \bibnamefont {Xia}}, \bibinfo {author} {\bibfnamefont {L.-J.}\ \bibnamefont {Zhang}}, \ and\
  \bibinfo {author} {\bibfnamefont {Y.-Q.}\ \bibnamefont {Lu}},\ }\bibfield  {title} {\enquote {\bibinfo {title} {High-dimensional entanglement generation based on a pancharatnam{\textendash}berry phase metasurface},}\ }\href {\doibase 10.1364/PRJ.470663} {\bibfield  {journal} {\bibinfo  {journal} {Photonics Research}\ }\textbf {\bibinfo {volume} {10}},\ \bibinfo {pages} {2702--2707} (\bibinfo {year} {2022})}\BibitemShut {NoStop}%
\bibitem [{\citenamefont {Lung}\ \emph {et~al.}(2020)\citenamefont {Lung}, \citenamefont {Wang}, \citenamefont {Kamali}, \citenamefont {Zhang}, \citenamefont {Rahmani}, \citenamefont {Neshev},\ and\ \citenamefont {Sukhorukov}}]{lung2020ComplexBirefringentDielectric}%
  \BibitemOpen
  \bibfield  {author} {\bibinfo {author} {\bibfnamefont {S.}~\bibnamefont {Lung}}, \bibinfo {author} {\bibfnamefont {K.}~\bibnamefont {Wang}}, \bibinfo {author} {\bibfnamefont {K.~Z.}\ \bibnamefont {Kamali}}, \bibinfo {author} {\bibfnamefont {J.}~\bibnamefont {Zhang}}, \bibinfo {author} {\bibfnamefont {M.}~\bibnamefont {Rahmani}}, \bibinfo {author} {\bibfnamefont {D.~N.}\ \bibnamefont {Neshev}}, \ and\ \bibinfo {author} {\bibfnamefont {A.~A.}\ \bibnamefont {Sukhorukov}},\ }\bibfield  {title} {\enquote {\bibinfo {title} {Complex-birefringent dielectric metasurfaces for arbitrary polarization-pair transformations},}\ }\href {\doibase 10.1021/acsphotonics.0c01044} {\bibfield  {journal} {\bibinfo  {journal} {ACS Photonics}\ }\textbf {\bibinfo {volume} {7}},\ \bibinfo {pages} {3015--3022} (\bibinfo {year} {2020})}\BibitemShut {NoStop}%
\bibitem [{\citenamefont {Jha}\ \emph {et~al.}(2018)\citenamefont {Jha}, \citenamefont {Shitrit}, \citenamefont {Kim}, \citenamefont {Ren}, \citenamefont {Wang},\ and\ \citenamefont {Zhang}}]{jha2018MetasurfaceMediatedQuantum}%
  \BibitemOpen
  \bibfield  {author} {\bibinfo {author} {\bibfnamefont {P.~K.}\ \bibnamefont {Jha}}, \bibinfo {author} {\bibfnamefont {N.}~\bibnamefont {Shitrit}}, \bibinfo {author} {\bibfnamefont {J.}~\bibnamefont {Kim}}, \bibinfo {author} {\bibfnamefont {X.}~\bibnamefont {Ren}}, \bibinfo {author} {\bibfnamefont {Y.}~\bibnamefont {Wang}}, \ and\ \bibinfo {author} {\bibfnamefont {X.}~\bibnamefont {Zhang}},\ }\bibfield  {title} {\enquote {\bibinfo {title} {Metasurface-mediated quantum entanglement},}\ }\href {\doibase 10.1021/acsphotonics.7b01241} {\bibfield  {journal} {\bibinfo  {journal} {ACS Photonics}\ }\textbf {\bibinfo {volume} {5}},\ \bibinfo {pages} {971--976} (\bibinfo {year} {2018})}\BibitemShut {NoStop}%
\bibitem [{\citenamefont {Asano}\ \emph {et~al.}(2016)\citenamefont {Asano}, \citenamefont {Bechu}, \citenamefont {Tame}, \citenamefont {Kaya~{\"O}zdemir}, \citenamefont {Ikuta}, \citenamefont {G{\"u}ney}, \citenamefont {Yamamoto}, \citenamefont {Yang}, \citenamefont {Wegener},\ and\ \citenamefont {Imoto}}]{asano2016DistillationPhoton}%
  \BibitemOpen
  \bibfield  {author} {\bibinfo {author} {\bibfnamefont {M.}~\bibnamefont {Asano}}, \bibinfo {author} {\bibfnamefont {M.}~\bibnamefont {Bechu}}, \bibinfo {author} {\bibfnamefont {M.}~\bibnamefont {Tame}}, \bibinfo {author} {\bibfnamefont {{\c S}.}~\bibnamefont {Kaya~{\"O}zdemir}}, \bibinfo {author} {\bibfnamefont {R.}~\bibnamefont {Ikuta}}, \bibinfo {author} {\bibfnamefont {D.~{\"O}.}\ \bibnamefont {G{\"u}ney}}, \bibinfo {author} {\bibfnamefont {T.}~\bibnamefont {Yamamoto}}, \bibinfo {author} {\bibfnamefont {L.}~\bibnamefont {Yang}}, \bibinfo {author} {\bibfnamefont {M.}~\bibnamefont {Wegener}}, \ and\ \bibinfo {author} {\bibfnamefont {N.}~\bibnamefont {Imoto}},\ }\bibfield  {title} {\enquote {\bibinfo {title} {Distillation of photon entanglement using a plasmonic metamaterial},}\ }\href {\doibase 10.1038/srep18313} {\bibfield  {journal} {\bibinfo  {journal} {Scientific Reports}\ }\textbf {\bibinfo {volume} {5}},\ \bibinfo {pages} {18313} (\bibinfo {year} {2016})}\BibitemShut {NoStop}%
\bibitem [{\citenamefont {Gao}\ \emph {et~al.}(2022)\citenamefont {Gao}, \citenamefont {Wang}, \citenamefont {Jiang}, \citenamefont {Peng}, \citenamefont {Wang}, \citenamefont {Qi}, \citenamefont {Fan}, \citenamefont {Tang},\ and\ \citenamefont {Wang}}]{gao2022MultichannelDistribution}%
  \BibitemOpen
  \bibfield  {author} {\bibinfo {author} {\bibfnamefont {Y.-J.}\ \bibnamefont {Gao}}, \bibinfo {author} {\bibfnamefont {Z.}~\bibnamefont {Wang}}, \bibinfo {author} {\bibfnamefont {Y.}~\bibnamefont {Jiang}}, \bibinfo {author} {\bibfnamefont {R.-W.}\ \bibnamefont {Peng}}, \bibinfo {author} {\bibfnamefont {Z.-Y.}\ \bibnamefont {Wang}}, \bibinfo {author} {\bibfnamefont {D.-X.}\ \bibnamefont {Qi}}, \bibinfo {author} {\bibfnamefont {R.-H.}\ \bibnamefont {Fan}}, \bibinfo {author} {\bibfnamefont {W.-J.}\ \bibnamefont {Tang}}, \ and\ \bibinfo {author} {\bibfnamefont {M.}~\bibnamefont {Wang}},\ }\bibfield  {title} {\enquote {\bibinfo {title} {Multichannel distribution and transformation of entangled photons with dielectric metasurfaces},}\ }\href {\doibase 10.1103/PhysRevLett.129.023601} {\bibfield  {journal} {\bibinfo  {journal} {Physical Review Letters}\ }\textbf {\bibinfo {volume} {129}},\ \bibinfo {pages} {023601} (\bibinfo {year} {2022})}\BibitemShut {NoStop}%
\bibitem [{\citenamefont {Zhu}\ \emph {et~al.}(2020)\citenamefont {Zhu}, \citenamefont {Liu}, \citenamefont {Sain}, \citenamefont {Wang}, \citenamefont {Schlickriede}, \citenamefont {Tang}, \citenamefont {Deng}, \citenamefont {Li}, \citenamefont {Yang}, \citenamefont {Holynski}, \citenamefont {Zhang}, \citenamefont {Zentgraf}, \citenamefont {Bongs}, \citenamefont {Lien},\ and\ \citenamefont {Li}}]{zhu2020DielectricMetasurface}%
  \BibitemOpen
  \bibfield  {author} {\bibinfo {author} {\bibfnamefont {L.}~\bibnamefont {Zhu}}, \bibinfo {author} {\bibfnamefont {X.}~\bibnamefont {Liu}}, \bibinfo {author} {\bibfnamefont {B.}~\bibnamefont {Sain}}, \bibinfo {author} {\bibfnamefont {M.}~\bibnamefont {Wang}}, \bibinfo {author} {\bibfnamefont {C.}~\bibnamefont {Schlickriede}}, \bibinfo {author} {\bibfnamefont {Y.}~\bibnamefont {Tang}}, \bibinfo {author} {\bibfnamefont {J.}~\bibnamefont {Deng}}, \bibinfo {author} {\bibfnamefont {K.}~\bibnamefont {Li}}, \bibinfo {author} {\bibfnamefont {J.}~\bibnamefont {Yang}}, \bibinfo {author} {\bibfnamefont {M.}~\bibnamefont {Holynski}}, \bibinfo {author} {\bibfnamefont {S.}~\bibnamefont {Zhang}}, \bibinfo {author} {\bibfnamefont {T.}~\bibnamefont {Zentgraf}}, \bibinfo {author} {\bibfnamefont {K.}~\bibnamefont {Bongs}}, \bibinfo {author} {\bibfnamefont {Y.-H.}\ \bibnamefont {Lien}}, \ and\ \bibinfo {author} {\bibfnamefont {G.}~\bibnamefont {Li}},\ }\bibfield  {title} {\enquote {\bibinfo {title} {A dielectric metasurface
  optical chip for the generation of cold atoms},}\ }\href {\doibase 10.1126/sciadv.abb6667} {\bibfield  {journal} {\bibinfo  {journal} {Science Advances}\ }\textbf {\bibinfo {volume} {6}},\ \bibinfo {pages} {eabb6667} (\bibinfo {year} {2020})}\BibitemShut {NoStop}%
\bibitem [{\citenamefont {Jin}\ \emph {et~al.}(2023)\citenamefont {Jin}, \citenamefont {Zhang}, \citenamefont {Liu}, \citenamefont {Liang}, \citenamefont {Liu}, \citenamefont {Hu}, \citenamefont {Li}, \citenamefont {Wang}, \citenamefont {Yang}, \citenamefont {Zhu},\ and\ \citenamefont {Li}}]{jin2023CentimeterScaleDielectric}%
  \BibitemOpen
  \bibfield  {author} {\bibinfo {author} {\bibfnamefont {M.}~\bibnamefont {Jin}}, \bibinfo {author} {\bibfnamefont {X.}~\bibnamefont {Zhang}}, \bibinfo {author} {\bibfnamefont {X.}~\bibnamefont {Liu}}, \bibinfo {author} {\bibfnamefont {C.}~\bibnamefont {Liang}}, \bibinfo {author} {\bibfnamefont {J.}~\bibnamefont {Liu}}, \bibinfo {author} {\bibfnamefont {Z.}~\bibnamefont {Hu}}, \bibinfo {author} {\bibfnamefont {K.}~\bibnamefont {Li}}, \bibinfo {author} {\bibfnamefont {G.}~\bibnamefont {Wang}}, \bibinfo {author} {\bibfnamefont {J.}~\bibnamefont {Yang}}, \bibinfo {author} {\bibfnamefont {L.}~\bibnamefont {Zhu}}, \ and\ \bibinfo {author} {\bibfnamefont {G.}~\bibnamefont {Li}},\ }\bibfield  {title} {\enquote {\bibinfo {title} {A centimeter-scale dielectric metasurface for the generation of cold atoms},}\ }\href {\doibase 10.1021/acs.nanolett.3c00791} {\bibfield  {journal} {\bibinfo  {journal} {Nano Letters}\ }\textbf {\bibinfo {volume} {23}},\ \bibinfo {pages} {4008--4013} (\bibinfo {year} {2023})}\BibitemShut
  {NoStop}%
\bibitem [{\citenamefont {Hsu}\ \emph {et~al.}(2022)\citenamefont {Hsu}, \citenamefont {Zhu}, \citenamefont {Thiele}, \citenamefont {Brown}, \citenamefont {Papp}, \citenamefont {Agrawal},\ and\ \citenamefont {Regal}}]{hsu2022SingleAtomTrapping}%
  \BibitemOpen
  \bibfield  {author} {\bibinfo {author} {\bibfnamefont {T.-W.}\ \bibnamefont {Hsu}}, \bibinfo {author} {\bibfnamefont {W.}~\bibnamefont {Zhu}}, \bibinfo {author} {\bibfnamefont {T.}~\bibnamefont {Thiele}}, \bibinfo {author} {\bibfnamefont {M.~O.}\ \bibnamefont {Brown}}, \bibinfo {author} {\bibfnamefont {S.~B.}\ \bibnamefont {Papp}}, \bibinfo {author} {\bibfnamefont {A.}~\bibnamefont {Agrawal}}, \ and\ \bibinfo {author} {\bibfnamefont {C.~A.}\ \bibnamefont {Regal}},\ }\bibfield  {title} {\enquote {\bibinfo {title} {Single-atom trapping in a metasurface-lens optical tweezer},}\ }\href {\doibase 10.1103/PRXQuantum.3.030316} {\bibfield  {journal} {\bibinfo  {journal} {PRX Quantum}\ }\textbf {\bibinfo {volume} {3}},\ \bibinfo {pages} {030316} (\bibinfo {year} {2022})}\BibitemShut {NoStop}%
\bibitem [{\citenamefont {McGehee}\ \emph {et~al.}(2021)\citenamefont {McGehee}, \citenamefont {Zhu}, \citenamefont {Barker}, \citenamefont {Westly}, \citenamefont {Yulaev}, \citenamefont {Klimov}, \citenamefont {Agrawal}, \citenamefont {Eckel}, \citenamefont {Aksyuk},\ and\ \citenamefont {McClelland}}]{mcgehee2021MagnetoopticalTrapping}%
  \BibitemOpen
  \bibfield  {author} {\bibinfo {author} {\bibfnamefont {W.~R.}\ \bibnamefont {McGehee}}, \bibinfo {author} {\bibfnamefont {W.}~\bibnamefont {Zhu}}, \bibinfo {author} {\bibfnamefont {D.~S.}\ \bibnamefont {Barker}}, \bibinfo {author} {\bibfnamefont {D.}~\bibnamefont {Westly}}, \bibinfo {author} {\bibfnamefont {A.}~\bibnamefont {Yulaev}}, \bibinfo {author} {\bibfnamefont {N.}~\bibnamefont {Klimov}}, \bibinfo {author} {\bibfnamefont {A.}~\bibnamefont {Agrawal}}, \bibinfo {author} {\bibfnamefont {S.}~\bibnamefont {Eckel}}, \bibinfo {author} {\bibfnamefont {V.}~\bibnamefont {Aksyuk}}, \ and\ \bibinfo {author} {\bibfnamefont {J.~J.}\ \bibnamefont {McClelland}},\ }\bibfield  {title} {\enquote {\bibinfo {title} {Magneto-optical trapping using planar optics},}\ }\href {\doibase 10.1088/1367-2630/abdce3} {\bibfield  {journal} {\bibinfo  {journal} {New Journal of Physics}\ }\textbf {\bibinfo {volume} {23}},\ \bibinfo {pages} {013021} (\bibinfo {year} {2021})}\BibitemShut {NoStop}%
\bibitem [{\citenamefont {Zhu}\ \emph {et~al.}(2023)\citenamefont {Zhu}, \citenamefont {Chen}, \citenamefont {Song}, \citenamefont {Yang}, \citenamefont {Koksal}, \citenamefont {Wang}, \citenamefont {Ferdinand}, \citenamefont {Jammi}, \citenamefont {Spektor}, \citenamefont {Papp},\ and\ \citenamefont {Agrawal}}]{zhu2023ConstructingMagnetooptical}%
  \BibitemOpen
  \bibfield  {author} {\bibinfo {author} {\bibfnamefont {W.}~\bibnamefont {Zhu}}, \bibinfo {author} {\bibfnamefont {L.}~\bibnamefont {Chen}}, \bibinfo {author} {\bibfnamefont {J.}~\bibnamefont {Song}}, \bibinfo {author} {\bibfnamefont {J.}~\bibnamefont {Yang}}, \bibinfo {author} {\bibfnamefont {O.}~\bibnamefont {Koksal}}, \bibinfo {author} {\bibfnamefont {Z.}~\bibnamefont {Wang}}, \bibinfo {author} {\bibfnamefont {A.}~\bibnamefont {Ferdinand}}, \bibinfo {author} {\bibfnamefont {S.}~\bibnamefont {Jammi}}, \bibinfo {author} {\bibfnamefont {G.}~\bibnamefont {Spektor}}, \bibinfo {author} {\bibfnamefont {S.~B.}\ \bibnamefont {Papp}}, \ and\ \bibinfo {author} {\bibfnamefont {A.}~\bibnamefont {Agrawal}},\ }\bibfield  {title} {\enquote {\bibinfo {title} {Constructing a magneto-optical trap using transmissive metasurfaces},}\ }in\ \href {\doibase 10.1364/CLEO_SI.2023.SM2G.7} {\emph {\bibinfo {booktitle} {CLEO 2023 (2023), Paper SM2G.7}}}\ (\bibinfo  {publisher} {Optica Publishing Group},\ \bibinfo {year} {2023})\ p.\
  \bibinfo {pages} {SM2G.7}\BibitemShut {NoStop}%
\bibitem [{\citenamefont {Shi}\ \emph {et~al.}(2022)\citenamefont {Shi}, \citenamefont {Song}, \citenamefont {Toftul}, \citenamefont {Zhu}, \citenamefont {Yu}, \citenamefont {Zhu}, \citenamefont {Tsai}, \citenamefont {Kivshar},\ and\ \citenamefont {Liu}}]{shi2022OpticalManipulation}%
  \BibitemOpen
  \bibfield  {author} {\bibinfo {author} {\bibfnamefont {Y.}~\bibnamefont {Shi}}, \bibinfo {author} {\bibfnamefont {Q.}~\bibnamefont {Song}}, \bibinfo {author} {\bibfnamefont {I.}~\bibnamefont {Toftul}}, \bibinfo {author} {\bibfnamefont {T.}~\bibnamefont {Zhu}}, \bibinfo {author} {\bibfnamefont {Y.}~\bibnamefont {Yu}}, \bibinfo {author} {\bibfnamefont {W.}~\bibnamefont {Zhu}}, \bibinfo {author} {\bibfnamefont {D.~P.}\ \bibnamefont {Tsai}}, \bibinfo {author} {\bibfnamefont {Y.}~\bibnamefont {Kivshar}}, \ and\ \bibinfo {author} {\bibfnamefont {A.~Q.}\ \bibnamefont {Liu}},\ }\bibfield  {title} {\enquote {\bibinfo {title} {Optical manipulation with metamaterial structures},}\ }\href {\doibase 10.1063/5.0091280} {\bibfield  {journal} {\bibinfo  {journal} {Applied Physics Reviews}\ }\textbf {\bibinfo {volume} {9}},\ \bibinfo {pages} {031303} (\bibinfo {year} {2022})}\BibitemShut {NoStop}%
\bibitem [{\citenamefont {Zhou}\ \emph {et~al.}(2017)\citenamefont {Zhou}, \citenamefont {Liu}, \citenamefont {Kats},\ and\ \citenamefont {Yu}}]{zhou2017OpticalMetasurface}%
  \BibitemOpen
  \bibfield  {author} {\bibinfo {author} {\bibfnamefont {M.}~\bibnamefont {Zhou}}, \bibinfo {author} {\bibfnamefont {J.}~\bibnamefont {Liu}}, \bibinfo {author} {\bibfnamefont {M.~A.}\ \bibnamefont {Kats}}, \ and\ \bibinfo {author} {\bibfnamefont {Z.}~\bibnamefont {Yu}},\ }\bibfield  {title} {\enquote {\bibinfo {title} {Optical metasurface based on the resonant scattering in electronic transitions},}\ }\href {\doibase 10.1021/acsphotonics.7b00219} {\bibfield  {journal} {\bibinfo  {journal} {ACS Photonics}\ }\textbf {\bibinfo {volume} {4}},\ \bibinfo {pages} {1279--1285} (\bibinfo {year} {2017})}\BibitemShut {NoStop}%
\bibitem [{\citenamefont {Wang}\ \emph {et~al.}(2017)\citenamefont {Wang}, \citenamefont {Zhao}, \citenamefont {Kan},\ and\ \citenamefont {Huang}}]{wang2017DesignMetasurface}%
  \BibitemOpen
  \bibfield  {author} {\bibinfo {author} {\bibfnamefont {B.~X.}\ \bibnamefont {Wang}}, \bibinfo {author} {\bibfnamefont {C.~Y.}\ \bibnamefont {Zhao}}, \bibinfo {author} {\bibfnamefont {Y.~H.}\ \bibnamefont {Kan}}, \ and\ \bibinfo {author} {\bibfnamefont {T.~C.}\ \bibnamefont {Huang}},\ }\bibfield  {title} {\enquote {\bibinfo {title} {Design of metasurface polarizers based on two-dimensional cold atomic arrays},}\ }\href {\doibase 10.1364/OE.25.018760} {\bibfield  {journal} {\bibinfo  {journal} {Optics Express}\ }\textbf {\bibinfo {volume} {25}},\ \bibinfo {pages} {18760--18773} (\bibinfo {year} {2017})}\BibitemShut {NoStop}%
\bibitem [{\citenamefont {Facchinetti}\ and\ \citenamefont {Ruostekoski}(2018)}]{facchinetti2018InteractionLight}%
  \BibitemOpen
  \bibfield  {author} {\bibinfo {author} {\bibfnamefont {G.}~\bibnamefont {Facchinetti}}\ and\ \bibinfo {author} {\bibfnamefont {J.}~\bibnamefont {Ruostekoski}},\ }\bibfield  {title} {\enquote {\bibinfo {title} {Interaction of light with planar lattices of atoms: Reflection, transmission, and cooperative magnetometry},}\ }\href {\doibase 10.1103/PhysRevA.97.023833} {\bibfield  {journal} {\bibinfo  {journal} {Physical Review A}\ }\textbf {\bibinfo {volume} {97}},\ \bibinfo {pages} {023833} (\bibinfo {year} {2018})}\BibitemShut {NoStop}%
\bibitem [{\citenamefont {Bekenstein}\ \emph {et~al.}(2020)\citenamefont {Bekenstein}, \citenamefont {Pikovski}, \citenamefont {Pichler}, \citenamefont {Shahmoon}, \citenamefont {Yelin},\ and\ \citenamefont {Lukin}}]{bekenstein2020QuantumMetasurfaces}%
  \BibitemOpen
  \bibfield  {author} {\bibinfo {author} {\bibfnamefont {R.}~\bibnamefont {Bekenstein}}, \bibinfo {author} {\bibfnamefont {I.}~\bibnamefont {Pikovski}}, \bibinfo {author} {\bibfnamefont {H.}~\bibnamefont {Pichler}}, \bibinfo {author} {\bibfnamefont {E.}~\bibnamefont {Shahmoon}}, \bibinfo {author} {\bibfnamefont {S.~F.}\ \bibnamefont {Yelin}}, \ and\ \bibinfo {author} {\bibfnamefont {M.~D.}\ \bibnamefont {Lukin}},\ }\bibfield  {title} {\enquote {\bibinfo {title} {Quantum metasurfaces with atom arrays},}\ }\href {\doibase 10.1038/s41567-020-0845-5} {\bibfield  {journal} {\bibinfo  {journal} {Nature Physics}\ }\textbf {\bibinfo {volume} {16}},\ \bibinfo {pages} {676--681} (\bibinfo {year} {2020})}\BibitemShut {NoStop}%
\bibitem [{\citenamefont {Rui}\ \emph {et~al.}(2020)\citenamefont {Rui}, \citenamefont {Wei}, \citenamefont {Rubio-Abadal}, \citenamefont {Hollerith}, \citenamefont {Zeiher}, \citenamefont {Stamper-Kurn}, \citenamefont {Gross},\ and\ \citenamefont {Bloch}}]{rui2020SubradiantOptical}%
  \BibitemOpen
  \bibfield  {author} {\bibinfo {author} {\bibfnamefont {J.}~\bibnamefont {Rui}}, \bibinfo {author} {\bibfnamefont {D.}~\bibnamefont {Wei}}, \bibinfo {author} {\bibfnamefont {A.}~\bibnamefont {Rubio-Abadal}}, \bibinfo {author} {\bibfnamefont {S.}~\bibnamefont {Hollerith}}, \bibinfo {author} {\bibfnamefont {J.}~\bibnamefont {Zeiher}}, \bibinfo {author} {\bibfnamefont {D.~M.}\ \bibnamefont {Stamper-Kurn}}, \bibinfo {author} {\bibfnamefont {C.}~\bibnamefont {Gross}}, \ and\ \bibinfo {author} {\bibfnamefont {I.}~\bibnamefont {Bloch}},\ }\bibfield  {title} {\enquote {\bibinfo {title} {A subradiant optical mirror formed by a single structured atomic layer},}\ }\href {\doibase 10.1038/s41586-020-2463-x} {\bibfield  {journal} {\bibinfo  {journal} {Nature}\ }\textbf {\bibinfo {volume} {583}},\ \bibinfo {pages} {369--374} (\bibinfo {year} {2020})}\BibitemShut {NoStop}%
\bibitem [{\citenamefont {Alaee}\ \emph {et~al.}(2020)\citenamefont {Alaee}, \citenamefont {Gurlek}, \citenamefont {Albooyeh}, \citenamefont {Martín-Cano},\ and\ \citenamefont {Sandoghdar}}]{alaee2020QuantumMetamaterials}%
  \BibitemOpen
  \bibfield  {author} {\bibinfo {author} {\bibfnamefont {R.}~\bibnamefont {Alaee}}, \bibinfo {author} {\bibfnamefont {B.}~\bibnamefont {Gurlek}}, \bibinfo {author} {\bibfnamefont {M.}~\bibnamefont {Albooyeh}}, \bibinfo {author} {\bibfnamefont {D.}~\bibnamefont {Martín-Cano}}, \ and\ \bibinfo {author} {\bibfnamefont {V.}~\bibnamefont {Sandoghdar}},\ }\bibfield  {title} {\enquote {\bibinfo {title} {Quantum metamaterials with magnetic response at optical frequencies},}\ }\href {\doibase 10.1103/PhysRevLett.125.063601} {\bibfield  {journal} {\bibinfo  {journal} {Physical Review Letters}\ }\textbf {\bibinfo {volume} {125}},\ \bibinfo {pages} {063601} (\bibinfo {year} {2020})}\BibitemShut {NoStop}%
\bibitem [{\citenamefont {Ballantine}\ and\ \citenamefont {Ruostekoski}(2020)}]{ballantine2020OpticalMagnetism}%
  \BibitemOpen
  \bibfield  {author} {\bibinfo {author} {\bibfnamefont {K.~E.}\ \bibnamefont {Ballantine}}\ and\ \bibinfo {author} {\bibfnamefont {J.}~\bibnamefont {Ruostekoski}},\ }\bibfield  {title} {\enquote {\bibinfo {title} {Optical magnetism and huygens' surfaces in arrays of atoms induced by cooperative responses},}\ }\href {\doibase 10.1103/PhysRevLett.125.143604} {\bibfield  {journal} {\bibinfo  {journal} {Physical Review Letters}\ }\textbf {\bibinfo {volume} {125}},\ \bibinfo {pages} {143604} (\bibinfo {year} {2020})}\BibitemShut {NoStop}%
\bibitem [{\citenamefont {Ballantine}\ and\ \citenamefont {Ruostekoski}(2021{\natexlab{a}})}]{ballantine2021CooperativeOptical}%
  \BibitemOpen
  \bibfield  {author} {\bibinfo {author} {\bibfnamefont {K.~E.}\ \bibnamefont {Ballantine}}\ and\ \bibinfo {author} {\bibfnamefont {J.}~\bibnamefont {Ruostekoski}},\ }\bibfield  {title} {\enquote {\bibinfo {title} {Cooperative optical wavefront engineering with atomic arrays},}\ }\href {\doibase 10.1515/nanoph-2021-0059} {\bibfield  {journal} {\bibinfo  {journal} {Nanophotonics}\ }\textbf {\bibinfo {volume} {10}},\ \bibinfo {pages} {1901--1909} (\bibinfo {year} {2021}{\natexlab{a}})}\BibitemShut {NoStop}%
\bibitem [{\citenamefont {Ballantine}\ and\ \citenamefont {Ruostekoski}(2021{\natexlab{b}})}]{ballantine2021QuantumSinglePhoton}%
  \BibitemOpen
  \bibfield  {author} {\bibinfo {author} {\bibfnamefont {K.~E.}\ \bibnamefont {Ballantine}}\ and\ \bibinfo {author} {\bibfnamefont {J.}~\bibnamefont {Ruostekoski}},\ }\bibfield  {title} {\enquote {\bibinfo {title} {Quantum single-photon control, storage, and entanglement generation with planar atomic arrays},}\ }\href {\doibase 10.1103/PRXQuantum.2.040362} {\bibfield  {journal} {\bibinfo  {journal} {PRX Quantum}\ }\textbf {\bibinfo {volume} {2}},\ \bibinfo {pages} {040362} (\bibinfo {year} {2021}{\natexlab{b}})}\BibitemShut {NoStop}%
\bibitem [{\citenamefont {Fernández-Fernández}\ and\ \citenamefont {González-Tudela}(2022)}]{fernandez-fernandez2022TunableDirectional}%
  \BibitemOpen
  \bibfield  {author} {\bibinfo {author} {\bibfnamefont {D.}~\bibnamefont {Fernández-Fernández}}\ and\ \bibinfo {author} {\bibfnamefont {A.}~\bibnamefont {González-Tudela}},\ }\bibfield  {title} {\enquote {\bibinfo {title} {Tunable directional emission and collective dissipation with quantum metasurfaces},}\ }\href {\doibase 10.1103/PhysRevLett.128.113601} {\bibfield  {journal} {\bibinfo  {journal} {Physical Review Letters}\ }\textbf {\bibinfo {volume} {128}},\ \bibinfo {pages} {113601} (\bibinfo {year} {2022})}\BibitemShut {NoStop}%
\bibitem [{\citenamefont {Levin}, \citenamefont {Israeli},\ and\ \citenamefont {Bekenstein}(2023)}]{levin2023ClusterStates}%
  \BibitemOpen
  \bibfield  {author} {\bibinfo {author} {\bibfnamefont {Y.}~\bibnamefont {Levin}}, \bibinfo {author} {\bibfnamefont {U.}~\bibnamefont {Israeli}}, \ and\ \bibinfo {author} {\bibfnamefont {R.}~\bibnamefont {Bekenstein}},\ }\bibfield  {title} {\enquote {\bibinfo {title} {Cluster states generation with quantum metasurface},}\ }in\ \href {\doibase 10.1364/QUANTUM.2023.QTu4A.5} {\emph {\bibinfo {booktitle} {Optica Quantum 2.0 Conference and Exhibition (2023), Paper QTu4A.5}}}\ (\bibinfo  {publisher} {Optica Publishing Group},\ \bibinfo {year} {2023})\ p.\ \bibinfo {pages} {QTu4A.5}\BibitemShut {NoStop}%
\bibitem [{\citenamefont {Jenkins}\ and\ \citenamefont {Ruostekoski}(2012)}]{janne2012}%
  \BibitemOpen
  \bibfield  {author} {\bibinfo {author} {\bibfnamefont {S.}~\bibnamefont {Jenkins}}\ and\ \bibinfo {author} {\bibfnamefont {J.}~\bibnamefont {Ruostekoski}},\ }\bibfield  {title} {\enquote {\bibinfo {title} {Controlled manipulation of light by cooperative response of atoms in an optical lattice},}\ }\href {\doibase 10.1103/PhysRevA.64.052312} {\bibfield  {journal} {\bibinfo  {journal} {Physical Review A}\ }\textbf {\bibinfo {volume} {86}},\ \bibinfo {pages} {031602(R)} (\bibinfo {year} {2012})}\BibitemShut {NoStop}%
\bibitem [{\citenamefont {Chen}\ \emph {et~al.}(2023)\citenamefont {Chen}, \citenamefont {You}, \citenamefont {Gu}, \citenamefont {Ma},\ and\ \citenamefont {Cui}}]{chen2023AnalogQuantum}%
  \BibitemOpen
  \bibfield  {author} {\bibinfo {author} {\bibfnamefont {L.}~\bibnamefont {Chen}}, \bibinfo {author} {\bibfnamefont {J.~W.}\ \bibnamefont {You}}, \bibinfo {author} {\bibfnamefont {Z.}~\bibnamefont {Gu}}, \bibinfo {author} {\bibfnamefont {Q.}~\bibnamefont {Ma}}, \ and\ \bibinfo {author} {\bibfnamefont {T.~J.}\ \bibnamefont {Cui}},\ }\bibfield  {title} {\enquote {\bibinfo {title} {Analog quantum bit based on pancharatnam-berry phase metasurfaces},}\ }\href {\doibase 10.1002/qute.202300135} {\bibfield  {journal} {\bibinfo  {journal} {Advanced Quantum Technologies}\ }\textbf {\bibinfo {volume} {6}},\ \bibinfo {pages} {2300135} (\bibinfo {year} {2023})}\BibitemShut {NoStop}%
\bibitem [{\citenamefont {Tanuwijaya}\ \emph {et~al.}(2024)\citenamefont {Tanuwijaya}, \citenamefont {Liang}, \citenamefont {Xi}, \citenamefont {Wong}, \citenamefont {Yung}, \citenamefont {Tam},\ and\ \citenamefont {Li}}]{tanuwijaya2024metasurface}%
  \BibitemOpen
  \bibfield  {author} {\bibinfo {author} {\bibfnamefont {R.~S.}\ \bibnamefont {Tanuwijaya}}, \bibinfo {author} {\bibfnamefont {H.}~\bibnamefont {Liang}}, \bibinfo {author} {\bibfnamefont {J.}~\bibnamefont {Xi}}, \bibinfo {author} {\bibfnamefont {W.~C.}\ \bibnamefont {Wong}}, \bibinfo {author} {\bibfnamefont {T.~K.}\ \bibnamefont {Yung}}, \bibinfo {author} {\bibfnamefont {W.~Y.}\ \bibnamefont {Tam}}, \ and\ \bibinfo {author} {\bibfnamefont {J.}~\bibnamefont {Li}},\ }\bibfield  {title} {\enquote {\bibinfo {title} {Metasurface for programmable quantum algorithms with classical and quantum light},}\ }\href {\doibase 10.1515/nanoph-2023-0844} {\bibfield  {journal} {\bibinfo  {journal} {Nanophotonics}\ } (\bibinfo {year} {2024}),\ 10.1515/nanoph-2023-0844}\BibitemShut {NoStop}%
\bibitem [{\citenamefont {Lee}\ \emph {et~al.}(2021)\citenamefont {Lee}, \citenamefont {Lawrie}, \citenamefont {Pooser}, \citenamefont {Lee}, \citenamefont {Rockstuhl},\ and\ \citenamefont {Tame}}]{lee2021QuantumPlasmonic}%
  \BibitemOpen
  \bibfield  {author} {\bibinfo {author} {\bibfnamefont {C.}~\bibnamefont {Lee}}, \bibinfo {author} {\bibfnamefont {B.}~\bibnamefont {Lawrie}}, \bibinfo {author} {\bibfnamefont {R.}~\bibnamefont {Pooser}}, \bibinfo {author} {\bibfnamefont {K.-G.}\ \bibnamefont {Lee}}, \bibinfo {author} {\bibfnamefont {C.}~\bibnamefont {Rockstuhl}}, \ and\ \bibinfo {author} {\bibfnamefont {M.}~\bibnamefont {Tame}},\ }\bibfield  {title} {\enquote {\bibinfo {title} {Quantum plasmonic sensors},}\ }\href {\doibase 10.1021/acs.chemrev.0c01028} {\bibfield  {journal} {\bibinfo  {journal} {Chemical Reviews}\ }\textbf {\bibinfo {volume} {121}},\ \bibinfo {pages} {4743--4804} (\bibinfo {year} {2021})}\BibitemShut {NoStop}%
\bibitem [{\citenamefont {Dowran}\ \emph {et~al.}(2018)\citenamefont {Dowran}, \citenamefont {Kumar}, \citenamefont {Lawrie}, \citenamefont {Pooser},\ and\ \citenamefont {Marino}}]{dowran2018QuantumenhancedPlasmonica}%
  \BibitemOpen
  \bibfield  {author} {\bibinfo {author} {\bibfnamefont {M.}~\bibnamefont {Dowran}}, \bibinfo {author} {\bibfnamefont {A.}~\bibnamefont {Kumar}}, \bibinfo {author} {\bibfnamefont {B.~J.}\ \bibnamefont {Lawrie}}, \bibinfo {author} {\bibfnamefont {R.~C.}\ \bibnamefont {Pooser}}, \ and\ \bibinfo {author} {\bibfnamefont {A.~M.}\ \bibnamefont {Marino}},\ }\bibfield  {title} {\enquote {\bibinfo {title} {Quantum-enhanced plasmonic sensing},}\ }\href {\doibase 10.1364/OPTICA.5.000628} {\bibfield  {journal} {\bibinfo  {journal} {Optica}\ }\textbf {\bibinfo {volume} {5}},\ \bibinfo {pages} {628--633} (\bibinfo {year} {2018})}\BibitemShut {NoStop}%
\bibitem [{\citenamefont {Kim}\ \emph {et~al.}(2021)\citenamefont {Kim}, \citenamefont {Choi}, \citenamefont {Trusheim},\ and\ \citenamefont {Englund}}]{kim2021AbsorptionBasedDiamond}%
  \BibitemOpen
  \bibfield  {author} {\bibinfo {author} {\bibfnamefont {L.}~\bibnamefont {Kim}}, \bibinfo {author} {\bibfnamefont {H.}~\bibnamefont {Choi}}, \bibinfo {author} {\bibfnamefont {M.~E.}\ \bibnamefont {Trusheim}}, \ and\ \bibinfo {author} {\bibfnamefont {D.~R.}\ \bibnamefont {Englund}},\ }\bibfield  {title} {\enquote {\bibinfo {title} {Absorption-based diamond spin microscopy on a plasmonic quantum metasurface},}\ }\href {\doibase 10.1021/acsphotonics.1c01005} {\bibfield  {journal} {\bibinfo  {journal} {ACS Photonics}\ }\textbf {\bibinfo {volume} {8}},\ \bibinfo {pages} {3218--3225} (\bibinfo {year} {2021})}\BibitemShut {NoStop}%
\bibitem [{\citenamefont {Altepeter}, \citenamefont {Jeffrey},\ and\ \citenamefont {Kwiat}(2005)}]{altepeter2005PhotonicState}%
  \BibitemOpen
  \bibfield  {author} {\bibinfo {author} {\bibfnamefont {J.~B.}\ \bibnamefont {Altepeter}}, \bibinfo {author} {\bibfnamefont {E.~R.}\ \bibnamefont {Jeffrey}}, \ and\ \bibinfo {author} {\bibfnamefont {P.~G.}\ \bibnamefont {Kwiat}},\ }\bibfield  {title} {\enquote {\bibinfo {title} {Photonic state tomography},}\ }in\ \href {\doibase 10.1016/S1049-250X(05)52003-2} {\emph {\bibinfo {booktitle} {Advances In Atomic, Molecular, and Optical Physics}}},\ Vol.~\bibinfo {volume} {52},\ \bibinfo {editor} {edited by\ \bibinfo {editor} {\bibfnamefont {P.~R.}\ \bibnamefont {Berman}}\ and\ \bibinfo {editor} {\bibfnamefont {C.~C.}\ \bibnamefont {Lin}}}\ (\bibinfo  {publisher} {Academic Press},\ \bibinfo {year} {2005})\ pp.\ \bibinfo {pages} {105--159}\BibitemShut {NoStop}%
\bibitem [{\citenamefont {James}\ \emph {et~al.}(2001)\citenamefont {James}, \citenamefont {Kwiat}, \citenamefont {Munro},\ and\ \citenamefont {White}}]{james2001MeasurementQubits}%
  \BibitemOpen
  \bibfield  {author} {\bibinfo {author} {\bibfnamefont {D.~F.~V.}\ \bibnamefont {James}}, \bibinfo {author} {\bibfnamefont {P.~G.}\ \bibnamefont {Kwiat}}, \bibinfo {author} {\bibfnamefont {W.~J.}\ \bibnamefont {Munro}}, \ and\ \bibinfo {author} {\bibfnamefont {A.~G.}\ \bibnamefont {White}},\ }\bibfield  {title} {\enquote {\bibinfo {title} {Measurement of qubits},}\ }\href {\doibase 10.1103/PhysRevA.64.052312} {\bibfield  {journal} {\bibinfo  {journal} {Physical Review A}\ }\textbf {\bibinfo {volume} {64}},\ \bibinfo {pages} {052312} (\bibinfo {year} {2001})}\BibitemShut {NoStop}%
\bibitem [{\citenamefont {Wang}\ \emph {et~al.}(2022{\natexlab{a}})\citenamefont {Wang}, \citenamefont {Jiang}, \citenamefont {Gao}, \citenamefont {Fan}, \citenamefont {Qi}, \citenamefont {Zhong}, \citenamefont {Zhang}, \citenamefont {Peng},\ and\ \citenamefont {Wang}}]{wang2022ImplementQuantum}%
  \BibitemOpen
  \bibfield  {author} {\bibinfo {author} {\bibfnamefont {Z.}~\bibnamefont {Wang}}, \bibinfo {author} {\bibfnamefont {Y.}~\bibnamefont {Jiang}}, \bibinfo {author} {\bibfnamefont {Y.-J.}\ \bibnamefont {Gao}}, \bibinfo {author} {\bibfnamefont {R.-H.}\ \bibnamefont {Fan}}, \bibinfo {author} {\bibfnamefont {D.-X.}\ \bibnamefont {Qi}}, \bibinfo {author} {\bibfnamefont {R.}~\bibnamefont {Zhong}}, \bibinfo {author} {\bibfnamefont {H.-L.}\ \bibnamefont {Zhang}}, \bibinfo {author} {\bibfnamefont {R.-W.}\ \bibnamefont {Peng}}, \ and\ \bibinfo {author} {\bibfnamefont {M.}~\bibnamefont {Wang}},\ }\bibfield  {title} {\enquote {\bibinfo {title} {Implement quantum tomography of polarization-entangled states via nondiffractive metasurfaces},}\ }\href {\doibase 10.1063/5.0102539} {\bibfield  {journal} {\bibinfo  {journal} {Applied Physics Letters}\ }\textbf {\bibinfo {volume} {121}},\ \bibinfo {pages} {081703} (\bibinfo {year} {2022}{\natexlab{a}})}\BibitemShut {NoStop}%
\bibitem [{\citenamefont {Gao}\ \emph {et~al.}(2023)\citenamefont {Gao}, \citenamefont {Su}, \citenamefont {Song}, \citenamefont {Genevet},\ and\ \citenamefont {Dorfman}}]{gao2023MetasurfaceComplete}%
  \BibitemOpen
  \bibfield  {author} {\bibinfo {author} {\bibfnamefont {Z.}~\bibnamefont {Gao}}, \bibinfo {author} {\bibfnamefont {Z.}~\bibnamefont {Su}}, \bibinfo {author} {\bibfnamefont {Q.}~\bibnamefont {Song}}, \bibinfo {author} {\bibfnamefont {P.}~\bibnamefont {Genevet}}, \ and\ \bibinfo {author} {\bibfnamefont {K.~E.}\ \bibnamefont {Dorfman}},\ }\bibfield  {title} {\enquote {\bibinfo {title} {Metasurface for complete measurement of polarization bell state:},}\ }\href {\doibase 10.1515/nanoph-2022-0593} {\bibfield  {journal} {\bibinfo  {journal} {Nanophotonics}\ }\textbf {\bibinfo {volume} {12}},\ \bibinfo {pages} {569--577} (\bibinfo {year} {2023})}\BibitemShut {NoStop}%
\bibitem [{\citenamefont {Chen}\ \emph {et~al.}(2017)\citenamefont {Chen}, \citenamefont {Zhou}, \citenamefont {Mi}, \citenamefont {Liu}, \citenamefont {Luo},\ and\ \citenamefont {Wen}}]{chen2017DielectricMetasurfaces}%
  \BibitemOpen
  \bibfield  {author} {\bibinfo {author} {\bibfnamefont {S.}~\bibnamefont {Chen}}, \bibinfo {author} {\bibfnamefont {X.}~\bibnamefont {Zhou}}, \bibinfo {author} {\bibfnamefont {C.}~\bibnamefont {Mi}}, \bibinfo {author} {\bibfnamefont {Z.}~\bibnamefont {Liu}}, \bibinfo {author} {\bibfnamefont {H.}~\bibnamefont {Luo}}, \ and\ \bibinfo {author} {\bibfnamefont {S.}~\bibnamefont {Wen}},\ }\bibfield  {title} {\enquote {\bibinfo {title} {Dielectric metasurfaces for quantum weak measurements},}\ }\href {\doibase 10.1063/1.4982164} {\bibfield  {journal} {\bibinfo  {journal} {Applied Physics Letters}\ }\textbf {\bibinfo {volume} {110}},\ \bibinfo {pages} {161115} (\bibinfo {year} {2017})}\BibitemShut {NoStop}%
\bibitem [{\citenamefont {Ren}\ \emph {et~al.}(2023)\citenamefont {Ren}, \citenamefont {Zhang}, \citenamefont {Zheng}, \citenamefont {Ying}, \citenamefont {Xu}, \citenamefont {Rahmani},\ and\ \citenamefont {Whaley}}]{ren2023error}%
  \BibitemOpen
  \bibfield  {author} {\bibinfo {author} {\bibfnamefont {H.}~\bibnamefont {Ren}}, \bibinfo {author} {\bibfnamefont {Y.}~\bibnamefont {Zhang}}, \bibinfo {author} {\bibfnamefont {Z.}~\bibnamefont {Zheng}}, \bibinfo {author} {\bibfnamefont {C.}~\bibnamefont {Ying}}, \bibinfo {author} {\bibfnamefont {L.}~\bibnamefont {Xu}}, \bibinfo {author} {\bibfnamefont {M.}~\bibnamefont {Rahmani}}, \ and\ \bibinfo {author} {\bibfnamefont {K.~B.}\ \bibnamefont {Whaley}},\ }\href {https://arxiv.org/abs/2308.08755} {\enquote {\bibinfo {title} {Error mitigated metasurface-based randomized measurement schemes},}\ } (\bibinfo {year} {2023}),\ \Eprint {http://arxiv.org/abs/2308.08755} {arXiv:2308.08755 [quant-ph]} \BibitemShut {NoStop}%
\bibitem [{\citenamefont {An}\ \emph {et~al.}(2023)\citenamefont {An}, \citenamefont {Liu}, \citenamefont {Zhang}, \citenamefont {Li}, \citenamefont {Zhou}, \citenamefont {Yuan}, \citenamefont {Wang}, \citenamefont {Zhang}, \citenamefont {Wang},\ and\ \citenamefont {Lu}}]{an2023efficient}%
  \BibitemOpen
  \bibfield  {author} {\bibinfo {author} {\bibfnamefont {K.}~\bibnamefont {An}}, \bibinfo {author} {\bibfnamefont {Z.}~\bibnamefont {Liu}}, \bibinfo {author} {\bibfnamefont {T.}~\bibnamefont {Zhang}}, \bibinfo {author} {\bibfnamefont {S.}~\bibnamefont {Li}}, \bibinfo {author} {\bibfnamefont {Y.}~\bibnamefont {Zhou}}, \bibinfo {author} {\bibfnamefont {X.}~\bibnamefont {Yuan}}, \bibinfo {author} {\bibfnamefont {L.}~\bibnamefont {Wang}}, \bibinfo {author} {\bibfnamefont {W.}~\bibnamefont {Zhang}}, \bibinfo {author} {\bibfnamefont {G.}~\bibnamefont {Wang}}, \ and\ \bibinfo {author} {\bibfnamefont {H.}~\bibnamefont {Lu}},\ }\href {https://arxiv.org/abs/2308.07067v1} {\enquote {\bibinfo {title} {Efficient characterizations of multiphoton states with ultra-thin integrated photonics},}\ } (\bibinfo {year} {2023}),\ \Eprint {http://arxiv.org/abs/2308.07067} {arXiv:2308.07067 [quant-ph]} \BibitemShut {NoStop}%
\bibitem [{\citenamefont {Zhou}\ \emph {et~al.}(2020)\citenamefont {Zhou}, \citenamefont {Liu}, \citenamefont {Qian}, \citenamefont {Li}, \citenamefont {Luo}, \citenamefont {Wen}, \citenamefont {Zhou}, \citenamefont {Guo}, \citenamefont {Shi},\ and\ \citenamefont {Liu}}]{zhou2020MetasurfaceEnabled}%
  \BibitemOpen
  \bibfield  {author} {\bibinfo {author} {\bibfnamefont {J.}~\bibnamefont {Zhou}}, \bibinfo {author} {\bibfnamefont {S.}~\bibnamefont {Liu}}, \bibinfo {author} {\bibfnamefont {H.}~\bibnamefont {Qian}}, \bibinfo {author} {\bibfnamefont {Y.}~\bibnamefont {Li}}, \bibinfo {author} {\bibfnamefont {H.}~\bibnamefont {Luo}}, \bibinfo {author} {\bibfnamefont {S.}~\bibnamefont {Wen}}, \bibinfo {author} {\bibfnamefont {Z.}~\bibnamefont {Zhou}}, \bibinfo {author} {\bibfnamefont {G.}~\bibnamefont {Guo}}, \bibinfo {author} {\bibfnamefont {B.}~\bibnamefont {Shi}}, \ and\ \bibinfo {author} {\bibfnamefont {Z.}~\bibnamefont {Liu}},\ }\bibfield  {title} {\enquote {\bibinfo {title} {Metasurface enabled quantum edge detection},}\ }\href {\doibase 10.1126/sciadv.abc4385} {\bibfield  {journal} {\bibinfo  {journal} {Science Advances}\ }\textbf {\bibinfo {volume} {6}},\ \bibinfo {pages} {eabc4385} (\bibinfo {year} {2020})}\BibitemShut {NoStop}%
\bibitem [{\citenamefont {Zhang}\ \emph {et~al.}(2023)\citenamefont {Zhang}, \citenamefont {Ren}, \citenamefont {Ma},\ and\ \citenamefont {Sukhorukov}}]{zhang2023QuantumImaging}%
  \BibitemOpen
  \bibfield  {author} {\bibinfo {author} {\bibfnamefont {J.}~\bibnamefont {Zhang}}, \bibinfo {author} {\bibfnamefont {J.}~\bibnamefont {Ren}}, \bibinfo {author} {\bibfnamefont {J.}~\bibnamefont {Ma}}, \ and\ \bibinfo {author} {\bibfnamefont {A.~A.}\ \bibnamefont {Sukhorukov}},\ }\bibfield  {title} {\enquote {\bibinfo {title} {Quantum imaging with a nonlinear metasurface photon-pair source},}\ }in\ \href {\doibase 10.1109/ACP/POEM59049.2023.10369751} {\emph {\bibinfo {booktitle} {2023 Asia Communications and Photonics Conference/2023 International Photonics and Optoelectronics Meetings (ACP/POEM)}}}\ (\bibinfo  {publisher} {IEEE},\ \bibinfo {year} {2023})\ pp.\ \bibinfo {pages} {1--3}\BibitemShut {NoStop}%
\bibitem [{\citenamefont {Pan}\ \emph {et~al.}(2022)\citenamefont {Pan}, \citenamefont {Fu}, \citenamefont {Zheng}, \citenamefont {Chen}, \citenamefont {Zang}, \citenamefont {Duan}, \citenamefont {Li}, \citenamefont {Qiu},\ and\ \citenamefont {Hu}}]{pan2022DielectricMetalens}%
  \BibitemOpen
  \bibfield  {author} {\bibinfo {author} {\bibfnamefont {M.}~\bibnamefont {Pan}}, \bibinfo {author} {\bibfnamefont {Y.}~\bibnamefont {Fu}}, \bibinfo {author} {\bibfnamefont {M.}~\bibnamefont {Zheng}}, \bibinfo {author} {\bibfnamefont {H.}~\bibnamefont {Chen}}, \bibinfo {author} {\bibfnamefont {Y.}~\bibnamefont {Zang}}, \bibinfo {author} {\bibfnamefont {H.}~\bibnamefont {Duan}}, \bibinfo {author} {\bibfnamefont {Q.}~\bibnamefont {Li}}, \bibinfo {author} {\bibfnamefont {M.}~\bibnamefont {Qiu}}, \ and\ \bibinfo {author} {\bibfnamefont {Y.}~\bibnamefont {Hu}},\ }\bibfield  {title} {\enquote {\bibinfo {title} {Dielectric metalens for miniaturized imaging systems: Progress and challenges},}\ }\href {\doibase 10.1038/s41377-022-00885-7} {\bibfield  {journal} {\bibinfo  {journal} {Light: Science \& Applications}\ }\textbf {\bibinfo {volume} {11}},\ \bibinfo {pages} {195} (\bibinfo {year} {2022})}\BibitemShut {NoStop}%
\bibitem [{\citenamefont {He}, \citenamefont {Wang},\ and\ \citenamefont {Luo}(2022)}]{he2022ComputingMetasurfaces}%
  \BibitemOpen
  \bibfield  {author} {\bibinfo {author} {\bibfnamefont {S.}~\bibnamefont {He}}, \bibinfo {author} {\bibfnamefont {R.}~\bibnamefont {Wang}}, \ and\ \bibinfo {author} {\bibfnamefont {H.}~\bibnamefont {Luo}},\ }\bibfield  {title} {\enquote {\bibinfo {title} {Computing metasurfaces for all-optical image processing: A brief review},}\ }\href {\doibase 10.1515/nanoph-2021-0823} {\bibfield  {journal} {\bibinfo  {journal} {Nanophotonics}\ }\textbf {\bibinfo {volume} {11}},\ \bibinfo {pages} {1083--1108} (\bibinfo {year} {2022})}\BibitemShut {NoStop}%
\bibitem [{\citenamefont {Altuzarra}\ \emph {et~al.}(2019)\citenamefont {Altuzarra}, \citenamefont {Lyons}, \citenamefont {Yuan}, \citenamefont {Simpson}, \citenamefont {Roger}, \citenamefont {{Ben-Benjamin}},\ and\ \citenamefont {Faccio}}]{altuzarra2019ImagingPolarizationsensitivea}%
  \BibitemOpen
  \bibfield  {author} {\bibinfo {author} {\bibfnamefont {C.}~\bibnamefont {Altuzarra}}, \bibinfo {author} {\bibfnamefont {A.}~\bibnamefont {Lyons}}, \bibinfo {author} {\bibfnamefont {G.}~\bibnamefont {Yuan}}, \bibinfo {author} {\bibfnamefont {C.}~\bibnamefont {Simpson}}, \bibinfo {author} {\bibfnamefont {T.}~\bibnamefont {Roger}}, \bibinfo {author} {\bibfnamefont {J.~S.}\ \bibnamefont {{Ben-Benjamin}}}, \ and\ \bibinfo {author} {\bibfnamefont {D.}~\bibnamefont {Faccio}},\ }\bibfield  {title} {\enquote {\bibinfo {title} {Imaging of polarization-sensitive metasurfaces with quantum entanglement},}\ }\href {\doibase 10.1103/PhysRevA.99.020101} {\bibfield  {journal} {\bibinfo  {journal} {Physical Review A}\ }\textbf {\bibinfo {volume} {99}},\ \bibinfo {pages} {020101} (\bibinfo {year} {2019})}\BibitemShut {NoStop}%
\bibitem [{\citenamefont {Vega}\ \emph {et~al.}(2021)\citenamefont {Vega}, \citenamefont {Pertsch}, \citenamefont {Setzpfandt},\ and\ \citenamefont {Sukhorukov}}]{vega2021MetasurfaceAssistedQuantum}%
  \BibitemOpen
  \bibfield  {author} {\bibinfo {author} {\bibfnamefont {A.}~\bibnamefont {Vega}}, \bibinfo {author} {\bibfnamefont {T.}~\bibnamefont {Pertsch}}, \bibinfo {author} {\bibfnamefont {F.}~\bibnamefont {Setzpfandt}}, \ and\ \bibinfo {author} {\bibfnamefont {A.~A.}\ \bibnamefont {Sukhorukov}},\ }\bibfield  {title} {\enquote {\bibinfo {title} {Metasurface-assisted quantum ghost discrimination of polarization objects},}\ }\href {\doibase 10.1103/PhysRevApplied.16.064032} {\bibfield  {journal} {\bibinfo  {journal} {Physical Review Applied}\ }\textbf {\bibinfo {volume} {16}},\ \bibinfo {pages} {064032} (\bibinfo {year} {2021})}\BibitemShut {NoStop}%
\bibitem [{\citenamefont {Yung}\ \emph {et~al.}(2022{\natexlab{b}})\citenamefont {Yung}, \citenamefont {Liang}, \citenamefont {Xi}, \citenamefont {Tam},\ and\ \citenamefont {Li}}]{yung2022JonesmatrixImaging}%
  \BibitemOpen
  \bibfield  {author} {\bibinfo {author} {\bibfnamefont {T.~K.}\ \bibnamefont {Yung}}, \bibinfo {author} {\bibfnamefont {H.}~\bibnamefont {Liang}}, \bibinfo {author} {\bibfnamefont {J.}~\bibnamefont {Xi}}, \bibinfo {author} {\bibfnamefont {W.~Y.}\ \bibnamefont {Tam}}, \ and\ \bibinfo {author} {\bibfnamefont {J.}~\bibnamefont {Li}},\ }\bibfield  {title} {\enquote {\bibinfo {title} {Jones-matrix imaging based on two-photon interference},}\ }\href {\doibase 10.1515/nanoph-2022-0499} {\bibfield  {journal} {\bibinfo  {journal} {Nanophotonics}\ }\textbf {\bibinfo {volume} {12}},\ \bibinfo {pages} {579--588} (\bibinfo {year} {2022}{\natexlab{b}})}\BibitemShut {NoStop}%
\bibitem [{\citenamefont {Liu}\ \emph {et~al.}(2023{\natexlab{c}})\citenamefont {Liu}, \citenamefont {Zhu}, \citenamefont {Zhou}, \citenamefont {Zou}, \citenamefont {Qin}, \citenamefont {Wang}, \citenamefont {Zhu},\ and\ \citenamefont {Wang}}]{liu2023MetasurfacesEnableda}%
  \BibitemOpen
  \bibfield  {author} {\bibinfo {author} {\bibfnamefont {J.}~\bibnamefont {Liu}}, \bibinfo {author} {\bibfnamefont {X.}~\bibnamefont {Zhu}}, \bibinfo {author} {\bibfnamefont {Y.}~\bibnamefont {Zhou}}, \bibinfo {author} {\bibfnamefont {X.}~\bibnamefont {Zou}}, \bibinfo {author} {\bibfnamefont {Z.}~\bibnamefont {Qin}}, \bibinfo {author} {\bibfnamefont {S.}~\bibnamefont {Wang}}, \bibinfo {author} {\bibfnamefont {S.}~\bibnamefont {Zhu}}, \ and\ \bibinfo {author} {\bibfnamefont {Z.}~\bibnamefont {Wang}},\ }\bibfield  {title} {\enquote {\bibinfo {title} {Metasurfaces enabled polarization-multiplexing heralded single photon imaging},}\ }\href {\doibase 10.1364/OE.482426} {\bibfield  {journal} {\bibinfo  {journal} {Optics Express}\ }\textbf {\bibinfo {volume} {31}},\ \bibinfo {pages} {6217--6227} (\bibinfo {year} {2023}{\natexlab{c}})}\BibitemShut {NoStop}%
\bibitem [{\citenamefont {Liu}\ \emph {et~al.}(2024)\citenamefont {Liu}, \citenamefont {Yang}, \citenamefont {Shou}, \citenamefont {Chen}, \citenamefont {Shu}, \citenamefont {Chen}, \citenamefont {Wen},\ and\ \citenamefont {Luo}}]{PhysRevLett.132.043601}%
  \BibitemOpen
  \bibfield  {author} {\bibinfo {author} {\bibfnamefont {J.}~\bibnamefont {Liu}}, \bibinfo {author} {\bibfnamefont {Q.}~\bibnamefont {Yang}}, \bibinfo {author} {\bibfnamefont {Y.}~\bibnamefont {Shou}}, \bibinfo {author} {\bibfnamefont {S.}~\bibnamefont {Chen}}, \bibinfo {author} {\bibfnamefont {W.}~\bibnamefont {Shu}}, \bibinfo {author} {\bibfnamefont {G.}~\bibnamefont {Chen}}, \bibinfo {author} {\bibfnamefont {S.}~\bibnamefont {Wen}}, \ and\ \bibinfo {author} {\bibfnamefont {H.}~\bibnamefont {Luo}},\ }\bibfield  {title} {\enquote {\bibinfo {title} {Metasurface-assisted quantum nonlocal weak-measurement microscopy},}\ }\href {\doibase 10.1103/PhysRevLett.132.043601} {\bibfield  {journal} {\bibinfo  {journal} {Phys. Rev. Lett.}\ }\textbf {\bibinfo {volume} {132}},\ \bibinfo {pages} {043601} (\bibinfo {year} {2024})}\BibitemShut {NoStop}%
\bibitem [{\citenamefont {Meng}\ \emph {et~al.}(2021)\citenamefont {Meng}, \citenamefont {Chen}, \citenamefont {Lu}, \citenamefont {Ding}, \citenamefont {Cusano}, \citenamefont {Fan}, \citenamefont {Hu}, \citenamefont {Wang}, \citenamefont {Xie}, \citenamefont {Liu}, \citenamefont {Yang}, \citenamefont {Liu}, \citenamefont {Gong}, \citenamefont {Xiao}, \citenamefont {Sun}, \citenamefont {Zhang}, \citenamefont {Yuan},\ and\ \citenamefont {Ni}}]{meng2021OpticalMetawaveguides}%
  \BibitemOpen
  \bibfield  {author} {\bibinfo {author} {\bibfnamefont {Y.}~\bibnamefont {Meng}}, \bibinfo {author} {\bibfnamefont {Y.}~\bibnamefont {Chen}}, \bibinfo {author} {\bibfnamefont {L.}~\bibnamefont {Lu}}, \bibinfo {author} {\bibfnamefont {Y.}~\bibnamefont {Ding}}, \bibinfo {author} {\bibfnamefont {A.}~\bibnamefont {Cusano}}, \bibinfo {author} {\bibfnamefont {J.~A.}\ \bibnamefont {Fan}}, \bibinfo {author} {\bibfnamefont {Q.}~\bibnamefont {Hu}}, \bibinfo {author} {\bibfnamefont {K.}~\bibnamefont {Wang}}, \bibinfo {author} {\bibfnamefont {Z.}~\bibnamefont {Xie}}, \bibinfo {author} {\bibfnamefont {Z.}~\bibnamefont {Liu}}, \bibinfo {author} {\bibfnamefont {Y.}~\bibnamefont {Yang}}, \bibinfo {author} {\bibfnamefont {Q.}~\bibnamefont {Liu}}, \bibinfo {author} {\bibfnamefont {M.}~\bibnamefont {Gong}}, \bibinfo {author} {\bibfnamefont {Q.}~\bibnamefont {Xiao}}, \bibinfo {author} {\bibfnamefont {S.}~\bibnamefont {Sun}}, \bibinfo {author} {\bibfnamefont {M.}~\bibnamefont {Zhang}}, \bibinfo {author} {\bibfnamefont
  {X.}~\bibnamefont {Yuan}}, \ and\ \bibinfo {author} {\bibfnamefont {X.}~\bibnamefont {Ni}},\ }\bibfield  {title} {\enquote {\bibinfo {title} {Optical meta-waveguides for integrated photonics and beyond},}\ }\href {\doibase 10.1038/s41377-021-00655-x} {\bibfield  {journal} {\bibinfo  {journal} {Light: Science \& Applications}\ }\textbf {\bibinfo {volume} {10}},\ \bibinfo {pages} {235} (\bibinfo {year} {2021})}\BibitemShut {NoStop}%
\bibitem [{\citenamefont {Wang}\ \emph {et~al.}(2022{\natexlab{b}})\citenamefont {Wang}, \citenamefont {Xiao}, \citenamefont {Liao}, \citenamefont {Li}, \citenamefont {Song}, \citenamefont {Chen}, \citenamefont {Uddin}, \citenamefont {Mao}, \citenamefont {Wang}, \citenamefont {Zhou}, \citenamefont {Yuan}, \citenamefont {Jiang}, \citenamefont {Fontaine}, \citenamefont {Agrawal}, \citenamefont {Willner}, \citenamefont {Hu},\ and\ \citenamefont {Gu}}]{wang2022MetasurfaceIntegrated}%
  \BibitemOpen
  \bibfield  {author} {\bibinfo {author} {\bibfnamefont {Z.}~\bibnamefont {Wang}}, \bibinfo {author} {\bibfnamefont {Y.}~\bibnamefont {Xiao}}, \bibinfo {author} {\bibfnamefont {K.}~\bibnamefont {Liao}}, \bibinfo {author} {\bibfnamefont {T.}~\bibnamefont {Li}}, \bibinfo {author} {\bibfnamefont {H.}~\bibnamefont {Song}}, \bibinfo {author} {\bibfnamefont {H.}~\bibnamefont {Chen}}, \bibinfo {author} {\bibfnamefont {S.~M.~Z.}\ \bibnamefont {Uddin}}, \bibinfo {author} {\bibfnamefont {D.}~\bibnamefont {Mao}}, \bibinfo {author} {\bibfnamefont {F.}~\bibnamefont {Wang}}, \bibinfo {author} {\bibfnamefont {Z.}~\bibnamefont {Zhou}}, \bibinfo {author} {\bibfnamefont {B.}~\bibnamefont {Yuan}}, \bibinfo {author} {\bibfnamefont {W.}~\bibnamefont {Jiang}}, \bibinfo {author} {\bibfnamefont {N.~K.}\ \bibnamefont {Fontaine}}, \bibinfo {author} {\bibfnamefont {A.}~\bibnamefont {Agrawal}}, \bibinfo {author} {\bibfnamefont {A.~E.}\ \bibnamefont {Willner}}, \bibinfo {author} {\bibfnamefont {X.}~\bibnamefont {Hu}}, \ and\ \bibinfo
  {author} {\bibfnamefont {T.}~\bibnamefont {Gu}},\ }\bibfield  {title} {\enquote {\bibinfo {title} {Metasurface on integrated photonic platform: From mode converters to machine learning},}\ }\href {\doibase 10.1515/nanoph-2022-0294} {\bibfield  {journal} {\bibinfo  {journal} {Nanophotonics}\ }\textbf {\bibinfo {volume} {11}},\ \bibinfo {pages} {3531--3546} (\bibinfo {year} {2022}{\natexlab{b}})}\BibitemShut {NoStop}%
\bibitem [{\citenamefont {Kudyshev}, \citenamefont {Shalaev},\ and\ \citenamefont {Boltasseva}(2021)}]{kudyshev2021MachineLearning}%
  \BibitemOpen
  \bibfield  {author} {\bibinfo {author} {\bibfnamefont {Z.~A.}\ \bibnamefont {Kudyshev}}, \bibinfo {author} {\bibfnamefont {V.~M.}\ \bibnamefont {Shalaev}}, \ and\ \bibinfo {author} {\bibfnamefont {A.}~\bibnamefont {Boltasseva}},\ }\bibfield  {title} {\enquote {\bibinfo {title} {Machine learning for integrated quantum photonics},}\ }\href {\doibase 10.1021/acsphotonics.0c00960} {\bibfield  {journal} {\bibinfo  {journal} {ACS Photonics}\ }\textbf {\bibinfo {volume} {8}},\ \bibinfo {pages} {34--46} (\bibinfo {year} {2021})}\BibitemShut {NoStop}%
\bibitem [{\citenamefont {Lee}\ \emph {et~al.}(2017)\citenamefont {Lee}, \citenamefont {Tian}, \citenamefont {Yang}, \citenamefont {Mustonen}, \citenamefont {Martinez}, \citenamefont {Dai}, \citenamefont {Kauppinen}, \citenamefont {Malowicki}, \citenamefont {Kumar},\ and\ \citenamefont {Sun}}]{lee2017PhotonPairGeneration}%
  \BibitemOpen
  \bibfield  {author} {\bibinfo {author} {\bibfnamefont {K.~F.}\ \bibnamefont {Lee}}, \bibinfo {author} {\bibfnamefont {Y.}~\bibnamefont {Tian}}, \bibinfo {author} {\bibfnamefont {H.}~\bibnamefont {Yang}}, \bibinfo {author} {\bibfnamefont {K.}~\bibnamefont {Mustonen}}, \bibinfo {author} {\bibfnamefont {A.}~\bibnamefont {Martinez}}, \bibinfo {author} {\bibfnamefont {Q.}~\bibnamefont {Dai}}, \bibinfo {author} {\bibfnamefont {E.~I.}\ \bibnamefont {Kauppinen}}, \bibinfo {author} {\bibfnamefont {J.}~\bibnamefont {Malowicki}}, \bibinfo {author} {\bibfnamefont {P.}~\bibnamefont {Kumar}}, \ and\ \bibinfo {author} {\bibfnamefont {Z.}~\bibnamefont {Sun}},\ }\bibfield  {title} {\enquote {\bibinfo {title} {Photon-pair generation with a 100 nm thick carbon nanotube film},}\ }\href {\doibase 10.1002/adma.201605978} {\bibfield  {journal} {\bibinfo  {journal} {Advanced Materials}\ }\textbf {\bibinfo {volume} {29}},\ \bibinfo {pages} {1605978} (\bibinfo {year} {2017})}\BibitemShut {NoStop}%
\bibitem [{\citenamefont {Guo}\ \emph {et~al.}(2023)\citenamefont {Guo}, \citenamefont {Qi}, \citenamefont {Zhang}, \citenamefont {Gao}, \citenamefont {Hu}, \citenamefont {Zhou}, \citenamefont {Zang}, \citenamefont {Zhao}, \citenamefont {Wang}, \citenamefont {Yan}, \citenamefont {Xu}, \citenamefont {Wu}, \citenamefont {Eda}, \citenamefont {Xiao}, \citenamefont {Yang}, \citenamefont {Gou}, \citenamefont {Feng}, \citenamefont {Guo}, \citenamefont {Zhou}, \citenamefont {Ren}, \citenamefont {Qiu}, \citenamefont {Pennycook},\ and\ \citenamefont {Wee}}]{guo2023UltrathinQuantum}%
  \BibitemOpen
  \bibfield  {author} {\bibinfo {author} {\bibfnamefont {Q.}~\bibnamefont {Guo}}, \bibinfo {author} {\bibfnamefont {X.-Z.}\ \bibnamefont {Qi}}, \bibinfo {author} {\bibfnamefont {L.}~\bibnamefont {Zhang}}, \bibinfo {author} {\bibfnamefont {M.}~\bibnamefont {Gao}}, \bibinfo {author} {\bibfnamefont {S.}~\bibnamefont {Hu}}, \bibinfo {author} {\bibfnamefont {W.}~\bibnamefont {Zhou}}, \bibinfo {author} {\bibfnamefont {W.}~\bibnamefont {Zang}}, \bibinfo {author} {\bibfnamefont {X.}~\bibnamefont {Zhao}}, \bibinfo {author} {\bibfnamefont {J.}~\bibnamefont {Wang}}, \bibinfo {author} {\bibfnamefont {B.}~\bibnamefont {Yan}}, \bibinfo {author} {\bibfnamefont {M.}~\bibnamefont {Xu}}, \bibinfo {author} {\bibfnamefont {Y.-K.}\ \bibnamefont {Wu}}, \bibinfo {author} {\bibfnamefont {G.}~\bibnamefont {Eda}}, \bibinfo {author} {\bibfnamefont {Z.}~\bibnamefont {Xiao}}, \bibinfo {author} {\bibfnamefont {S.~A.}\ \bibnamefont {Yang}}, \bibinfo {author} {\bibfnamefont {H.}~\bibnamefont {Gou}}, \bibinfo {author} {\bibfnamefont {Y.~P.}\
  \bibnamefont {Feng}}, \bibinfo {author} {\bibfnamefont {G.-C.}\ \bibnamefont {Guo}}, \bibinfo {author} {\bibfnamefont {W.}~\bibnamefont {Zhou}}, \bibinfo {author} {\bibfnamefont {X.-F.}\ \bibnamefont {Ren}}, \bibinfo {author} {\bibfnamefont {C.-W.}\ \bibnamefont {Qiu}}, \bibinfo {author} {\bibfnamefont {S.~J.}\ \bibnamefont {Pennycook}}, \ and\ \bibinfo {author} {\bibfnamefont {A.~T.~S.}\ \bibnamefont {Wee}},\ }\bibfield  {title} {\enquote {\bibinfo {title} {Ultrathin quantum light source with van der waals nbocl2 crystal},}\ }\href {\doibase 10.1038/s41586-022-05393-7} {\bibfield  {journal} {\bibinfo  {journal} {Nature}\ }\textbf {\bibinfo {volume} {613}},\ \bibinfo {pages} {53--59} (\bibinfo {year} {2023})}\BibitemShut {NoStop}%
\bibitem [{\citenamefont {Weissflog}\ \emph {et~al.}(2023)\citenamefont {Weissflog}, \citenamefont {Fedotova}, \citenamefont {Tang}, \citenamefont {Santos}, \citenamefont {Laudert}, \citenamefont {Shinde}, \citenamefont {Abtahi}, \citenamefont {Afsharnia}, \citenamefont {Pérez}, \citenamefont {Ritter}, \citenamefont {Qin}, \citenamefont {Janousek}, \citenamefont {Shradha}, \citenamefont {Staude}, \citenamefont {Saravi}, \citenamefont {Pertsch}, \citenamefont {Setzpfandt}, \citenamefont {Lu},\ and\ \citenamefont {Eilenberger}}]{weissflog2023TunableTransition}%
  \BibitemOpen
  \bibfield  {author} {\bibinfo {author} {\bibfnamefont {M.~A.}\ \bibnamefont {Weissflog}}, \bibinfo {author} {\bibfnamefont {A.}~\bibnamefont {Fedotova}}, \bibinfo {author} {\bibfnamefont {Y.}~\bibnamefont {Tang}}, \bibinfo {author} {\bibfnamefont {E.~A.}\ \bibnamefont {Santos}}, \bibinfo {author} {\bibfnamefont {B.}~\bibnamefont {Laudert}}, \bibinfo {author} {\bibfnamefont {S.}~\bibnamefont {Shinde}}, \bibinfo {author} {\bibfnamefont {F.}~\bibnamefont {Abtahi}}, \bibinfo {author} {\bibfnamefont {M.}~\bibnamefont {Afsharnia}}, \bibinfo {author} {\bibfnamefont {I.~P.}\ \bibnamefont {Pérez}}, \bibinfo {author} {\bibfnamefont {S.}~\bibnamefont {Ritter}}, \bibinfo {author} {\bibfnamefont {H.}~\bibnamefont {Qin}}, \bibinfo {author} {\bibfnamefont {J.}~\bibnamefont {Janousek}}, \bibinfo {author} {\bibfnamefont {S.}~\bibnamefont {Shradha}}, \bibinfo {author} {\bibfnamefont {I.}~\bibnamefont {Staude}}, \bibinfo {author} {\bibfnamefont {S.}~\bibnamefont {Saravi}}, \bibinfo {author} {\bibfnamefont {T.}~\bibnamefont
  {Pertsch}}, \bibinfo {author} {\bibfnamefont {F.}~\bibnamefont {Setzpfandt}}, \bibinfo {author} {\bibfnamefont {Y.}~\bibnamefont {Lu}}, \ and\ \bibinfo {author} {\bibfnamefont {F.}~\bibnamefont {Eilenberger}},\ }\href {https://arxiv.org/abs/2311.16036v1} {\enquote {\bibinfo {title} {A tunable transition metal dichalcogenide entangled photon-pair source},}\ } (\bibinfo {year} {2023}),\ \Eprint {http://arxiv.org/abs/2311.16036} {arXiv:2311.16036 [quant-ph]} \BibitemShut {NoStop}%
\bibitem [{\citenamefont {Braun}\ \emph {et~al.}(2024)\citenamefont {Braun}, \citenamefont {Bajo}, \citenamefont {Trovatello}, \citenamefont {Jenke}, \citenamefont {Cerullo}, \citenamefont {Schuck}, \citenamefont {Walther},\ and\ \citenamefont {Rozema}}]{Braun2024spontaneousparametric}%
  \BibitemOpen
  \bibfield  {author} {\bibinfo {author} {\bibfnamefont {B.}~\bibnamefont {Braun}}, \bibinfo {author} {\bibfnamefont {J.}~\bibnamefont {Bajo}}, \bibinfo {author} {\bibfnamefont {C.}~\bibnamefont {Trovatello}}, \bibinfo {author} {\bibfnamefont {P.~K.}\ \bibnamefont {Jenke}}, \bibinfo {author} {\bibfnamefont {G.}~\bibnamefont {Cerullo}}, \bibinfo {author} {\bibfnamefont {P.~J.}\ \bibnamefont {Schuck}}, \bibinfo {author} {\bibfnamefont {P.}~\bibnamefont {Walther}}, \ and\ \bibinfo {author} {\bibfnamefont {L.~A.}\ \bibnamefont {Rozema}},\ }\bibfield  {title} {\enquote {\bibinfo {title} {Spontaneous parametric downconversion in ultra-thin 3r-stacked transition-metal dichalcogenides},}\ }in\ \href@noop {} {\emph {\bibinfo {booktitle} {9th International Topical Meeting on Nanophotonics and Metamaterials}}}\ (\bibinfo {year} {2024})\BibitemShut {NoStop}%
\bibitem [{\citenamefont {Lemos}\ \emph {et~al.}(2014)\citenamefont {Lemos}, \citenamefont {Borish}, \citenamefont {Cole}, \citenamefont {Ramelow}, \citenamefont {Lapkiewicz},\ and\ \citenamefont {Zeilinger}}]{lemos2014QuantumImaginga}%
  \BibitemOpen
  \bibfield  {author} {\bibinfo {author} {\bibfnamefont {G.~B.}\ \bibnamefont {Lemos}}, \bibinfo {author} {\bibfnamefont {V.}~\bibnamefont {Borish}}, \bibinfo {author} {\bibfnamefont {G.~D.}\ \bibnamefont {Cole}}, \bibinfo {author} {\bibfnamefont {S.}~\bibnamefont {Ramelow}}, \bibinfo {author} {\bibfnamefont {R.}~\bibnamefont {Lapkiewicz}}, \ and\ \bibinfo {author} {\bibfnamefont {A.}~\bibnamefont {Zeilinger}},\ }\bibfield  {title} {\enquote {\bibinfo {title} {Quantum imaging with undetected photons},}\ }\href {\doibase 10.1038/nature13586} {\bibfield  {journal} {\bibinfo  {journal} {Nature}\ }\textbf {\bibinfo {volume} {512}},\ \bibinfo {pages} {409--412} (\bibinfo {year} {2014})}\BibitemShut {NoStop}%
\bibitem [{\citenamefont {Gorlach}\ \emph {et~al.}(2020)\citenamefont {Gorlach}, \citenamefont {Neufeld}, \citenamefont {Rivera}, \citenamefont {Cohen},\ and\ \citenamefont {Kaminer}}]{gorlach2020quantum}%
  \BibitemOpen
  \bibfield  {author} {\bibinfo {author} {\bibfnamefont {A.}~\bibnamefont {Gorlach}}, \bibinfo {author} {\bibfnamefont {O.}~\bibnamefont {Neufeld}}, \bibinfo {author} {\bibfnamefont {N.}~\bibnamefont {Rivera}}, \bibinfo {author} {\bibfnamefont {O.}~\bibnamefont {Cohen}}, \ and\ \bibinfo {author} {\bibfnamefont {I.}~\bibnamefont {Kaminer}},\ }\bibfield  {title} {\enquote {\bibinfo {title} {The quantum-optical nature of high harmonic generation},}\ }\href@noop {} {\bibfield  {journal} {\bibinfo  {journal} {Nature Communications}\ }\textbf {\bibinfo {volume} {11}},\ \bibinfo {pages} {4598} (\bibinfo {year} {2020})}\BibitemShut {NoStop}%
\bibitem [{\citenamefont {Gorlach}\ \emph {et~al.}(2023)\citenamefont {Gorlach}, \citenamefont {Tzur}, \citenamefont {Birk}, \citenamefont {Kr{\"u}ger}, \citenamefont {Rivera}, \citenamefont {Cohen},\ and\ \citenamefont {Kaminer}}]{gorlach2023high}%
  \BibitemOpen
  \bibfield  {author} {\bibinfo {author} {\bibfnamefont {A.}~\bibnamefont {Gorlach}}, \bibinfo {author} {\bibfnamefont {M.~E.}\ \bibnamefont {Tzur}}, \bibinfo {author} {\bibfnamefont {M.}~\bibnamefont {Birk}}, \bibinfo {author} {\bibfnamefont {M.}~\bibnamefont {Kr{\"u}ger}}, \bibinfo {author} {\bibfnamefont {N.}~\bibnamefont {Rivera}}, \bibinfo {author} {\bibfnamefont {O.}~\bibnamefont {Cohen}}, \ and\ \bibinfo {author} {\bibfnamefont {I.}~\bibnamefont {Kaminer}},\ }\bibfield  {title} {\enquote {\bibinfo {title} {High-harmonic generation driven by quantum light},}\ }\href@noop {} {\bibfield  {journal} {\bibinfo  {journal} {Nature Physics}\ }\textbf {\bibinfo {volume} {19}},\ \bibinfo {pages} {1689--1696} (\bibinfo {year} {2023})}\BibitemShut {NoStop}%
\bibitem [{\citenamefont {Forbes}, \citenamefont {de~Oliveira},\ and\ \citenamefont {Dennis}(2021)}]{forbes2021StructuredLighta}%
  \BibitemOpen
  \bibfield  {author} {\bibinfo {author} {\bibfnamefont {A.}~\bibnamefont {Forbes}}, \bibinfo {author} {\bibfnamefont {M.}~\bibnamefont {de~Oliveira}}, \ and\ \bibinfo {author} {\bibfnamefont {M.~R.}\ \bibnamefont {Dennis}},\ }\bibfield  {title} {\enquote {\bibinfo {title} {Structured light},}\ }\href {\doibase 10.1038/s41566-021-00780-4} {\bibfield  {journal} {\bibinfo  {journal} {Nature Photonics}\ }\textbf {\bibinfo {volume} {15}},\ \bibinfo {pages} {253--262} (\bibinfo {year} {2021})}\BibitemShut {NoStop}%
\bibitem [{\citenamefont {Bliokh}\ \emph {et~al.}(2023)\citenamefont {Bliokh}, \citenamefont {Karimi}, \citenamefont {Padgett}, \citenamefont {Alonso}, \citenamefont {Dennis}, \citenamefont {Dudley}, \citenamefont {Forbes}, \citenamefont {Zahedpour}, \citenamefont {Hancock}, \citenamefont {Milchberg}, \citenamefont {Rotter}, \citenamefont {Nori}, \citenamefont {Özdemir}, \citenamefont {Bender}, \citenamefont {Cao}, \citenamefont {Corkum}, \citenamefont {Hernández-García}, \citenamefont {Ren}, \citenamefont {Kivshar}, \citenamefont {Silveirinha}, \citenamefont {Engheta}, \citenamefont {Rauschenbeutel}, \citenamefont {Schneeweiss}, \citenamefont {Volz}, \citenamefont {Leykam}, \citenamefont {Smirnova}, \citenamefont {Rong}, \citenamefont {Wang}, \citenamefont {Hasman}, \citenamefont {Picardi}, \citenamefont {Zayats}, \citenamefont {Rodríguez-Fortuño}, \citenamefont {Yang}, \citenamefont {Ren}, \citenamefont {Khanikaev}, \citenamefont {Alù}, \citenamefont {Brasselet}, \citenamefont {Shats}, \citenamefont
  {Verbeeck}, \citenamefont {Schattschneider}, \citenamefont {Sarenac}, \citenamefont {Cory}, \citenamefont {Pushin}, \citenamefont {Birk}, \citenamefont {Gorlach}, \citenamefont {Kaminer}, \citenamefont {Cardano}, \citenamefont {Marrucci}, \citenamefont {Krenn},\ and\ \citenamefont {Marquardt}}]{bliokh2023RoadmapStructured}%
  \BibitemOpen
  \bibfield  {author} {\bibinfo {author} {\bibfnamefont {K.~Y.}\ \bibnamefont {Bliokh}}, \bibinfo {author} {\bibfnamefont {E.}~\bibnamefont {Karimi}}, \bibinfo {author} {\bibfnamefont {M.~J.}\ \bibnamefont {Padgett}}, \bibinfo {author} {\bibfnamefont {M.~A.}\ \bibnamefont {Alonso}}, \bibinfo {author} {\bibfnamefont {M.~R.}\ \bibnamefont {Dennis}}, \bibinfo {author} {\bibfnamefont {A.}~\bibnamefont {Dudley}}, \bibinfo {author} {\bibfnamefont {A.}~\bibnamefont {Forbes}}, \bibinfo {author} {\bibfnamefont {S.}~\bibnamefont {Zahedpour}}, \bibinfo {author} {\bibfnamefont {S.~W.}\ \bibnamefont {Hancock}}, \bibinfo {author} {\bibfnamefont {H.~M.}\ \bibnamefont {Milchberg}}, \bibinfo {author} {\bibfnamefont {S.}~\bibnamefont {Rotter}}, \bibinfo {author} {\bibfnamefont {F.}~\bibnamefont {Nori}}, \bibinfo {author} {\bibfnamefont {{\c S}.~K.}\ \bibnamefont {Özdemir}}, \bibinfo {author} {\bibfnamefont {N.}~\bibnamefont {Bender}}, \bibinfo {author} {\bibfnamefont {H.}~\bibnamefont {Cao}}, \bibinfo {author} {\bibfnamefont
  {P.~B.}\ \bibnamefont {Corkum}}, \bibinfo {author} {\bibfnamefont {C.}~\bibnamefont {Hernández-García}}, \bibinfo {author} {\bibfnamefont {H.}~\bibnamefont {Ren}}, \bibinfo {author} {\bibfnamefont {Y.}~\bibnamefont {Kivshar}}, \bibinfo {author} {\bibfnamefont {M.~G.}\ \bibnamefont {Silveirinha}}, \bibinfo {author} {\bibfnamefont {N.}~\bibnamefont {Engheta}}, \bibinfo {author} {\bibfnamefont {A.}~\bibnamefont {Rauschenbeutel}}, \bibinfo {author} {\bibfnamefont {P.}~\bibnamefont {Schneeweiss}}, \bibinfo {author} {\bibfnamefont {J.}~\bibnamefont {Volz}}, \bibinfo {author} {\bibfnamefont {D.}~\bibnamefont {Leykam}}, \bibinfo {author} {\bibfnamefont {D.~A.}\ \bibnamefont {Smirnova}}, \bibinfo {author} {\bibfnamefont {K.}~\bibnamefont {Rong}}, \bibinfo {author} {\bibfnamefont {B.}~\bibnamefont {Wang}}, \bibinfo {author} {\bibfnamefont {E.}~\bibnamefont {Hasman}}, \bibinfo {author} {\bibfnamefont {M.~F.}\ \bibnamefont {Picardi}}, \bibinfo {author} {\bibfnamefont {A.~V.}\ \bibnamefont {Zayats}}, \bibinfo {author}
  {\bibfnamefont {F.~J.}\ \bibnamefont {Rodríguez-Fortuño}}, \bibinfo {author} {\bibfnamefont {C.}~\bibnamefont {Yang}}, \bibinfo {author} {\bibfnamefont {J.}~\bibnamefont {Ren}}, \bibinfo {author} {\bibfnamefont {A.~B.}\ \bibnamefont {Khanikaev}}, \bibinfo {author} {\bibfnamefont {A.}~\bibnamefont {Alù}}, \bibinfo {author} {\bibfnamefont {E.}~\bibnamefont {Brasselet}}, \bibinfo {author} {\bibfnamefont {M.}~\bibnamefont {Shats}}, \bibinfo {author} {\bibfnamefont {J.}~\bibnamefont {Verbeeck}}, \bibinfo {author} {\bibfnamefont {P.}~\bibnamefont {Schattschneider}}, \bibinfo {author} {\bibfnamefont {D.}~\bibnamefont {Sarenac}}, \bibinfo {author} {\bibfnamefont {D.~G.}\ \bibnamefont {Cory}}, \bibinfo {author} {\bibfnamefont {D.~A.}\ \bibnamefont {Pushin}}, \bibinfo {author} {\bibfnamefont {M.}~\bibnamefont {Birk}}, \bibinfo {author} {\bibfnamefont {A.}~\bibnamefont {Gorlach}}, \bibinfo {author} {\bibfnamefont {I.}~\bibnamefont {Kaminer}}, \bibinfo {author} {\bibfnamefont {F.}~\bibnamefont {Cardano}}, \bibinfo
  {author} {\bibfnamefont {L.}~\bibnamefont {Marrucci}}, \bibinfo {author} {\bibfnamefont {M.}~\bibnamefont {Krenn}}, \ and\ \bibinfo {author} {\bibfnamefont {F.}~\bibnamefont {Marquardt}},\ }\bibfield  {title} {\enquote {\bibinfo {title} {Roadmap on structured waves},}\ }\href {\doibase 10.1088/2040-8986/acea92} {\bibfield  {journal} {\bibinfo  {journal} {Journal of Optics}\ }\textbf {\bibinfo {volume} {25}},\ \bibinfo {pages} {103001} (\bibinfo {year} {2023})}\BibitemShut {NoStop}%
\bibitem [{\citenamefont {Nape}\ \emph {et~al.}(2023)\citenamefont {Nape}, \citenamefont {Sephton}, \citenamefont {Ornelas}, \citenamefont {Moodley},\ and\ \citenamefont {Forbes}}]{forbes2023}%
  \BibitemOpen
  \bibfield  {author} {\bibinfo {author} {\bibfnamefont {I.}~\bibnamefont {Nape}}, \bibinfo {author} {\bibfnamefont {B.}~\bibnamefont {Sephton}}, \bibinfo {author} {\bibfnamefont {P.}~\bibnamefont {Ornelas}}, \bibinfo {author} {\bibfnamefont {C.}~\bibnamefont {Moodley}}, \ and\ \bibinfo {author} {\bibfnamefont {A.}~\bibnamefont {Forbes}},\ }\bibfield  {title} {\enquote {\bibinfo {title} {Quantum structured light in high dimensions},}\ }\href {\doibase 10.1103/PhysRevA.64.052312} {\bibfield  {journal} {\bibinfo  {journal} {APL Photonics}\ }\textbf {\bibinfo {volume} {8}},\ \bibinfo {pages} {051101} (\bibinfo {year} {2023})}\BibitemShut {NoStop}%
\bibitem [{\citenamefont {Bao}\ \emph {et~al.}(2023)\citenamefont {Bao}, \citenamefont {Yu}, \citenamefont {Anderegg}, \citenamefont {Chae}, \citenamefont {Ketterle}, \citenamefont {Ni},\ and\ \citenamefont {Doyle}}]{bao2023dipolar}%
  \BibitemOpen
  \bibfield  {author} {\bibinfo {author} {\bibfnamefont {Y.}~\bibnamefont {Bao}}, \bibinfo {author} {\bibfnamefont {S.~S.}\ \bibnamefont {Yu}}, \bibinfo {author} {\bibfnamefont {L.}~\bibnamefont {Anderegg}}, \bibinfo {author} {\bibfnamefont {E.}~\bibnamefont {Chae}}, \bibinfo {author} {\bibfnamefont {W.}~\bibnamefont {Ketterle}}, \bibinfo {author} {\bibfnamefont {K.-K.}\ \bibnamefont {Ni}}, \ and\ \bibinfo {author} {\bibfnamefont {J.~M.}\ \bibnamefont {Doyle}},\ }\bibfield  {title} {\enquote {\bibinfo {title} {Dipolar spin-exchange and entanglement between molecules in an optical tweezer array},}\ }\href@noop {} {\bibfield  {journal} {\bibinfo  {journal} {Science}\ }\textbf {\bibinfo {volume} {382}},\ \bibinfo {pages} {1138--1143} (\bibinfo {year} {2023})}\BibitemShut {NoStop}%
\bibitem [{\citenamefont {Holland}, \citenamefont {Lu},\ and\ \citenamefont {Cheuk}(2023)}]{holland2023demand}%
  \BibitemOpen
  \bibfield  {author} {\bibinfo {author} {\bibfnamefont {C.~M.}\ \bibnamefont {Holland}}, \bibinfo {author} {\bibfnamefont {Y.}~\bibnamefont {Lu}}, \ and\ \bibinfo {author} {\bibfnamefont {L.~W.}\ \bibnamefont {Cheuk}},\ }\bibfield  {title} {\enquote {\bibinfo {title} {On-demand entanglement of molecules in a reconfigurable optical tweezer array},}\ }\href@noop {} {\bibfield  {journal} {\bibinfo  {journal} {Science}\ }\textbf {\bibinfo {volume} {382}},\ \bibinfo {pages} {1143--1147} (\bibinfo {year} {2023})}\BibitemShut {NoStop}%
\bibitem [{\citenamefont {{Kort-Kamp}}, \citenamefont {Azad},\ and\ \citenamefont {Dalvit}(2021)}]{kort-kamp2021SpaceTimeQuantum}%
  \BibitemOpen
  \bibfield  {author} {\bibinfo {author} {\bibfnamefont {W.~J.~M.}\ \bibnamefont {{Kort-Kamp}}}, \bibinfo {author} {\bibfnamefont {A.~K.}\ \bibnamefont {Azad}}, \ and\ \bibinfo {author} {\bibfnamefont {D.~A.~R.}\ \bibnamefont {Dalvit}},\ }\bibfield  {title} {\enquote {\bibinfo {title} {Space-time quantum metasurfaces},}\ }\href {\doibase 10.1103/PhysRevLett.127.043603} {\bibfield  {journal} {\bibinfo  {journal} {Physical Review Letters}\ }\textbf {\bibinfo {volume} {127}},\ \bibinfo {pages} {043603} (\bibinfo {year} {2021})}\BibitemShut {NoStop}%
\bibitem [{\citenamefont {Uriri}\ \emph {et~al.}(2018)\citenamefont {Uriri}, \citenamefont {Tashima}, \citenamefont {Zhang}, \citenamefont {Asano}, \citenamefont {Bechu}, \citenamefont {G{\"u}ney}, \citenamefont {Yamamoto}, \citenamefont {{\"O}zdemir}, \citenamefont {Wegener},\ and\ \citenamefont {Tame}}]{uriri2018ActiveControl}%
  \BibitemOpen
  \bibfield  {author} {\bibinfo {author} {\bibfnamefont {S.~A.}\ \bibnamefont {Uriri}}, \bibinfo {author} {\bibfnamefont {T.}~\bibnamefont {Tashima}}, \bibinfo {author} {\bibfnamefont {X.}~\bibnamefont {Zhang}}, \bibinfo {author} {\bibfnamefont {M.}~\bibnamefont {Asano}}, \bibinfo {author} {\bibfnamefont {M.}~\bibnamefont {Bechu}}, \bibinfo {author} {\bibfnamefont {D.~{\"O}.}\ \bibnamefont {G{\"u}ney}}, \bibinfo {author} {\bibfnamefont {T.}~\bibnamefont {Yamamoto}}, \bibinfo {author} {\bibfnamefont {{\c S}.~K.}\ \bibnamefont {{\"O}zdemir}}, \bibinfo {author} {\bibfnamefont {M.}~\bibnamefont {Wegener}}, \ and\ \bibinfo {author} {\bibfnamefont {M.~S.}\ \bibnamefont {Tame}},\ }\bibfield  {title} {\enquote {\bibinfo {title} {Active control of a plasmonic metamaterial for quantum state engineering},}\ }\href {\doibase 10.1103/PhysRevA.97.053810} {\bibfield  {journal} {\bibinfo  {journal} {Physical Review A}\ }\textbf {\bibinfo {volume} {97}},\ \bibinfo {pages} {053810} (\bibinfo {year} {2018})}\BibitemShut
  {NoStop}%
\bibitem [{\citenamefont {Roger}\ \emph {et~al.}(2015)\citenamefont {Roger}, \citenamefont {Vezzoli}, \citenamefont {Bolduc}, \citenamefont {Valente}, \citenamefont {Heitz}, \citenamefont {Jeffers}, \citenamefont {Soci}, \citenamefont {Leach}, \citenamefont {Couteau}, \citenamefont {Zheludev},\ and\ \citenamefont {Faccio}}]{roger2015CoherentPerfect}%
  \BibitemOpen
  \bibfield  {author} {\bibinfo {author} {\bibfnamefont {T.}~\bibnamefont {Roger}}, \bibinfo {author} {\bibfnamefont {S.}~\bibnamefont {Vezzoli}}, \bibinfo {author} {\bibfnamefont {E.}~\bibnamefont {Bolduc}}, \bibinfo {author} {\bibfnamefont {J.}~\bibnamefont {Valente}}, \bibinfo {author} {\bibfnamefont {J.~J.~F.}\ \bibnamefont {Heitz}}, \bibinfo {author} {\bibfnamefont {J.}~\bibnamefont {Jeffers}}, \bibinfo {author} {\bibfnamefont {C.}~\bibnamefont {Soci}}, \bibinfo {author} {\bibfnamefont {J.}~\bibnamefont {Leach}}, \bibinfo {author} {\bibfnamefont {C.}~\bibnamefont {Couteau}}, \bibinfo {author} {\bibfnamefont {N.~I.}\ \bibnamefont {Zheludev}}, \ and\ \bibinfo {author} {\bibfnamefont {D.}~\bibnamefont {Faccio}},\ }\bibfield  {title} {\enquote {\bibinfo {title} {Coherent perfect absorption in deeply subwavelength films in the single-photon regime},}\ }\href {\doibase 10.1038/ncomms8031} {\bibfield  {journal} {\bibinfo  {journal} {Nature Communications}\ }\textbf {\bibinfo {volume} {6}},\ \bibinfo {pages}
  {7031} (\bibinfo {year} {2015})}\BibitemShut {NoStop}%
\bibitem [{\citenamefont {Altuzarra}\ \emph {et~al.}(2017)\citenamefont {Altuzarra}, \citenamefont {Vezzoli}, \citenamefont {Valente}, \citenamefont {Gao}, \citenamefont {Soci}, \citenamefont {Faccio},\ and\ \citenamefont {Couteau}}]{altuzarra2017CoherentPerfect}%
  \BibitemOpen
  \bibfield  {author} {\bibinfo {author} {\bibfnamefont {C.}~\bibnamefont {Altuzarra}}, \bibinfo {author} {\bibfnamefont {S.}~\bibnamefont {Vezzoli}}, \bibinfo {author} {\bibfnamefont {J.}~\bibnamefont {Valente}}, \bibinfo {author} {\bibfnamefont {W.}~\bibnamefont {Gao}}, \bibinfo {author} {\bibfnamefont {C.}~\bibnamefont {Soci}}, \bibinfo {author} {\bibfnamefont {D.}~\bibnamefont {Faccio}}, \ and\ \bibinfo {author} {\bibfnamefont {C.}~\bibnamefont {Couteau}},\ }\bibfield  {title} {\enquote {\bibinfo {title} {Coherent perfect absorption in metamaterials with entangled photons},}\ }\href {\doibase 10.1021/acsphotonics.7b00514} {\bibfield  {journal} {\bibinfo  {journal} {ACS Photonics}\ }\textbf {\bibinfo {volume} {4}},\ \bibinfo {pages} {2124--2128} (\bibinfo {year} {2017})}\BibitemShut {NoStop}%
\bibitem [{\citenamefont {Lyons}\ \emph {et~al.}(2019)\citenamefont {Lyons}, \citenamefont {Oren}, \citenamefont {Roger}, \citenamefont {Savinov}, \citenamefont {Valente}, \citenamefont {Vezzoli}, \citenamefont {Zheludev}, \citenamefont {Segev},\ and\ \citenamefont {Faccio}}]{lyons2019CoherentMetamaterial}%
  \BibitemOpen
  \bibfield  {author} {\bibinfo {author} {\bibfnamefont {A.}~\bibnamefont {Lyons}}, \bibinfo {author} {\bibfnamefont {D.}~\bibnamefont {Oren}}, \bibinfo {author} {\bibfnamefont {T.}~\bibnamefont {Roger}}, \bibinfo {author} {\bibfnamefont {V.}~\bibnamefont {Savinov}}, \bibinfo {author} {\bibfnamefont {J.}~\bibnamefont {Valente}}, \bibinfo {author} {\bibfnamefont {S.}~\bibnamefont {Vezzoli}}, \bibinfo {author} {\bibfnamefont {N.~I.}\ \bibnamefont {Zheludev}}, \bibinfo {author} {\bibfnamefont {M.}~\bibnamefont {Segev}}, \ and\ \bibinfo {author} {\bibfnamefont {D.}~\bibnamefont {Faccio}},\ }\bibfield  {title} {\enquote {\bibinfo {title} {Coherent metamaterial absorption of two-photon states with 40\% efficiency},}\ }\href {\doibase 10.1103/PhysRevA.99.011801} {\bibfield  {journal} {\bibinfo  {journal} {Physical Review A}\ }\textbf {\bibinfo {volume} {99}},\ \bibinfo {pages} {011801} (\bibinfo {year} {2019})}\BibitemShut {NoStop}%
\bibitem [{\citenamefont {Ruostekoski}(2023)}]{janne2023}%
  \BibitemOpen
  \bibfield  {author} {\bibinfo {author} {\bibfnamefont {J.}~\bibnamefont {Ruostekoski}},\ }\bibfield  {title} {\enquote {\bibinfo {title} {Cooperative quantum-optical planar arrays of atoms},}\ }\href {\doibase 10.1103/PhysRevA.64.052312} {\bibfield  {journal} {\bibinfo  {journal} {Physical Review A}\ }\textbf {\bibinfo {volume} {108}},\ \bibinfo {pages} {030101} (\bibinfo {year} {2023})}\BibitemShut {NoStop}%
\bibitem [{\citenamefont {Knall}\ \emph {et~al.}(2022)\citenamefont {Knall}, \citenamefont {Knaut}, \citenamefont {Assumpcao}, \citenamefont {Stroganov}, \citenamefont {Gong}, \citenamefont {Huan}, \citenamefont {Machielse}, \citenamefont {Levonian}, \citenamefont {Riedinger}, \citenamefont {Park}, \citenamefont {Lončar}, \citenamefont {Bhaskar},\ and\ \citenamefont {Lukin}}]{knall2022}%
  \BibitemOpen
  \bibfield  {author} {\bibinfo {author} {\bibfnamefont {E.}~\bibnamefont {Knall}}, \bibinfo {author} {\bibfnamefont {R.}~\bibnamefont {Knaut}, \bibfnamefont {C.M. andd~Bekenstein}}, \bibinfo {author} {\bibfnamefont {D.}~\bibnamefont {Assumpcao}}, \bibinfo {author} {\bibfnamefont {P.}~\bibnamefont {Stroganov}}, \bibinfo {author} {\bibfnamefont {W.}~\bibnamefont {Gong}}, \bibinfo {author} {\bibfnamefont {P.}~\bibnamefont {Huan}, \bibfnamefont {Y.Q.~Stas}}, \bibinfo {author} {\bibfnamefont {B.}~\bibnamefont {Machielse}, \bibfnamefont {Chalupnik}}, \bibinfo {author} {\bibfnamefont {D.}~\bibnamefont {Levonian}, \bibfnamefont {Suleymanzade}}, \bibinfo {author} {\bibfnamefont {R.}~\bibnamefont {Riedinger}}, \bibinfo {author} {\bibfnamefont {H.}~\bibnamefont {Park}}, \bibinfo {author} {\bibfnamefont {M.}~\bibnamefont {Lončar}}, \bibinfo {author} {\bibfnamefont {M.}~\bibnamefont {Bhaskar}}, \ and\ \bibinfo {author} {\bibfnamefont {M.}~\bibnamefont {Lukin}},\ }\bibfield  {title} {\enquote {\bibinfo {title} {Efficient
  source of shaped single photons based on an integrated diamond nanophotonic system},}\ }\href {\doibase 10.1103/PhysRevA.64.052312} {\bibfield  {journal} {\bibinfo  {journal} {Physical Review Letters}\ }\textbf {\bibinfo {volume} {129}},\ \bibinfo {pages} {053603} (\bibinfo {year} {2022})}\BibitemShut {NoStop}%
\end{thebibliography}%

\end{document}